\documentclass[journal]{IEEEtran}
\usepackage{amsmath}
\usepackage{cite}
\usepackage{amssymb}
\usepackage{amsfonts}
\usepackage{algorithm}
\usepackage{dsfont}
\usepackage{graphicx}
\usepackage{epsfig}
\usepackage{subfigure}
\usepackage{psfrag}
\usepackage{xcolor}
\usepackage{multirow}

\title{{Uplink-Downlink Duality Between Multiple-Access and Broadcast Channels with Compressing Relays}
\author{\IEEEauthorblockN{Liang~Liu, \IEEEmembership{Member, IEEE}, Ya-Feng Liu, \IEEEmembership{Senior Member, IEEE}, Pratik Patil, and Wei Yu, \IEEEmembership{Fellow, IEEE}}
\thanks{
Manuscript received August 24, 2020; revised April 20, 2021; accepted July 28, 2021.
This work was supported
in part by Natural Sciences and Engineering Research Council (NSERC) of Canada via
the Canada Research Chairs program,
in part by The Hong Kong
Polytechnic University under Grant P0030001, and
in part by the National Natural Science Foundation of China (NSFC) under Grant 12022116, Grant 11688101, Grant 12021001, Grant 11991020, and Grant 11991021.
The article was presented in part at the {\it IEEE International Symposium on Information Theory (ISIT)}, July 2016. ({\it Corresponding author: Ya-Feng Liu})}
\thanks{Liang Liu is with the Department of Electronic and Information Engineering, The Hong Kong Polytechnic University, Hong Kong SAR, China (e-mail: liang-eie.liu@polyu.edu.hk).}
\thanks{Ya-Feng Liu is with the State Key Laboratory of Scientific and Engineering Computing, Institute of Computational Mathematics and Scientific/Engineering Computing, Academy of Mathematics and Systems Science, Chinese Academy of Sciences, Beijing 100190, China (e-mail:yafliu@lsec.cc.ac.cn).}
\thanks{Pratik Patil was with The Edward S. Rogers Sr. Department of Electrical and Computer Engineering
the University of Toronto; he is now with the Department of Statistics and Data Science and the Machine Learning Department, Carnegie Mellon University, Pittsburgh, PA 15213 USA (e-mail: pratik@cmu.edu).}
\thanks{Wei Yu is with The Edward S. Rogers Sr. Department of Electrical and Computer Engineering, University of Toronto, 10 King's College Road, Toronto, Ontario M5S3G4, Canada
(e-mail: weiyu@ece.utoronto.ca).}
}
}

\begin{document}
\maketitle \thispagestyle{empty} 

%
%

\newtheorem{definition}{Definition}
\newtheorem{assumption}{Assumption}
\newtheorem{lemma}{Lemma}
\newtheorem{example}{Example}
\newtheorem{theorem}{Theorem}
\newtheorem{proposition}{Proposition}
\newtheorem{conjecture}{Conjecture}
\newtheorem{remark}{Remark}
\newcommand{\mv}[1]{\mbox{\boldmath{$ #1 $}}}

\begin{abstract}
Uplink-downlink duality refers to the fact that under a sum-power constraint,
the capacity regions of a Gaussian multiple-access channel and a Gaussian
broadcast channel with Hermitian transposed channel matrices are identical. This
paper generalizes this result to a cooperative cellular network, in which remote access-points are deployed as relays in serving the users under the coordination of a central processor
(CP). In this model, the users and the relays are connected over noisy wireless links,
while the relays and the CP are connected over noiseless but rate-limited
fronthaul links.  Based on a Lagrangian technique, this paper
establishes a duality relationship between such a multiple-access relay channel and broadcast relay channel,
under the assumption that the relays use compression-based strategies.
Specifically, we show that under the same total transmit power
constraint and individual fronthaul rate constraints, the achievable rate
regions of the Gaussian multiple-access and broadcast relay channels are
identical, when either independent compression or Wyner-Ziv and multivariate compression strategies are used.
The key observations are that if the beamforming vectors at the relays are fixed, the sum-power
minimization problems under the achievable rate and fronthaul constraints
in both the uplink and the downlink can be transformed into either a linear programming
or a semidefinite programming problem depending on the compression
technique, and that the uplink and downlink problems are Lagrangian duals of each other.
Moreover, the dual variables corresponding to the downlink
rate constraints become the uplink powers; the dual
variables corresponding to the downlink fronthaul constraints become the uplink quantization noises. This duality relationship enables an efficient algorithm for optimizing the downlink transmission and relaying strategies based on the uplink.
\end{abstract}


\vspace{-10pt}
\section{Introduction}\label{sec:Introduction}

There is a curious uplink-downlink duality between the Gaussian multiple-access
channel with a multiple-antenna receiver and the Gaussian broadcast channel with
a multiple-antenna transmitter --- under the same total power constraint, the uplink
and downlink achievable rate regions with linear processing, or alternatively the uplink and downlink capacity regions with optimal nonlinear processing, are
identical \cite{duality1}.  While the traditional multiple-access and broadcast
channel models are suited for an isolated wireless cellular system with a single
base-station
(BS), in this paper, we are motivated by a generalization of this model to
cooperative cellular networks in which BSs cooperate over \emph{rate-limited}
digital links to a central processor (CP) in communicating with the users.
In this model, the BSs in effect act as remote radio-heads with finite-capacity
fronthaul links and function as relays between the CP and the users.  The aim
of this paper is to establish an uplink-downlink duality between achievable
rate regions of such a
Gaussian multiple-access \emph{relay} channel and a Gaussian broadcast
\emph{relay} channel.

\begin{figure}[t]
  \centering
  \includegraphics[width=9cm]{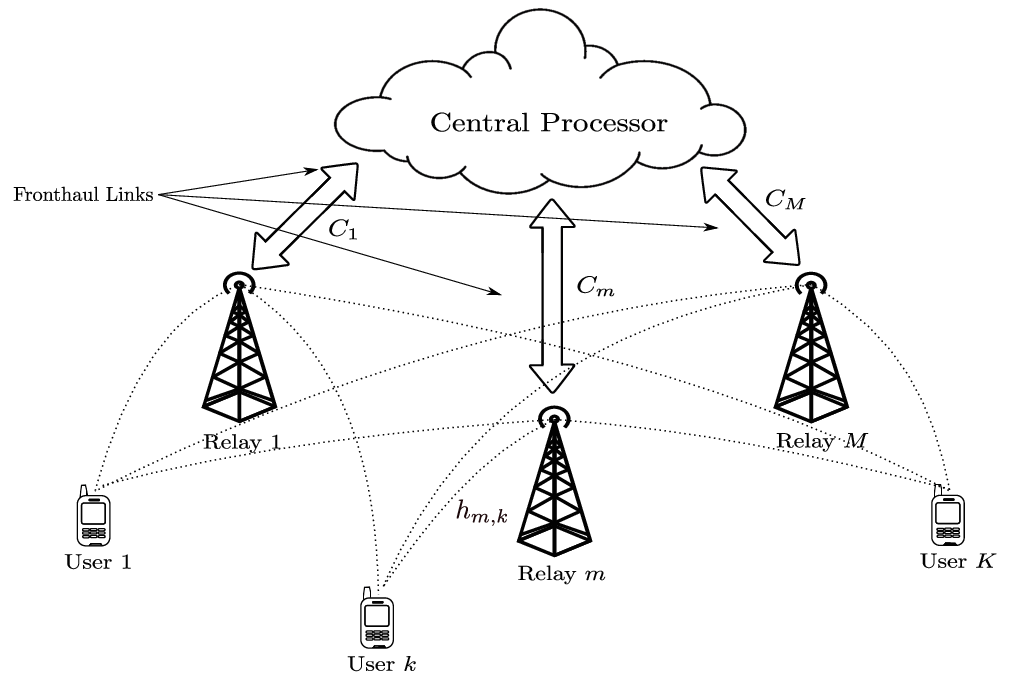} 
  \caption{Cooperative cellular network in which the relay-like BSs are connected to the CP via rate-limited fronthaul links and the user messages are jointly encoded/decoded at the CP.}
\label{fig1}
\vspace{-10pt}
\end{figure}

The centralized cooperative communication architecture, in which multiple
relay-like BSs cooperatively serve the users under the coordination of a
CP, is an appealing solution to the ever-increasing demand
for mobile broadband in future wireless communication networks, because of its
ability to mitigate intercell interference. Under the above architecture, the
users and the relay-like BSs are connected by noisy wireless channels,
while the relay-like BSs and the CP are connected by noiseless fronthaul
links of finite rate limit, as shown in Fig.~\ref{fig1}. In the uplink, the
users transmit their signals and the relay-like BSs forward their received signals
to the CP for joint information decoding, while in the downlink, the CP jointly
encodes the user messages and sends them to the relay-like BSs to transmit to
the users. Because the CP can jointly decode and encode the user messages, this
cooperative architecture can effectively utilize cross-cell links to enhance
message transmission, instead of treating signals from neighboring cells as
interference, thus enabling a significant improvement in the overall network
throughput. In the literature, there are several terminologies to describe the
above centralized cellular architecture, including coordinated multipoint
(CoMP) \cite{Irmer11}, distributed antenna system (DAS) \cite{Kerpez96}, cloud
radio access network (C-RAN) \cite{Simeone16}, cell-free massive multiple-input
multiple-output (MIMO) \cite{Ngo17}, etc. All of these systems can be modeled
by the multiple-access relay channel in the uplink and the broadcast relay
channel in the downlink as discussed above, where the hop between the users and
the relays is wireless, while the hop between the relays and the CP is digital.

When the capacities of all the fronthaul links are infinite, the above model
reduces to the traditional multiple-access and broadcast channels. In this
case, the relays can simply be treated as remote antennas of a single virtual
BS over the entire network.
From the existing literature, we know that there exists a duality relationship
between the multiple-access channel and the broadcast channel
\cite{duality6,duality7,duality8,duality11,duality9,duality3,duality1,duality4},
which states that given the same sum-power constraint, any rate-tuple
achievable in the uplink is also achievable in the downlink, and vice versa.
This duality holds under both linear and nonlinear receiving/precoding at
the BS. We now ask the following question: If the capacity of the fronthaul
links is finite, does a similar uplink-downlink duality relationship hold? The answer to
this question depends on not only the joint processing scheme at the CP, but
also the relaying strategies employed at the BSs. In practical implementations,
a variety of ways of jointly optimizing the utilization of fronthaul and the
wireless links have been proposed, e.g., for
CoMP \cite{Marsch11,Yu13}, DAS \cite{Gerhard07}, C-RAN
\cite{Zhou14,Yu16,Zhou16,Liu15,Liang15,Simeone13}, and cell-free massive MIMO
\cite{Bashar18,Marzetta17a,Marzetta17b}.
This paper provides an affirmative answer to the above question in the sense
that if the compression-based strategies are utilized over the fronthaul links,
then indeed the achievable rate regions of the Gaussian multiple-access relay
channel and the Gaussian broadcast relay channel are identical under the same
sum-power constraint and individual fronthaul capacity constraints. This
duality relationship holds with either independent compression \cite{Zhou14,Liu15,Liang15,dai2014sparse} or
Wyner-Ziv/multivariate compression \cite{Zhou14,Yu16,Zhou16,Simeone13} at the relays, and with either linear or nonlinear
processing at the CP.

\vspace{-5pt}
\subsection{Prior Works on Uplink-Downlink Duality}

\subsubsection{Linear Encoding and Linear Decoding}
The uplink-downlink duality between the multiple-access channel and the broadcast channel is first established in the case of linear encoding/decoding while treating interference as noise. Assuming single-antenna users and a multiple-antenna BS, the main result is that any signal-to-interference-plus-noise ratio (SINR) tuple that is achievable in the multiple-access channel can also be achieved in the broadcast channel under the same sum-power constraint, and vice versa. This duality has been proved in \cite{duality6,duality7,duality8,duality11,duality3} as follows. First, it is shown that if the receive beamforming vectors in the multiple-access channel and the transmit beamforming vectors in the broadcast channel are identical, then the feasibility conditions to ensure that an SINR-tuple can be achieved for both the multiple-access and the broadcast channels via power control are the same. The key observation is that a feasible power control solution exists if and only if the spectral radius of a so-called interference matrix is less than one \cite{powercontrol}, while given the same receive/transmit beamforming vectors and SINR targets, the spectral radii of the interference matrices of the multiple-access and the broadcast channels are the same.
Moreover, it is shown that given the same transmit/receive beamforming vectors, the minimum total transmit power to achieve a set of feasible SINR targets in the uplink is the same as that to achieve the same set of SINR targets in the downlink. As a result, the achievable SINR regions of the multiple-access and broadcast channels are identical under the same power constraint.


An alternative approach to proving the above duality result for the case of linear encoding and linear decoding is based on the Lagrangian duality technique \cite{duality9}. Specifically, given the receive/transmit beamforming vectors, the power control problems of minimizing the total transmit power subject to the users' individual SINR constraints in the multiple-access channel and the broadcast channel are both convex. Moreover, if the receive beamforming vectors and users' SINR targets in the multiple-access channel are the same as the transmit beamforming vectors and users' SINR targets in the broadcast channel, then the Lagrangian dual of the uplink sum-power minimization problem can be shown to be equivalent to the downlink sum-power minimization problem, and vice versa. This shows that the achievable SINR regions of the multiple-access channel and the broadcast channel are identical under the same sum-power constraint.

\subsubsection{Nonlinear Encoding and Nonlinear Decoding}
The uplink-downlink duality between the multiple-access channel and the broadcast channel is also established in the case of nonlinear encoding/decoding. Assuming again the case of single-antenna users, the main result is that the sum capacity (and also the capacity region) of the multiple-access channel, which is achieved by successive interference cancellation \cite{Cover}, is the same as the sum capacity (or the capacity region) of the broadcast channel, which is achieved by dirty-paper coding \cite{DPC}. Similar to the linear encoding/decoding case \cite{duality6,duality7,duality8,duality11}, it is shown in \cite{duality3} that if the decoding order for successive interference cancellation is the reverse of the encoding order for dirty-paper coding, and the uplink receive beamforming vectors are the same as the downlink transmit beamforming vectors, then the feasibility conditions to ensure that an SINR-tuple can be achieved in both the multiple-access channel and the broadcast channel via power control are the same. Moreover, it is shown in \cite{duality3} that if each user achieves the same SINR in the multiple-access channel and the broadcast channel, then the total transmit power in the uplink is the same as in the downlink. As a result, under the same sum-power constraint, the capacity region of the multiple-access channel is the same as the capacity region of the broadcast channel achieved by dirty-paper coding. This duality result can be extended \cite{duality1} even to the case where the users are equipped with multiple antennas, using a clever choice of the transmit covariance matrix for the broadcast channel to achieve each achievable rate-tuple in the multiple-access channel and vice versa. 

For the sum-capacity problem, there is also an alternative approach of establishing uplink-downlink duality based on the Lagrangian duality of a minimax problem. Along this line, \cite{duality4} shows that the sum capacity of the broadcast channel can be characterized by a minimax optimization problem, where the maximization is over the transmit covariance matrix, while the minimization is over the receive covariance matrix. Since this minimax problem is a convex problem, it is equivalent to its dual problem, which is another minimax problem in the multiple-access channel. Moreover, the optimal value of this new minimax problem is shown to be exactly the sum capacity of the multiple-access channel. As a result, the sum capacities of the multiple-access channel and the broadcast channel are the same.

\subsubsection{Other Duality Results}
Apart from the above results, the duality relationship is also established for the multiple-access channel and the broadcast channel under different power constraints and encoding/decoding strategies. For example, \cite{duality10} shows that the power minimization problem in the broadcast channel under the per-antenna power constraints is equivalent to the minimax problem in the multiple-access channel with an uncertain noise. Moreover, \cite{duality12} shows that the rate region of the broadcast channel achieved by dirty-paper coding and under multiple power constraints is the same as that of the multiple-access channel achieved by successive interference cancellation and under a weighted sum-power constraint. This result is generalized in \cite{An1,An2} to the uplink and downlink interference networks under multiple linear constraints. Further, \cite{duality14} shows that any sum rate achievable via integer-forcing in the MIMO multiple-access channel can be achieved via integer-forcing in the MIMO broadcast channel with the same sum-power and vice versa. In \cite{Tsung-Hui18}, the duality between the multiple-access channel and the broadcast channel is extended to the scenario with a full-duplex BS. 
Moreover, the duality relationship is also investigated for the multiple-access channel and the broadcast channel with amplify-and-forward relays. It is shown in \cite{duality16,duality13,duality15} that for both the two-hop and multihop relay scenarios, the user rate regions are the same in the uplink and downlink under the same sum-power constraint. This result can also be generalized to cases with multiple linear constraints \cite{An3}.
Finally, duality is used in \cite{Yafeng13} to characterize the polynomial-time solvability of a power control problem in the multiple-input single-output (MISO) and single-input multiple-output (SIMO) interference channels.

\subsection{Overview of Main Results}

This paper establishes a duality relationship between the multiple-access relay channel and the broadcast relay channel when the compression-based relay strategies are used over the rate-limited fronthaul links between the CP and the relays. Both the users and the relays are assumed to be equipped with a single antenna. In the uplink, each relay compresses its received signals from the users, and sends the compressed signal to the CP via the fronthaul link. The CP first decompresses the signals from the relays, then jointly decodes the user messages based on the decompressed signals. In the downlink, the CP jointly encodes the user messages, compresses the transmit signals for the relays, and sends the compressed signals to the relays via the fronthaul links. Then, each relay decompresses its received signal and transmits it to the users. Compared to the classic uplink-downlink duality results in the literature, the novel contributions of our work are as follows.

We show that under the same sum-power constraint and individual fronthaul capacity constraints, the achievable rate regions of the multiple-access channel and the broadcast channel are identical using compression-based relays, under the following four cases:
    \begin{itemize}
    \item[I:] In the uplink, the relays compress their received signals independently and the CP applies the linear decoding strategy. In the downlink, the CP applies the linear encoding strategy and compresses the signals for the relays independently.
    \item[II:] In the uplink, the relays compress their received signals independently and the CP applies the successive interference cancellation strategy. In the downlink, the CP applies the dirty-paper coding strategy and compresses the signals for the relays independently.
    \item[III:] In the uplink, the relays apply the Wyner-Ziv compression strategy to compress their received signals and the CP applies the linear decoding strategy. In the downlink, the CP applies the linear encoding strategy and the multivariate compression strategy to compress the signals for the relays.
    \item[IV:] In the uplink, the relays apply the Wyner-Ziv compression strategy to compress their received signals and the CP applies the successive interference cancellation strategy. In the downlink, the CP applies the dirty-paper coding strategy and the multivariate compression strategy to compress the signals for the relays.
    \end{itemize}
    Note that the conventional uplink-downlink duality relationship between the multiple-access channel and the broadcast channel \cite{duality6,duality7,duality8,duality11,duality9,duality3,duality1,duality4} is a special case of the duality relationship established in this work if we assume the fronthaul links all have infinite capacities.

For Cases I and II with independent compression for the relays, we provide an alternative proof for the duality relationship as compared to our previous work \cite{Liang16}. Specifically, the duality relationship is validated based on the Lagrangian duality \cite{Boyd04}, which provides a unified approach for all the cases. In particular, we show that given the transmit beamforming vectors in the downlink, the sum-power minimization problem subject to the individual user rate constraints and individual fronthaul capacity constraints is a linear program (LP) with strong duality. Then, it is shown that given the same receive and transmit beamforming vectors (and the reversed decoding order and encoding order for Case II) and under the same individual user rate constraints as well as individual fronthaul capacity constraints, the uplink sum-power minimization problem is equivalent to the Lagrangian dual of the downlink sum-power minimization problem. This approach is similar to that used in \cite{duality9} to verify the duality of the conventional multiple-access channel and broadcast channel without relays. However, interesting new insights can be obtained when relays are deployed between the CP and the users.
Specifically, the dual variables associated with the user rate constraints and
the fronthaul capacity constraints in the downlink sum-power minimization problem
play the role of uplink user transmit powers and uplink relay quantization noise
levels, respectively, in the dual problem.

For Cases III and IV, we establish a novel duality relationship between Wyner-Ziv compression and multivariate compression \cite{Gamal}. Intuitively, Wyner-Ziv compression over the noiseless fronthaul resembles successive interference cancellation in the noisy wireless channel in the sense that the decompressed signals can provide side information for decompressing the remaining signals. On the other hand, multivariate compression in the noiseless fronthaul resembles dirty-paper coding in the noisy wireless channel in the sense that it can control the interference caused by compression seen by the users. Despite the well-known duality between successive interference cancellation and dirty-paper coding, the relationship between these two compression strategies has not been established previously. We show in this work that if the decompression order in Wyner-Ziv compression is the reverse of the compression order in the multivariate compression, the uplink-downlink duality remains true between the multiple-access relay channel and the broadcast relay channel.

		To prove the above result, we use the Lagrangian duality approach similar to that taken in \cite{duality9,duality4}, rather than the approach of checking the feasibility conditions adopted in \cite{duality6,duality7,duality8,duality11,duality3}. This is because under the Wyner-Ziv and multivariate compression strategies, the fronthaul rates are complicated functions of the transmit powers and quantization noises. In this case, the fronthaul capacity constraints are no longer linear in the variables, and the feasibility condition proposed in \cite{powercontrol} for linear constraints does not work. Despite the complicated fronthaul rate expressions, we reveal that under the multivariate compression strategy in the broadcast relay channel, if the transmit beamforming vectors are fixed, the sum-power minimization problem subject to individual user rate constraints and individual fronthaul capacity constraints can be transformed into a convex optimization problem over the transmit powers as well as the compression noise covariance matrix. Then, we characterize the dual problem of this convex optimization. It turns out that if we interpret the dual variables associated with the user rate constraints as the uplink transmit powers and some diagonal elements of the dual matrices associated with the fronthaul capacity constraints as the uplink compression noise levels, then the Lagrangian dual of the broadcast relay channel problem is equivalent to the sum-power minimization problem in the multiple-access relay channel subject to individual user rate constraints as well as a single matrix inequality constraint. In contrast to Cases I and II with independent compression where the primal downlink problem is an LP and its dual problem directly has individual fronthaul capacity constraints, here the problem is a semidefinite program (SDP), and we need to take a further step to transform the single matrix inequality constraint into the individual fronthaul capacity constraints under the Wyner-Ziv compression strategy via a series of matrix operations. At the end, we show that at the optimal solution, all the user rate constraints and fronthaul capacity constraints in the dual problem are satisfied with equality, and moreover there exists a unique solution to this set of nonlinear equations. As a result, the dual of the broadcast relay channel problem is equivalent to 
the multiple-access relay channel problem.

\subsection{Organization}
The rest of this paper is organized as follows. Section \ref{sec:System Model} describes the system model and characterizes the achievable rate regions of the multiple-access relay channel and the broadcast relay channel under various encoding/decoding and compression/decompression strategies. In Section \ref{sec:Duality between Multiple-Access Channel and Broadcast Relay Channel}, the duality relationship between the multiple-access channel and the broadcast relay channel with compression-based relays is established. Sections \ref{sec 1} to \ref{sec 4} prove the duality for Cases I-IV, respectively, based on the Lagrangian duality theory.
We summarize the main duality relations in Section \ref{sec:Summary} and provide an application of duality in optimizing the broadcast relay channel via its dual multiple-access relay channel in Section \ref{sec:Application}.
We conclude this paper with Section \ref{sec:Conclusion}.

{\it Notation}: Scalars are denoted by lower-case letters; vectors are denoted by bold-face lower-case letters; matrices are denoted by bold-face upper-case letters. We use $\mv{I}$ to denote an identity matrix with an appropriate dimension, $\mv{0}_{x\times y}$ to denote a all-zero matrix with dimension of $x\times y$. For a matrix $\mv{A}$, $\mv{A}^{(x,y)}$ denotes the entry on the $x$-th row and the $y$-th column of $\mv{A}$, and $\mv{A}^{(x_1:y_1,x_2:y_2)}$ denotes a submatrix of $\mv{A}$ defined by
\begin{align*}
\mv{A}^{(x_1:y_1,x_2:y_2)}=\left[\begin{array}{ccc}\mv{A}^{(x_1,y_1)} & \cdots & \mv{A}^{(x_1,y_2)} \\ \vdots & \ddots & \vdots \\ \mv{A}^{(x_2,y_1)} & \cdots & \mv{A}^{(x_2,y_2)}\end{array}\right].
\end{align*}For a square full-rank matrix $\mv{S}$, $\mv{S}^{-1}$
denotes its inverse, and $\mv{S}\succeq \mv{0}$ or $\mv{S} \succ \mv{0}$ indicates that $\mv{S}$ is a positive semidefinite matrix or a positive definite matrix, respectively. For a matrix
$\mv{M}$ of an arbitrary size, $\mv{M}^{H}$, $\mv{M}^{T}$, and $\mv{M}^\ast$ denote the
conjugate transpose, transpose and conjugate of $\mv{M}$, respectively, and ${\rm rank}(\mv{M})$ denotes the rank of $\mv{M}$. We use ${\rm diag}(x_1,\ldots,x_K)$ to denote a diagonal matrix
with the diagonal elements given by $x_1,\ldots,x_K$. For two real vectors $\mv{x}=[x_1,\ldots,x_N]^T$ and $\mv{y}=[y_1,\ldots,y_N]^T$, $\mv{x}\geq \mv{y}$ means that $x_n\geq y_n$, $\forall n=1,\ldots,N$.

\section{System Model and Achievable Rate Regions}\label{sec:System Model}

\begin{figure*}
\begin{center}
\scalebox{0.5}{\includegraphics*{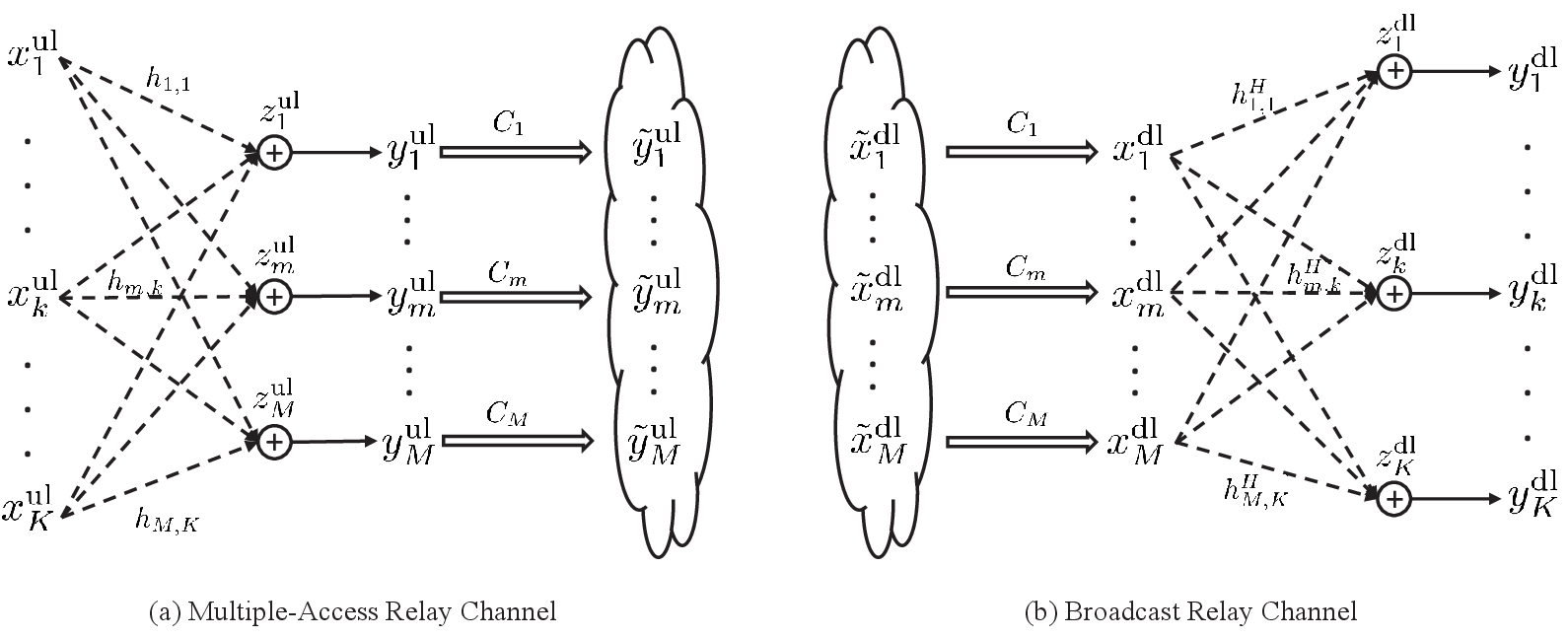}}
\end{center}
\caption{System model of the multiple-access relay channel and the broadcast relay channel.}\label{fig3}
\end{figure*}

The Gaussian multiple-access relay channel and the Gaussian broadcast relay channel considered in this paper consist of one CP, $M$ single-antenna relays, denoted by the set $\mathcal{M}=\{1,\ldots,M\}$, and $K$ single-antenna users, denoted by the set $\mathcal{K}=\{1,\ldots,K\}$, as shown in Fig.~\ref{fig3},
where each relay $m\in \mathcal{M}$ is connected to the users over the wireless channels and to the CP via the noiseless digital fronthaul link of capacity $C_m$ bits per symbol (bps). For the multiple-access relay channel, the overall channel from user $k\in \mathcal{K}$ to all the relays is denoted by
\begin{align}
\mv{h}_k=[h_{1,k},\ldots,h_{M,k}]^T, ~~~ \forall k\in \mathcal{K},
\end{align}where $h_{m,k}$ denotes the channel from user $k\in \mathcal{K}$ to relay $m\in \mathcal{M}$; in the dual broadcast relay channel, the overall channel from all the relays to user $k\in \mathcal{K}$ is the Hermitian transpose of the corresponding uplink channel, i.e., $\mv{h}_k^H$.
Further, we assume a sum-power constraint $P$ for all the users in the multiple-access relay channel,
and the same sum-power constraint $P$ for all the relays in the broadcast relay channel. In the following, we review the compression-based relaying strategies \cite{Simeone16, Yu_CRAN_book} for the multiple-access relay channel and the broadcast relay channel in detail.
These compression-based strategies are simplified versions of the more general relaying strategies for the multihop relay networks studied in \cite{Kim_NNC,Kim_DDF}.
In particular, the compression-based strategies considered in this paper take
the approach of separating the encoding/decoding of the relay codeword and
the encoding/decoding of the user messages, in contrast to the joint encoding/decoding
approach in \cite{Kim_NNC,Kim_DDF}.
In several specific cases \cite{Yu16,Yu19}, these simplified strategies can be shown to already achieve the capacity regions of the specific Gaussian multiple-access and broadcast relay channels to within a constant gap.

\subsection{Multiple-Access Channel with Compressing Relays}\label{sec:Multiple-Access Relay Channel}

The Gaussian multiple-access relay channel model is as shown in Fig.~\ref{fig3}(a). The discrete-time baseband channel between the users and the relays can be modelled as
\begin{align}\label{eqn:uplink signal}
\hspace{-8pt} \left[\begin{array}{c} y_{1}^{\rm ul} \\ \vdots \\ y_M^{\rm ul} \end{array} \right]\hspace{-2pt} =\hspace{-2pt} \left[\begin{array}{ccc}h_{1,1} & \cdots & h_{1,K} \\  \vdots & \ddots & \vdots \\ h_{M,1} & \cdots & h_{M,K} \end{array}\right]\left[\begin{array}{c}x_1^{\rm ul} \\ \vdots \\ x_K^{\rm ul} \end{array}\right]\hspace{-2pt}+\hspace{-2pt}\left[\begin{array}{c} z_1^{\rm ul} \\ \vdots \\ z_M^{\rm ul} \end{array}\right],\end{align}
where $x_k^{\rm ul}$ denotes the transmit signal of user $k$, $\forall k\in \mathcal{K}$, and $z_m^{\rm ul}\sim\mathcal{CN}(0,\sigma^2)$ denotes the additive white Gaussian noise (AWGN) at relay $m$, $\forall m\in \mathcal{M}$.

\setcounter{equation}{8}
\begin{figure*}
\normalsize
\begin{align}
 C_{\rho^{{\rm ul}}(m)}^{{\rm ul,WZ}}(\{p_k^{\rm ul}\},q_{\rho^{{\rm ul}}(1)}^{\rm ul},\ldots,q_{\rho^{{\rm ul}}(m)}^{\rm ul},\{\rho^{{\rm ul}}(m)\}) \nonumber & = I(\hat{y}_{\rho^{{\rm ul}}(m)}^{\rm ul};y_{\rho^{{\rm ul}}(m)}^{\rm ul}|\hat{y}_{\rho^{{\rm ul}}(1)}^{\rm ul},\ldots,\hat{y}_{\rho^{{\rm ul}}(m-1)}^{\rm ul}) \nonumber \\
	& = \log_2\frac{|\mv{\Gamma}_{\{\rho^{{\rm ul}}(m)\}}^{(1:m,1:m)}|}{|\mv{\Gamma}_{\{\rho^{{\rm ul}}(m)\}}^{(1:m-1,1:m-1)}|q_{\rho^{{\rm ul}}(m)}^{\rm ul}}   \label{eqn:uplink fronthaul rate}  \\
	& = \log_2 \frac{\mv{\Gamma}_{\{\rho^{{\rm ul}}(m)\}}^{(m,m)} - \mv{\Gamma}_{\{\rho^{{\rm ul}}(m)\}}^{(m,1:m-1)} (\mv{\Gamma}_{\{\rho^{{\rm ul}}(m)\}}^{(1:m-1,1:m-1)})^{-1} \mv{\Gamma}_{\{\rho^{{\rm ul}}(m)\}}^{(1:m-1,m)}}{q_{\rho^{{\rm ul}}(m)}^{\rm ul}}, ~~~ \forall m\in \mathcal{M}, \nonumber
\end{align} 
where 
\begin{align}
& \mv{\Gamma}_{\{\rho^{{\rm ul}}(m)\}}=\mathbb{E}\left[\tilde{\mv{y}}_{\{\rho^{{\rm ul}}(m)\}}^{\rm ul}\left(\tilde{\mv{y}}_{\{\rho^{{\rm ul}}(m)\}}^{\rm ul}\right)^H\right] =\sum\limits_{k=1}^K p_k^{\rm ul}\bar{\mv{h}}_{k,\{\rho^{{\rm ul}}(m)\}}(\bar{\mv{h}}_{k,\{\rho^{{\rm ul}}(m)\}})^H+\sigma^2\mv{I} +{\rm diag}(q_{\rho^{{\rm ul}}(1)}^{{\rm ul}},\ldots,q_{\rho^{{\rm ul}}(M)}^{{\rm ul}}), \label{eqn:Gamma} \\
& \tilde{\mv{y}}_{\{\rho^{{\rm ul}}(m)\}}^{\rm ul}=[\tilde{y}_{\rho^{{\rm ul}}(1)}^{\rm ul},\ldots,\tilde{y}_{\rho^{{\rm ul}}(M)}^{\rm ul}]^T, \\
& \bar{\mv{h}}_{k,\{\rho^{{\rm ul}}(m)\}}=[h_{\rho^{{\rm ul}}(1),k},\ldots,h_{\rho^{{\rm ul}}(M),k}]^T, ~~~ \forall k\in \mathcal{K}. \label{eqn:h_bar}
\end{align}
\hrulefill 
\end{figure*}
\setcounter{equation}{13}
\begin{figure*}[!t]
	\normalsize \begin{align}\label{eqn:uplink rate successive interference cancellation}
R_{\tau^{{\rm ul}}(k)}^{\rm ul,SIC}(\{p_k^{\rm ul},\mv{w}_k\},\{q_m^{\rm ul}\},\{\tau^{{\rm ul}}(k)\}) & = I(s_{\tau^{{\rm ul}}(k)}^{\rm ul};\tilde{s}_{\tau^{{\rm ul}}(k)}^{\rm ul}|s_{\tau^{{\rm ul}}(1)}^{\rm ul},\ldots,s_{\tau^{{\rm ul}}(k-1)}^{\rm ul}) \nonumber \\
& = \log_2\frac{\sum\limits_{i\geq k}p_{\tau^{{\rm ul}}(i)}^{\rm ul}|\mv{w}_{\tau^{{\rm ul}}(k)}^H\mv{h}_{\tau^{{\rm ul}}(i)}|^2+\sum\limits_{m=1}^Mq_m^{\rm ul}|w_{\tau^{{\rm ul}}(k),m}|^2+\sigma^2}{\sum\limits_{j> k}p_{\tau^{{\rm ul}}(j)}^{\rm ul}|\mv{w}_{\tau^{{\rm ul}}(k)}^H\mv{h}_{\tau^{{\rm ul}}(j)}|^2+\sum\limits_{m=1}^Mq_m^{\rm ul}|w_{\tau^{{\rm ul}}(k),m}|^2+\sigma^2}, ~ \forall k\in \mathcal{K}.
\end{align}\hrulefill 
\end{figure*}

\setcounter{equation}{2}

Transmission and relaying strategies for the multiple-access relay channel have been studied extensively in the literature \cite{Sanderovich08, Yu13, Zhou14, Zhou16, Yu16}.
In this paper, we assume the following strategy in which each user transmits using a Gaussian codebook, i.e.,
\begin{align}
x_k^{\rm ul}=\sqrt{p_k^{\rm ul}}s_k^{\rm ul}, ~~~ \forall k\in \mathcal{K},
\end{align}where $s_k^{\rm ul}\sim \mathcal{CN}(0,1)$ denotes the message of user $k$, and $p_k^{\rm ul}\geq 0$ denotes the transmit power of user $k$. As a result, the total transmit power of all the users is expressed as
\begin{align}\label{eqn:uplink sum-power}
P^{\rm ul}(\{p_k^{\rm ul}\})=\sum\limits_{k=1}^K\mathbb{E}[|x_k^{\rm ul}|^2]=\sum\limits_{k=1}^Kp_k^{\rm ul}.
\end{align}After receiving the wireless signals from the users, we assume that relay $m$ compresses $y_m^{\rm ul}$ and sends the compressed signals to the CP, $\forall m$. We assume a Gaussian compression codebook and model the quantization noise introduced in the compression process as an independent Gaussian random variable, i.e,\begin{align}\label{eqn:uplink compression}
\tilde{y}_m^{\rm ul}=y_m^{\rm ul}+e_m^{\rm ul}=\sum\limits_{k=1}^Kh_{m,k}x_k^{\rm ul}+z_m^{\rm ul}+e_m^{\rm ul}, ~~~ \forall m\in \mathcal{M},
\end{align}where $e_m^{\rm ul}\sim \mathcal{CN}(0,q_m^{\rm ul})$ denotes the compression noise at relay $m$, with $q_m^{\rm ul}\geq 0$ denoting its  variance. While the Gaussian compression codebook is not necessarily optimal \cite{Sanderovich08}, it gives tractable achievable rate regions. Note that $e_m^{{\rm ul}}$'s are independent of $y_m^{\rm ul}$'s and are independent across $m$. In other words, if we define $\mv{e}^{{\rm ul}}=[e_1^{{\rm ul}},\ldots,e_M^{{\rm ul}}]^T$, then it follows that
\begin{align}
\mathbb{E}\left[\mv{e}^{{\rm ul}}\left(\mv{e}^{{\rm ul}}\right)^H\right]={\rm diag}(q_1^{{\rm ul}},\ldots,q_M^{{\rm ul}})\succeq \mv{0}.
\end{align}

After receiving the compressed signals, the CP first decodes the compression codewords then decodes each user's message based on the beamformed signals, i.e.,
\begin{align}
\tilde{s}_k^{\rm ul}=\mv{w}_k^H\tilde{\mv{y}}^{\rm ul}, ~~~ \forall k\in \mathcal{K},
\end{align}where $\mv{w}_k=[w_{k,1},\ldots,w_{k,M}]^T$ with $\|\mv{w}_k\|^2=1$ denotes the receive beamforming vector for user $k$'s message, and $\tilde{\mv{y}}^{\rm ul}=[\tilde{y}_1^{\rm ul},\ldots,\tilde{y}_M^{\rm ul}]^T$ denotes the collective compressed signals from all the relays.

\subsubsection{Compression Strategies}
We consider two compression strategies at the relays in this work: the independent compression strategy and the Wyner-Ziv compression strategy. If independent compression is performed across the relays, the fronthaul rate for transmitting $\tilde{y}_m^{\rm ul}$ is expressed as
\begin{multline}
C_m^{\rm ul,IN}(\{p_k^{\rm ul}\},q_m^{\rm ul})=I(y_m^{\rm ul};\tilde{y}_m^{\rm ul}) \\  = \log_2\frac{\sum\limits_{k=1}^Kp_k^{\rm ul}|h_{m,k}|^2+q_m^{\rm ul}+\sigma^2}{q_m^{\rm ul}}, ~ \forall m\in \mathcal{M}. \label{eqn:uplink fronthaul rate independent}
\end{multline}
Alternatively, the relays can also perform the Wyner-Ziv compression strategy in a successive fashion, accounting for the fact that the compressed signals are to be decoded jointly at the CP. Given a decompression order $\rho^{{\rm ul}}(1),\ldots,\rho^{{\rm ul}}(M)$ at the CP in which the signal from relay $\rho^{{\rm ul}}(1)\in\mathcal{M}$ is decompressed first, the signal from relay $\rho^{{\rm ul}}(2)\in \mathcal{M}$ is decoded second (with $\rho^{\rm ul}(1)$ as side information), etc., the Wyner-Ziv compression rate
\cite{Simeone16} of relay $\rho^{{\rm ul}}(m)$ can be expressed as \eqref{eqn:uplink fronthaul rate}--\eqref{eqn:h_bar} on top of the page.

\setcounter{equation}{12}

In words, $\bar{\mv{h}}_{k,\{\rho^{{\rm ul}}(m)\}}$ denotes the collection of the channels from user $k$ to relays $\rho^{{\rm ul}}(1),\ldots,\rho^{{\rm ul}}(M)$, $\tilde{\mv{y}}_{\{\rho^{{\rm ul}}(m)\}}^{\rm ul}$ is the collection of the compressed received signals from relay $\rho^{{\rm ul}}(1)$ to relay $\rho^{{\rm ul}}(M)$, and $\mv{\Gamma}_{\{\rho^{{\rm ul}}(m)\}}$ is the covariance matrix of this re-ordered vector.

\subsubsection{Decoding Strategies}
We consider two decoding strategies at the CP: the linear decoding strategy by treating interference as noise and the nonlinear decoding strategy with successive interference cancellation. First, if the CP treats interference as noise, the achievable rate of user $k$ is expressed as
\begin{align}\label{eqn:uplink rate}
& R_k^{\rm ul,TIN}(\{p_k^{\rm ul},\mv{w}_k\},\{q_m^{\rm ul}\})=I(s_k^{\rm ul};\tilde{s}_k^{\rm ul})\nonumber \\  & = \log_2\frac{\sum\limits_{i=1}^Kp_i^{\rm ul}|\mv{w}_k^H\mv{h}_i|^2+\sum\limits_{m=1}^Mq_m^{\rm ul}|w_{k,m}|^2+\sigma^2}{\sum\limits_{j\neq k}p_j^{\rm ul}|\mv{w}_k^H\mv{h}_j|^2+\sum\limits_{m=1}^Mq_m^{\rm ul}|w_{k,m}|^2+\sigma^2}, ~ \forall k\in \mathcal{K}.
\end{align}
If the successive interference cancellation strategy is used, given a decoding order $\tau^{{\rm ul}}(1),\ldots,\tau^{{\rm ul}}(K)$ at the CP in which the message of user $\tau^{{\rm ul}}(1)$ is decoded first, the message of user $\tau^{{\rm ul}}(2)$ is decoded second, etc., the achievable rate of user $\tau^{{\rm ul}}(k)$ is expressed as \eqref{eqn:uplink rate successive interference cancellation} on top of the page.

\setcounter{equation}{14}

\begin{figure*}[!t]
	\normalsize
\begin{align}
&\mathcal{T}^{{\rm ul,IN}}(\{C_m\},P) =\big\{(\{p_k^{\rm ul},\mv{w}_k\},\{q_m^{\rm ul}\}): \nonumber \\
& \qquad P^{\rm ul}(\{p_k^{\rm ul}\})\leq P, C_m^{\rm ul}(\{p_k^{\rm ul}\},q_m^{\rm ul})\leq C_m, q_m^{{\rm ul}}\geq 0, \forall m\in \mathcal{M}, p_k^{{\rm ul}}\geq 0, \|\mv{w}_k\|^2=1, \forall k\in \mathcal{K}  \big\}, \label{TulIN}\\
&\mathcal{T}^{{\rm ul,WZ}}(\{C_m\},P,\{\rho^{{\rm ul}}(m)\}) =\big\{(\{p_k^{\rm ul},\mv{w}_k\},\{q_m^{\rm ul}\}): P^{\rm ul}(\{p_k^{\rm ul}\})\leq P, \nonumber \\
& \qquad
C_{\rho^{{\rm ul}}(m)}^{{\rm ul,WZ}}(\{p_k^{\rm ul}\},q_{\rho^{{\rm ul}}(1)}^{\rm ul},\ldots,q_{\rho^{{\rm ul}}(m)}^{\rm ul},\{\rho^{{\rm ul}}(j)\})\leq C_{\rho^{{\rm ul}}(m)}, q_m^{{\rm ul}}\geq 0,
\forall m\in \mathcal{M}, p_k^{{\rm ul}}\geq 0, \|\mv{w}_k\|^2=1, \forall k\in \mathcal{K}  \big\}.\label{TulWZ}
\end{align}\hrulefill 
\end{figure*}

\setcounter{equation}{20}

\begin{figure*}[!t]
	\normalsize\begin{align}
& \bar{\mathcal{R}}_{{\rm I}}^{\rm ul}(\{C_m\},P)  \triangleq \bigcup\limits_{\left(\{p_k^{\rm ul},\mv{w}_k\},\{q_m^{\rm ul}\}\right)\in \mathcal{T}^{\rm ul,IN}(\{C_m\},P)} \bigg\{(r_1^{\rm ul},\ldots,r_K^{\rm ul}): \nonumber \\
& \qquad \qquad \qquad \qquad \qquad \qquad \qquad \qquad \qquad \qquad \qquad r_k^{\rm ul} \le R_k^{\rm ul,TIN}(\{p_k^{\rm ul},\mv{w}_k\},\{q_m^{\rm ul}\}), \forall k\in \mathcal{K} \bigg\},\label{barR1}\\[3pt]
& \bar{\mathcal{R}}_{{\rm II}}^{\rm ul}(\{C_m\},P,\{\tau^{{\rm ul}}(k)\}) \triangleq
\bigcup\limits_{\left(\{p_k^{\rm ul},\mv{w}_k\},\{q_m^{\rm ul}\}\right)\in \mathcal{T}^{{\rm ul,IN}}(\{C_m\},P)} \bigg\{(r_1^{\rm ul},\ldots,r_K^{\rm ul}): \nonumber \\
& \qquad \qquad \qquad \qquad \qquad \qquad \qquad \qquad \qquad \qquad \qquad \quad r_{\tau^{{\rm ul}}(k)}^{\rm ul} \le  R_{\tau^{{\rm ul}}(k)}^{\rm ul,SIC}(\{p_k^{\rm ul},\mv{w}_k\},\{q_m^{\rm ul}\},\{\tau^{{\rm ul}}(k)\}), \forall k\in \mathcal{K} \bigg\}, \label{barR2}\\
& \bar{\mathcal{R}}_{{\rm III}}^{\rm ul}(\{C_m\},P,\{\rho^{{\rm ul}}(m)\}) \triangleq
\bigcup\limits_{\left(\{p_k^{\rm ul},\mv{w}_k\},\{q_m^{\rm ul}\}\right)\in \mathcal{T}^{{\rm ul,WZ}}(\{C_m\},P,\{\rho^{{\rm ul}}(m)\})} \bigg\{(r_1^{\rm ul},\ldots,r_K^{\rm ul}): \nonumber \\
& \qquad \qquad \qquad \qquad \qquad \qquad \qquad \qquad \qquad \qquad \qquad \qquad r_k^{\rm ul} \le R_k^{\rm ul,TIN}(\{p_k^{\rm ul},\mv{w}_k\},\{q_m^{\rm ul}\}),\forall k\in \mathcal{K} \bigg\}, \label{barR3} \\
& \bar{\mathcal{R}}_{{\rm IV}}^{\rm ul}(\{C_m\},P,\{\rho^{{\rm ul}}(m)\},\{\tau^{{\rm ul}}(k)\}) \triangleq
\bigcup\limits_{\left(\{p_k^{\rm ul},\mv{w}_i\},\{q_m^{\rm ul}\}\right)\in \mathcal{T}^{{\rm ul,WZ}}(\{C_m\},P,\{\rho^{{\rm ul}}(m)\})} \bigg\{(r_1^{\rm ul},\ldots,r_K^{\rm ul}): \nonumber \\ & \quad\quad\quad\quad\quad\quad\quad\quad\quad\quad\quad\quad\quad\quad\quad\quad\quad\quad\quad\quad\quad\quad\quad\quad\quad r_{\tau^{{\rm ul}}(k)}^{\rm ul} \le R_{\tau^{{\rm ul}}(k)}^{\rm ul,SIC}(\{p_k^{\rm ul},\mv{w}_k\},\{q_m^{\rm ul}\},\{\tau^{{\rm ul}}(k)\}),\forall k\in \mathcal{K} \bigg\}. \label{barR4}
\end{align} \hrulefill 
\end{figure*}

\setcounter{equation}{13}

\subsubsection{Achievable Rate Regions}
Given the individual fronthaul capacity constraints $C_m$'s and sum-power constraint $P$,
define $\mathcal{T}^{{\rm ul,IN}}(\{C_m\},P)$ and $\mathcal{T}^{{\rm ul,WZ}}(\{C_m\},P,\{\rho^{{\rm ul}}(m)\})$ as shown in \eqref{TulIN}--\eqref{TulWZ} on the next page
as the sets of feasible transmit powers, compression noise levels, and receive beamforming vectors for the cases of independent compression and Wyner-Ziv compression under the decompression order of $\rho^{{\rm ul}}(1),\ldots,\rho^{{\rm ul}}(M)$ at the CP, respectively. Then, for the considered multiple-access relay channel, the achievable rate regions for Case I: linear decoding at the CP and independent compression across the relays, Case II: successive interference cancellation at the CP and independent compression across the relays, Case III: linear decoding at the CP and Wyner-Ziv compression across the relays, and Case IV: successive interference cancellation at the CP and Wyner-Ziv compression across the relays, are respectively given by \cite{Sanderovich08,Yu16}
\setcounter{equation}{16}
\begin{align}
& \mathcal{R}_{{\rm I}}^{\rm ul}(\{C_m\},P)  \triangleq \text{co }
\bar{\mathcal{R}}_{{\rm I}}^{\rm ul}(\{C_m\},P),
 \label{eqn:uplink rate region 1} \\
& \mathcal{R}_{{\rm II}}^{\rm ul}(\{C_m\},P\})\triangleq \text{co} \bigcup\limits_{\{\tau^{{\rm ul}}(k)\}}\bar{\mathcal{R}}_{{\rm II}}^{\rm ul}(\{C_m\},P,\{\tau^{{\rm ul}}(k)\}), \label{eqn:uplink rate region 2} \\
& \mathcal{R}_{{\rm III}}^{\rm ul}(\{C_m\},P\})\triangleq \text{co} \bigcup\limits_{\{\rho^{{\rm ul}}(m)\}}\bar{\mathcal{R}}_{{\rm III}}^{\rm ul}(\{C_m\},P,\{\rho^{{\rm ul}}(m)\}), \label{eqn:uplink rate region} \\
& \mathcal{R}_{{\rm IV}}^{\rm ul}(\{C_m\},P\}) \nonumber \\ & ~ \triangleq \text{co} \bigcup\limits_{\left(\{\rho^{{\rm ul}}(m)\},\{\tau^{{\rm ul}}(k)\}\right)}\bar{\mathcal{R}}_{{\rm IV}}^{\rm ul}(\{C_m\},P,\{\rho^{{\rm ul}}(m)\},\{\tau^{{\rm ul}}(k)\}), \label{eqn:uplink rate region 4}
\end{align}
where ``co'' stands for the closure of convex hull operation and in (\ref{eqn:uplink rate region 1}), (\ref{eqn:uplink rate region 2}), (\ref{eqn:uplink rate region}), and (\ref{eqn:uplink rate region 4}), $\bar{\mathcal{R}}_{{\rm I}}^{\rm ul}(\{C_m\},P),$ $\bar{\mathcal{R}}_{{\rm II}}^{\rm ul}(\{C_m\},P,\{\tau^{{\rm ul}}(k)\}),$ $\bar{\mathcal{R}}_{{\rm III}}^{\rm ul}(\{C_m\},P,\{\rho^{{\rm ul}}(m)\}),$ and $\bar{\mathcal{R}}_{{\rm IV}}^{\rm ul}(\{C_m\},P,\{\rho^{{\rm ul}}(m)\},\{\tau^{{\rm ul}}(k)\})$ are expressed in \eqref{barR1}--\eqref{barR4} on top of the page and
denote the rate regions of Case I, and Case II given the decoding order $\tau^{{\rm ul}}(1),\ldots,\tau^{{\rm ul}}(K)$, and Case III given the decompression order $\rho^{{\rm ul}}(1),\ldots,\rho^{{\rm ul}}(M)$, and Case IV given the decoding order $\tau^{{\rm ul}}(1),\ldots,\tau^{{\rm ul}}(K)$ and the decompression order $\rho^{{\rm ul}}(1),\ldots,\rho^{{\rm ul}}(M)$, respectively.

As a remark, because the achievable rates under the proposed transmission and relaying strategies are not necessarily concave functions of $C_m$ and $P$, there is the potential to further enlarge the above rate region by taking the convex hull over different $C_m$'s and $P$'s.
For ease of presentation, the statements of the main results in this paper do not include this additional convex hull operation, but such an operation can be easily incorporated.

\begin{figure*}
\setcounter{equation}{32}
\begin{align}\label{eqn:downlink fronthaul rate}
C_{\rho^{{\rm dl}}(m)}^{\rm dl,MV}(\{p_k^{\rm dl},\mv{v}_k\},\mv{Q},\{\rho^{{\rm dl}}(m)\}) & = I(x_{\rho^{{\rm dl}}(m)}^{\rm dl};\tilde{x}_{\rho^{{\rm dl}}(m)}^{\rm dl}|\tilde{x}_{\rho^{{\rm dl}}(1)}^{\rm dl},\ldots,\tilde{x}_{\rho^{{\rm dl}}(m-1)}^{\rm dl}) \nonumber \\
& = \log_2 \frac{\sum_{k=1}^{K} p_k^{\rm dl} \left|v_{k,\rho^{{\rm dl}}(m)}\right|^2 + \mv{Q}_{\{\rho^{{\rm dl}}(m)\}}^{(m,m)}}{{\mv{Q}_{\{\rho^{{\rm dl}}(m)\}}^{(m,m)}}-\mv{Q}_{\{\rho^{{\rm dl}}(m)\}}^{(m,m+1:M)}(\mv{Q}_{\{\rho^{{\rm dl}}(m)\}}^{(m+1:M,m+1:M)})^{-1}\mv{Q}_{\{\rho^{{\rm dl}}(m)\}}^{(m+1:M,m)}}, ~~~ \forall m \in \mathcal{M},
\end{align}
where
\begin{align}
\mv{Q}_{\{\rho^{{\rm dl}}(m)\}} = \mathbb{E}\left[[e_{\rho^{{\rm dl}}(1)}^{{\rm dl}},\ldots,e_{\rho^{{\rm dl}}(M)}^{{\rm dl}}]^T[(e_{\rho^{{\rm dl}}(1)}^{{\rm dl}})^\ast,\ldots,(e_{\rho^{{\rm dl}}(M)}^{{\rm dl}})^\ast]\right]. \label{eqn:downlink compression covariance}
\end{align}
\setcounter{equation}{29}
\hrulefill
\end{figure*}

\setcounter{equation}{24}

\subsection{Broadcast Channel with Compressing Relays}\label{sec:Broadcast Relay Channel}

The Gaussian broadcast relay channel model is as shown in Fig.~\ref{fig3}(b). The discrete-time baseband channel model between the relays and the users is the dual of the broadcast relay channel given by
\begin{align}\label{eqn:downlink received signal}
\left[\begin{array}{c} y_{1}^{\rm dl} \\ \vdots \\ y_K^{\rm dl} \end{array} \right]\hspace{-2pt} =\hspace{-2pt} \left[\begin{array}{ccc}h_{1,1}^H & \cdots & h_{M,1}^H \\  \vdots & \ddots & \vdots \\ h_{1,K}^H & \cdots & h_{M,K}^H \end{array}\right]\left[\begin{array}{c}x_1^{\rm dl} \\ \vdots \\ x_M^{\rm dl} \end{array}\right]\hspace{-2pt}+\hspace{-2pt}\left[\begin{array}{c} z_1^{\rm dl} \\ \vdots \\ z_K^{\rm dl} \end{array}\right],
\end{align}where $x_m^{\rm dl}$ denotes the transmit signal of relay $m$, and $z_k^{\rm dl}\sim \mathcal{CN}(0,\sigma^2)$ denotes the AWGN at user $k$.

Transmission and relaying strategies for the broadcast relay channel have also been studied extensively in the literature. For example, the CP can choose to partially share the messages of each user with multiple BSs in order to enable cooperation \cite{Pratik_hybrid}.
This paper however focuses on a compression strategy in which the beamformed signals are precomputed at the CP, then compressed and forwarded to the relays \cite{Simeone13}, because of its potential to achieve the capacity region to within a constant gap \cite{Kim_DDF, Yu19}.

More specifically,
we use a Gaussian codebook for each user and define the beamformed signal intended for user $k$ to be transmitted across all the relays as $\mv{v}_k\sqrt{p_k^{\rm dl}}s_k^{\rm dl}$, $\forall k$, where $s_k^{\rm dl}\sim \mathcal{CN}(0,1)$ denotes the message for user $k$, $p_k^{\rm dl}\geq 0$ denotes the transmit power, and $\mv{v}_k=[v_{k,1},\ldots,v_{k,M}]^T$ with $\|\mv{v}_k\|^2=1$ denotes the transmit beamforming vector across the relays. The aggregate signal intended for all the relays is thus given by
\begin{align}
\tilde{\mv{x}}^{{\rm dl}}=[\tilde{x}_1^{{\rm dl}},\ldots,\tilde{x}_M^{{\rm dl}}]^T=\sum_{k=1}^K\mv{v}_k\sqrt{p_k^{\rm dl}}s_k^{\rm dl},
\end{align}
which is compressed then sent to the relays via fronthaul links.

Similar to (\ref{eqn:uplink compression}), the quantization noises are modelled as Gaussian random variables, i.e.,
\begin{align}\label{eqn:downlink compression}
x_m^{\rm dl}=\tilde{x}_m^{\rm dl}+e_m^{\rm dl}, ~~~ \forall m\in \mathcal{M},
\end{align}where $e_m^{\rm dl}\sim \mathcal{CN}(0,q_m^{\rm dl})$ denotes the quantization noise at relay $m$, with $q_m^{\rm dl}$ denoting its variance. Putting all the quantization noises across all the relays together as $\mv{e}^{\rm dl}=[e_1^{\rm dl},\ldots,e_M^{\rm dl}]^T$, we have the quantization noise covariance matrix
\begin{align}\label{eqn:downlink compression covariance 0}
\mv{Q}=\mathbb{E}\left[\mv{e}^{\rm dl}\left(\mv{e}^{\rm dl}\right)^H\right]\succeq \mv{0}.
\end{align}
The compressed versions of the beamformed signals are transmitted across the relays.
According to (\ref{eqn:downlink compression}), the transmit signal can be expressed as
\begin{align}\label{eqn:downlink signal}
\left[\begin{array}{c} x_1^{\rm dl} \\ \vdots \\ x_M^{\rm dl}  \end{array}\right]=\left[\begin{array}{c} \sum_{k=1}^Kv_{k,1}\sqrt{p_k^{\rm dl}}s_k^{\rm dl} \\ \vdots \\ \sum_{k=1}^Kv_{k,M}\sqrt{p_k^{\rm dl}}s_k^{\rm dl}  \end{array}\right]+\left[\begin{array}{c}e_1^{\rm dl}\\ \vdots \\ e_M^{\rm dl}  \end{array}\right].
\end{align}

Under the above model, the transmit power of all the relays is expressed as
\begin{align}\label{eqn:downlink sum-power}
P^{\rm dl}(\{p_k^{\rm dl}\},\mv{Q})=\sum\limits_{m=1}^M\mathbb{E}[|x_m^{\rm dl}|^2]=\sum\limits_{k=1}^Kp_k^{\rm dl}+{\rm tr}(\mv{Q}).
\end{align}

\subsubsection{Compression Strategies}
We consider two compression strategies in our considered broadcast relay channel: the independent compression strategy and the multivariate compression strategy. If the compression is done independently for the signals across different relays, then the covariance matrix of the compression noise given in (\ref{eqn:downlink compression covariance 0}) reduces to a diagonal matrix, i.e.,
\begin{align}\label{eqn:downlink compression covariance 1}
\mv{Q}= \mv{Q}_{{\rm diag}}\triangleq {\rm diag}([q_1^{{\rm dl}},\ldots,q_M^{{\rm dl}}]).
\end{align}As a result, the fronthaul rate for transmitting $x_m^{\rm dl}$ is expressed as
\begin{multline}
C_m^{\rm dl,IN}(\{p_k^{\rm dl},\mv{v}_k\},\mv{Q}_{{\rm diag}})=  I(\tilde{x}_m^{\rm dl};x_m^{\rm dl}) \\ =  \log_2 \frac{\sum\limits_{k=1}^Kp_k^{\rm dl}|v_{k,m}|^2+\mv{Q}_{{\rm diag}}^{(m,m)} }{\mv{Q}_{{\rm diag}}^{(m,m)}}, ~~~ \forall m\in \mathcal{M}. \label{eqn:downlink fronthaul rate 1}
\end{multline}
Alternatively, the CP can also use the multivariate compression strategy. Given a compression order at the CP $\rho^{{\rm dl}}(1),\ldots,\rho^{{\rm dl}}(M)$ in which the signal for relay $\rho^{{\rm dl}}(1)\in \mathcal{M}$ is compressed first, the signal for relay $\rho^{{\rm dl}}(2)$ is compressed second, etc., the compression rate for relay $\rho^{{\rm dl}}(m)$ can be expressed as (\ref{eqn:downlink fronthaul rate})--(\ref{eqn:downlink compression covariance}) on top of the page \cite{Simeone13}.

\setcounter{equation}{36}
\begin{figure*}[!t]
	\normalsize\begin{align}
& \mathcal{T}^{\rm dl,IN}(\{C_m\},P) = \big\{(\{p_k^{\rm dl},\mv{v}_k\},\mv{Q}_{{\rm diag}}): P^{\rm dl}(\{p_k^{\rm dl}\},\mv{Q}_{{\rm diag}})\leq P, \mv{Q}_{{\rm diag}}\succeq \mv{0}~{\rm is} ~ {\rm diagonal}, \nonumber \\ & \quad\quad\quad\quad\quad\quad\quad~~~~~~~~~~~~~~~~~~~~ C_m^{\rm dl,IN}(\{p_k^{\rm dl},\mv{v}_k\},\mv{Q}_{{\rm diag}}) \leq C_m, \forall m\in \mathcal{M}, p_k^{{\rm dl}}\geq 0, \|\mv{v}_k\|^2=1,\forall k\in \mathcal{K} \big\}, \label{TdlIN}\\
& \mathcal{T}^{\rm dl,MV}(\{C_m\},P,\{\rho^{{\rm dl}}(m)\}) = \big\{(\{p_k^{\rm dl},\mv{v}_k\},\mv{Q}\}): P^{\rm dl}(\{p_k^{\rm dl}\},\mv{Q})\leq P, \mv{Q}\succeq \mv{0}, \nonumber \\ & \quad\quad\quad\quad\quad\quad\quad~~~~~~~~~~~~~~~~~~~~ C_{\rho^{{\rm dl}}(m)}^{\rm dl,MV}(\{p_k^{\rm dl},\mv{v}_k\},\mv{Q},\{\rho^{{\rm dl}}(m)\}) \leq C_{\rho^{{\rm dl}}(m)}, \forall m\in \mathcal{M}, p_k^{{\rm dl}}\geq 0, \|\mv{v}_k\|^2=1,\forall k\in \mathcal{K} \big\}.\label{TdlMV}
\end{align}\hrulefill 
\end{figure*}

\setcounter{equation}{34}

\subsubsection{Encoding Strategies}
We consider two encoding strategies at the CP: the linear encoding strategy and the nonlinear encoding strategy via dirty-paper coding. If the CP employs linear encoding, the achievable rate of user $k$ can be expressed as
\begin{multline}\label{eqn:downlink rate 1}
R_k^{\rm dl,LIN}(\{p_k^{\rm dl},\mv{v}_k\},\mv{Q})=I(s_k^{\rm dl};y_k^{\rm dl}) \\  = \log_2\frac{\sum\limits_{i=1}^Kp_i^{\rm dl}|\mv{v}_i^H\mv{h}_k|^2+\mv{h}_k^H\mv{Q}\mv{h}_k+\sigma^2}{\sum\limits_{j\neq  k}p_j^{\rm dl}|\mv{v}_j^H\mv{h}_k|^2+\mv{h}_k^H\mv{Q}\mv{h}_k+\sigma^2}, ~ \forall k\in \mathcal{K}.
\end{multline}
If the dirty-paper coding strategy is used, given an encoding order $\tau^{{\rm dl}}(1),\ldots,\tau^{{\rm dl}}(K)$ at the CP in which the message of user $\tau^{{\rm dl}}(1)\in \mathcal{K}$ is decoded first, the message of user $\tau^{{\rm dl}}(2)\in \mathcal{K}$ is decoded second, etc., then the achievable rate of user $\tau^{{\rm dl}}(k)$ is expressed as
\begin{align}\label{eqn:downlink rate}
&~~~~R_{\tau^{{\rm dl}}(k)}^{\rm dl,DPC}\left(\{p_k^{\rm dl},\mv{v}_k\},\mv{Q},\{\tau^{{\rm dl}}(k)\}\right) \nonumber \\ &= I(s_{\tau^{{\rm dl}}(k)}^{\rm dl};y_{\tau^{{\rm dl}}(k)}^{\rm dl}|s_{\tau^{{\rm dl}}(1)}^{\rm dl},\ldots,s_{\tau^{{\rm dl}}(k-1)}^{\rm dl}) \nonumber \\
&= \log_2\frac{\sum\limits_{i\leq k}p_{\tau^{{\rm dl}}(i)}^{\rm dl}|\mv{v}_{\tau^{{\rm dl}}(i)}^H\mv{h}_{\tau^{{\rm dl}}(k)}|^2+\mv{h}_{\tau^{{\rm dl}}(k)}^H\mv{Q}\mv{h}_{\tau^{{\rm dl}}(k)}+\sigma^2}{\sum\limits_{j< k}p_{\tau^{{\rm dl}}(j)}^{\rm dl}|\mv{v}_{\tau^{{\rm dl}}(j)}^H\mv{h}_{\tau^{{\rm dl}}(k)}|^2+\mv{h}_{\tau^{{\rm dl}}(k)}^H\mv{Q}\mv{h}_{\tau^{{\rm dl}}(k)}+\sigma^2}, \nonumber \\
& \qquad\qquad\qquad\qquad\qquad\qquad\qquad\qquad\qquad \forall k \in \mathcal{K}.
\end{align}
In (\ref{eqn:downlink rate 1}) and (\ref{eqn:downlink rate}), if we set $\mv{Q}=\mv{Q}_{{\rm diag}}$ as shown in (\ref{eqn:downlink compression covariance 1}), then $R_k^{\rm dl,LIN}(\{p_k^{\rm dl},\mv{v}_k\},\mv{Q}_{{\rm diag}})$'s and $R_{\tau^{{\rm dl}}(k)}^{\rm dl,DPC}(\{p_k^{\rm dl},\mv{v}_k\},\mv{Q}_{{\rm diag}},\{\tau^{{\rm dl}}(k)\})$'s denote the user rates achieved by the independent compression strategy. If $\mv{Q}$ is a full matrix (i.e., non-diagonal), then $R_k^{\rm dl,LIN}(\{p_k^{\rm dl},\mv{v}_k\},\mv{Q})$'s and $R_{\tau^{{\rm dl}}(k)}^{\rm dl,DPC}\left(\{p_k^{\rm dl},\mv{v}_k\},\mv{Q},\{\tau^{{\rm dl}}(k)\}\right)$'s denote the user rates achieved by the multivariate compression strategy.

\subsubsection{Achievable Rate Regions}
Given the individual fronthaul capacity constraints $\{C_m\}$ and sum-power constraint $P$,
define $\mathcal{T}^{\rm dl,IN}(\{C_m\},P)$
and $\mathcal{T}^{\rm dl,MV}(\{C_m\},P,\{\rho^{{\rm dl}}(m)\})$
as shown in \eqref{TdlIN}--\eqref{TdlMV} on top of the page
as the sets of feasible transmit powers, beamforming vectors, and compression noise covariance matrices for the cases of independent compression and multivariate compression under the compression order of $\rho^{{\rm dl}}(1),\ldots,\rho^{{\rm dl}}(M)$, respectively. Then, in the broadcast relay channel, the achievable rate regions for Case I: linear encoding and independent compression at the CP, Case II: dirty-paper coding and independent compression at the CP, Case III: linear encoding and multivariate compression at the CP, and Case IV: dirty-paper coding and multivariate compression at the CP, are respectively given by \cite{Simeone13,Yu19}
\setcounter{equation}{38}
\begin{align}
& \mathcal{R}_{{\rm I}}^{\rm dl}(\{C_m\},P)  \triangleq \text{co}
 \mathcal{\bar R}_{{\rm I}}^{\rm dl}(\{C_m\},P)
\label{eqn:downlink rate region 1} \\
& \mathcal{R}_{{\rm II}}^{\rm dl}(\{C_m\},P\})\triangleq \text{co} \bigcup\limits_{\{\tau^{{\rm dl}}(k)\}}\bar{\mathcal{R}}_{{\rm II}}^{\rm dl}(\{C_m\},P,\{\tau^{{\rm dl}}(k)\}), \label{eqn:downlink rate region 2} \\
& \mathcal{R}_{{\rm III}}^{\rm dl}(\{C_m\},P\})\triangleq \text{co} \bigcup\limits_{\{\rho^{{\rm dl}}(m)\}}\bar{\mathcal{R}}_{{\rm III}}^{\rm dl}(\{C_m\},P,\{\rho^{{\rm dl}}(m)\}), \label{eqn:downlink rate region} \\
& \mathcal{R}_{{\rm IV}}^{\rm dl}(\{C_m\},P\})\triangleq \nonumber \\ & ~ \text{co} \bigcup\limits_{\left(\{\rho^{{\rm dl}}(m)\},\{\tau^{{\rm dl}}(k)\}\right)}\bar{\mathcal{R}}_{{\rm IV}}^{\rm dl}(\{C_m\},P,\{\rho^{{\rm dl}}(m)\},\{\tau^{{\rm dl}}(k)\}), \label{eqn:downlink rate region 4}
\end{align}
where the rate regions denoted by $\mathcal{\bar R}_{{\rm I}}^{\rm dl}(\{C_m\},P),$ $\mathcal{\bar R}_{{\rm II}}^{\rm dl}(\{C_m\},P,\{\tau^{{\rm dl}}(k)\}),$ $\mathcal{\bar R}_{{\rm III}}^{\rm dl}(\{C_m\},P,\{\rho^{{\rm dl}}(m)\}),$ and $\mathcal{\bar R}_{{\rm IV}}^{\rm dl}(\{C_m\},P,\{\rho^{{\rm dl}}(m)\},\{\tau^{{\rm dl}}(k)\})$ and expressed in \eqref{barRdl1}--\eqref{barRdl4} on the next page
\begin{figure*}[!t]
	\normalsize
\begin{align}
& \mathcal{\bar R}_{{\rm I}}^{\rm dl}(\{C_m\},P) \triangleq \bigcup\limits_{\left(\{p_k^{\rm dl},\mv{v}_k\},\mv{Q}_{{\rm diag}}\right)\in \mathcal{T}^{\rm dl,IN}(\{C_m\},P)}  \bigg\{ (r_1^{\rm dl},\ldots,r_K^{\rm dl}): \nonumber \\
& \qquad \qquad \qquad \qquad \qquad \qquad \qquad \qquad \qquad \qquad \qquad ~ r_k^{\rm dl} \le R_k^{\rm dl,LIN}(\{p_k^{\rm dl},\mv{v}_k\},\mv{Q}_{{\rm diag}}), \forall k\in \mathcal{K} \bigg\}, \label{barRdl1}\\
& \mathcal{\bar R}_{{\rm II}}^{\rm dl}(\{C_m\},P,\{\tau^{{\rm dl}}(k)\}) \triangleq \bigcup\limits_{\left(\{p_k^{\rm dl},\mv{v}_k\},\mv{Q}_{{\rm diag}}\right)\in \mathcal{T}^{\rm dl,IN}(\{C_m\},P)}  \bigg\{ (r_1^{\rm dl},\ldots,r_K^{\rm dl}): \nonumber \\
& \qquad \qquad \qquad \qquad \qquad \qquad \qquad \qquad \qquad \qquad \qquad \qquad r_{\tau^{{\rm dl}}(k)}^{\rm dl} \le R_{\tau^{{\rm dl}}(k)}^{\rm dl,DPC}(\{p_k^{\rm dl},\mv{v}_k\},\mv{Q}_{{\rm diag}},\{\tau^{{\rm dl}}(k)\}), \forall k\in \mathcal{K} \bigg\}, \label{barRdl2}\\
& \mathcal{\bar R}_{{\rm III}}^{\rm dl}(\{C_m\},P,\{\rho^{{\rm dl}}(m)\}) \triangleq \bigcup\limits_{\left(\{p_k^{\rm dl},\mv{v}_k\},\mv{Q}\right)\in \mathcal{T}^{\rm dl,MV}(\{C_m\},P,\{\rho^{{\rm dl}}(m)\})}  \bigg\{ (r_1^{\rm dl},\ldots,r_K^{\rm dl}): \nonumber \\
& \qquad \qquad \qquad \qquad \qquad \qquad \qquad \qquad \qquad \qquad \qquad \qquad \qquad r_k^{\rm dl} \le R_k^{\rm dl,LIN}(\{p_k^{\rm dl},\mv{v}_k\},\mv{Q}), \forall k\in \mathcal{K} \bigg\}, \label{barRdl3} \\
& \mathcal{\bar R}_{{\rm IV}}^{\rm dl}(\{C_m\},P,\{\rho^{{\rm dl}}(m)\},\{\tau^{{\rm dl}}(k)\}) \triangleq \bigcup\limits_{\left(\{p_k^{\rm dl},\mv{v}_k\},\mv{Q}\right)\in \mathcal{T}^{\rm dl,MV}(\{C_m\},P,\{\rho^{{\rm dl}}(m)\})}  \bigg\{ (r_1^{\rm dl},\ldots,r_K^{\rm dl}): \nonumber \\ & \quad\quad\quad\quad\quad\quad\quad\quad\quad\quad\quad\quad\quad\quad\quad\quad\quad\quad\quad\quad\quad\quad\quad\quad\quad\quad\quad r_{\tau^{{\rm dl}}(k)}^{\rm dl} \le R_{\tau^{{\rm dl}}(k)}^{\rm dl,DPC}(\{p_k^{\rm dl},\mv{v}_k\},\mv{Q},\{\tau^{{\rm dl}}(k)\}), \forall k\in \mathcal{K} \bigg\}.\label{barRdl4}
\end{align}\hrulefill 
\end{figure*}
are the rate regions of Case I, and Case II given the encoding order $\tau^{{\rm dl}}(1),\ldots,\tau^{{\rm dl}}(K)$, and Case III given the compression order $\rho^{{\rm dl}}(1),\ldots,\rho^{{\rm dl}}(M)$, and Case IV given the encoding order $\tau^{{\rm dl}}(1),\ldots,\tau^{{\rm dl}}(K)$ and the compression order $\rho^{{\rm dl}}(1),\ldots,\rho^{{\rm dl}}(M)$, respectively.

As a remark, similar to the multiple-access relay channel case, an additional closure of convex hull operation can be applied over the $C_m$'s and $P$'s to potentially enlarge the achievable rate region. The statements of main results in this paper do not include this extra convex hull operation for simplicity, but it can be easily incorporated in all the statements of the theorems and the proofs.

\section{Main Results}\label{sec:Duality between Multiple-Access Channel and Broadcast Relay Channel}

The main results of this work are the following set of theorems showing the duality relationships between the achievable rate regions of the multiple-access relay channel and the broadcast relay channel under the same sum-power constraint and individual fronthaul constraints. 

\begin{theorem}\label{theorem3}
Consider the multiple-access relay channel implementing independent compression across the relays as well as linear decoding at the CP and the broadcast relay channel implementing independent compression across the relays as well as linear encoding at the CP, where all the users and relays are equipped with a single antenna. Then, under the same sum-power constraint $P$ and individual fronthaul capacity constraints $C_m$'s, the achievable rate regions of the multiple-access relay channel defined in (\ref{eqn:uplink rate region 1}) and the broadcast relay channel defined in (\ref{eqn:downlink rate region 1}) are identical. In other words,
\begin{align}
\mathcal{R}_{{\rm I}}^{\rm ul}(\{C_m\},P)=\mathcal{R}_{{\rm I}}^{\rm dl}(\{C_m\},P).
\end{align}
\end{theorem}

\begin{theorem}\label{theorem2}
Consider the multiple-access relay channel implementing independent compression across the relays as well as successive interference cancellation at the CP and the broadcast relay channel implementing independent compression across the relays as well as dirty-paper coding at the CP, where all the users and relays are equipped with a single antenna. Then, under the same sum-power constraint $P$ and individual fronthaul capacity constraints $C_m$'s, the achievable rate regions of the multiple-access relay channel defined in (\ref{eqn:uplink rate region 2}) and the broadcast relay channel defined in (\ref{eqn:downlink rate region 2}) are identical. In other words,
\begin{align}
\mathcal{R}_{{\rm II}}^{\rm ul}(\{C_m\},P)=\mathcal{R}_{{\rm II}}^{\rm dl}(\{C_m\},P).
\end{align}
\end{theorem}

\begin{theorem}\label{theorem1}
Consider the multiple-access relay channel implementing Wyner-Ziv compression across the relays as well as linear decoding at the CP and the broadcast relay channel implementing multivariate compression across the relays as well as linear encoding at the CP, where all the users and relays are equipped with a single antenna. Then, under the same sum-power constraint $P$ and individual fronthaul capacity constraints $C_m$'s, the achievable rate regions of the multiple-access relay channel defined in (\ref{eqn:uplink rate region}) and the broadcast relay channel defined in (\ref{eqn:downlink rate region}) are identical. In other words,
\begin{align}
\mathcal{R}_{{\rm III}}^{\rm ul}(\{C_m\},P)=\mathcal{R}_{{\rm III}}^{\rm dl}(\{C_m\},P).
\end{align}
\end{theorem}

\begin{theorem}\label{theorem4}
Consider the multiple-access relay channel implementing Wyner-Ziv compression across the relays as well as successive interference cancellation at the CP and the broadcast relay channel implementing multivariate compression across the relays as well as dirty-paper coding at the CP, where all the users and relays are equipped with a single antenna. Then, under the same sum-power constraint $P$ and individual fronthaul capacity constraints $C_m$'s, the achievable rate regions of the multiple-access relay channel defined in (\ref{eqn:uplink rate region 4}) and the broadcast relay channel defined in (\ref{eqn:downlink rate region 4}) are identical. In other words,
\begin{align}
\mathcal{R}_{{\rm IV}}^{\rm ul}(\{C_m\},P)=\mathcal{R}_{{\rm IV}}^{\rm dl}(\{C_m\},P).
\end{align}
\end{theorem}

As mentioned earlier, when the fronthaul capacity of each relay is infinite, i.e., $C_m\rightarrow \infty$, $\forall m$, the quantization noises can be set to zero. As a result, the $M$ relays and the CP are equivalent to a virtual BS with $M$ antennas. In this case, the multiple-access relay channel and the broadcast relay channel reduce to the usual multiple-access channel and broadcast channel, respectively; therefore, the classic uplink-downlink duality directly applies. Our main results, i.e., Theorems \ref{theorem3} to \ref{theorem4}, are generalizations of the classic uplink-downlink duality result to the case with non-zero 
quantization noises.

We note that the exact capacity region characterizations of the multiple-access relay channel and the
broadcast relay channel are both still open problems. The duality results above
pertain to the specific compression-based relaying strategies. Although it is
possible to outperform these strategies for specific channel instances, the
compression strategies are important both in practical implementations
\cite{Simeone16} and due to its ability to approximately approach the
theoretical capacity regions under specific conditions for both the uplink
\cite{Yu16} and the downlink \cite{Yu19} as mentioned earlier.

It is worth noting that as shown in \cite{duality1,duality3}, with dirty paper coding in the broadcast channel and successive interference cancellation in the multiple-access channel (without relays), to achieve the same rate tuple with the same sum power in the downlink and the uplink, the encoding order and the decoding order should be reverse of each other. In this paper, we show a similar result for the broadcast channel and the multiple-access channel with compressing relays: if multivariate compression and Wyner-Ziv compression are used by the relays, then to achieve the same rate tuple with the same sum power and the same fronthaul capacities in the downlink and the uplink, the compression order and the decompression order should also be reverse of each other. Please note the clear analogy between dirty paper coding and multivariate compression (in the sense that they both cancel the interference caused by encoding/compression seen by the users) and successive interference cancellation and Wyner-Ziv compression (in the sense that the decoded/decompressed signals can be used for decoding/decompressing the remaining signals). Rigorous proofs of these results are provided in Sections \ref{sec 3} and \ref{sec 4}.


\section{Proof of Theorem \ref{theorem3}}\label{sec 1}
In this section, we prove the duality between the multiple-access relay channel with linear decoding at the CP as well as independent compression across the relays and the broadcast relay channel with linear encoding at the CP and independent compression across the relays. Suppose that the same beamforming vectors
\begin{align}\label{eqn:beamforming vector}
\mv{v}_k=\mv{w}_k=\bar{\mv{u}}_k=[\bar{u}_{k,1},\ldots,\bar{u}_{k,M}]^T, ~~~ \forall k\in \mathcal{K},
\end{align}with $\|\bar{\mv{u}}_k\|=1$, $\forall k\in \mathcal{K}$, are used in both the multiple-access relay channel and the broadcast relay channel. For simplicity, we assume that the beamforming vectors $\bar{\mv{u}}_k$'s satisfy the following condition:
\begin{align}\label{eqn:assumption}
\sum\limits_{k=1}^K|\bar{u}_{k,m}|^2>0, ~~~ \forall m\in \mathcal{M}.
\end{align}Condition (\ref{eqn:assumption}) indicates that all the relays are used for communications so that the fronthaul rates are properly defined. If we have $\sum_{k=1}^K|\bar{u}_{k,m}|^2=0$ for some $m$, then we can simply define $\bar{\mathcal{M}}=\{m:\sum_{k=1}^K|\bar{u}_{k,m}|^2>0\}$ and $\bar{M}=|\bar{\mathcal{M}}|$, respectively. In this case, the considered system is equivalent to a system merely consisting of $\bar{M}$ relays in the set of $\bar{\mathcal{M}}$, in which we have $\sum\limits_{k=1}^K|\bar{u}_{k,m}|^2>0$, $\forall m\in \bar{\mathcal{M}}$. As a result, condition (\ref{eqn:assumption}) does not affect the generality of our following results.

Let $\{R_k \ge 0, k \in \mathcal{K} \}$ be a set of user target rates and $\{C_m\geq 0, m \in \mathcal{M} \}$ be a set of fronthaul rate requirements for the relays. 
For the multiple-access relay channel as described in Section \ref{sec:Multiple-Access Relay Channel}, we fix the receive beamforming vectors as in (\ref{eqn:beamforming vector})
and formulate the transmit power minimization problem
subject to the individual rate constraints as well as
the individual fronthaul capacity constraints as follows:
\begin{align}\hspace{-8pt} \mathop{\mathrm{minimize}}_{\{p_k^{\rm ul}\},\{q_m^{\rm ul}\}} & ~ P^{\rm ul}(\{p_k^{\rm ul}\})  \label{eqn:problem3 1} \\
\hspace{-8pt} \mathrm {subject ~ to}  & ~ R_k^{\rm ul,TIN}(\{p_k^{\rm ul},\bar{\mv{u}}_k\},\{q_m^{\rm ul}\})\geq R_k, ~ \forall k\in \mathcal{K}, \label{eqn:uplink rate constraint} \\ \hspace{-8pt} & ~ C_m^{\rm ul,IN}(\{p_k^{\rm ul}\},q_m^{\rm ul})\leq C_m, ~ \forall m\in \mathcal{M}, \label{eqn:uplink fronthaul constraint3} \\ \hspace{-8pt} & ~ p_k^{\rm ul} \ge 0, ~ \forall k\in \mathcal{K}, \label{eqn:uplink nonnegative power} \\ \hspace{-8pt} & ~ q_m^{\rm ul} \ge 0, ~ \forall m\in \mathcal{M}. \label{eqn:uplink positive}
\end{align}
Similarly, for the broadcast relay channel as described in Section \ref{sec:Broadcast Relay Channel},
we fix the transmit beamforming vectors as in (\ref{eqn:beamforming vector}) and
formulate the transmit power minimization problem as
\begin{align}\hspace{-8pt} \mathop{\mathrm{minimize}}_{\{p_k^{\rm dl}\},\mv{Q}_{{\rm diag}}} & ~ P^{\rm dl}(\{p_k^{\rm dl}\},\mv{Q}_{{\rm diag}})  \label{eqn:problem3 2} \\
\hspace{-8pt} \mathrm {subject ~ to}  & ~ R_k^{\rm dl,LIN}(\{p_k^{\rm dl},\bar{\mv{u}}_k\},\mv{Q}_{{\rm diag}})\geq R_k, ~ \forall k\in \mathcal{K}, \label{eqn:downlink rate constraint} \\ \hspace{-8pt} & ~ C_m^{\rm dl,IN}(\{p_k^{\rm dl},\bar{\mv{u}}_k\},\mv{Q}_{{\rm diag}})\leq C_m, ~ \forall m\in \mathcal{M}, \label{eqn:downlink fronthaul constraint3} \\ \hspace{-8pt} & ~ p_k^{\rm dl} \ge 0, ~ \forall k\in \mathcal{K}, \label{eqn:nonnegative power} \\ \hspace{-8pt} & ~ \mv{Q}_{{\rm diag}}^{(m,m)} \ge 0, ~ \forall m\in \mathcal{M}. \label{eqn:positive quantization noise3}
\end{align}

Our aim is to show that given the same fixed beamformer and under the same set of rate
targets $\{R_k\}$, the optimization problems (\ref{eqn:problem3 1}) and
(\ref{eqn:problem3 2}) are equivalent in the sense that either both are
infeasible, or both are feasible and have the same minimum solution.
This would imply that under the same fronthaul capacity constraints $\{C_m\}$ and total transmit power constraint $P$, fixing the same beamformers, any achievable rate-tuple $\{R_k\}$ of the multiple-access relay channel is also achievable in the broadcast relay channel, and vice versa.
Then, by trying all beamforming vectors, this would imply that the achievable rate regions of the multiple-access relay channel under linear decoding as well as independent compression and the broadcast relay channel under linear encoding as well as independent compression are identical, i.e., $\mathcal{\bar R}_{{\rm I}}^{\rm ul}(\{C_m\},P)=\mathcal{\bar R}_{{\rm I}}^{\rm dl}(\{C_m\},P)$. 
Finally, by taking convex hull, we get
$\mathcal{R}_{{\rm I}}^{\rm ul}(\{C_m\},P)=\mathcal{R}_{{\rm I}}^{\rm dl}(\{C_m\},P)$. 

To show the equivalence of the optimization problems (\ref{eqn:problem3 1}) and
(\ref{eqn:problem3 2}) for the fixed beamformers, we take a set of $\{R_k\}$ and $\{C_m\}$ such
that both (\ref{eqn:problem3 1}) and (\ref{eqn:problem3 2}) are strictly feasible,
and show that (\ref{eqn:problem3 1}) and (\ref{eqn:problem3 2}) can both be transformed
into convex optimization problems. 
Further, we show that the two convex formulations are the
Lagrangian duals of each other, which implies that they must have the same minimum sum power.  Once
this is proved, we can further infer that the feasible rate regions of the two problems
are identical. This is because the feasible rate regions can be
equivalently viewed as the sets of rate-tuples for which the minimum
values of the optimization problems (\ref{eqn:problem3 1}) and
(\ref{eqn:problem3 2}) are less than infinity. As both (\ref{eqn:problem3 1})
and (\ref{eqn:problem3 2}) can be reformulated as convex problems,
the minimum powers in the two problems are convex functions of $\{R_k\}$ and $\{C_m\}$ \cite{Boyd04}.
This means that the minimum powers in the two problems
are continuous functions of $\{R_k\}$. Since the minimum powers of the two problems
are the same whenever $\{R_k\}$ is strictly feasible for both problems, as $\{R_k\}$
approaches the feasibility boundary of one problem,
the minimum powers for both problems must go to infinity at the same time,
implying that the same $\{R_k\}$ must be approaching the feasibility boundary of
the other problem as well; thus the two problems must have the same feasibility region.
This same argument also applies to the proofs of Theorems \ref{theorem2} to \ref{theorem4}.
Thus in the rest of the proofs of all four theorems, we only show that
given a set of user rates $\{R_k\}$ and fronthaul capacities $\{C_m\}$ that are strictly feasible in both the multiple-access relay channel and the broadcast relay channel under the fixed beamforming vectors (\ref{eqn:beamforming vector}), the minimum sum powers of the two problems
are the same.

We remark that our previous work \cite{Liang16} provides a different approach to validate the equivalence between (\ref{eqn:problem3 1}) and (\ref{eqn:problem3 2}) based on the classic power control technique. This paper uses the alternative approach of showing the equivalence between (\ref{eqn:problem3 1}) and (\ref{eqn:problem3 2}) based on a Lagrangian duality technique. This allows a unified approach for proving Theorems \ref{theorem3} to \ref{theorem4}.
The proof involves the following two steps.

\begin{figure*}[!t]
\normalsize
\setcounter{equation}{65}
\begin{align}
 L(\{p_k^{\rm dl},\beta_k\},\{q_m^{{\rm dl}},\lambda_m\})
= &  \sum\limits_{k=1}^Kp_k^{\rm dl}\sigma^2+\sum\limits_{m=1}^Mq_m^{{\rm dl}}\sigma^2-\sum\limits_{k=1}^K\beta_k\bigg(\frac{p_k^{{\rm dl}}|\bar{\mv{u}}_k^H\mv{h}_k|^2}{2^{R_k}-1} -\sum\limits_{j\neq  k}p_j^{\rm dl}|\bar{\mv{u}}_j^H\mv{h}_k|^2-\sum\limits_{m=1}^Mq_m^{{\rm dl}}|h_{k,m}|^2-\sigma^2\bigg) \nonumber \\
& \quad +\sum\limits_{m=1}^M\lambda_m\left(\sum\limits_{k=1}^Kp_k^{\rm dl}|{\bar{u}}_{k,m}|^2+q_m^{\rm dl}-2^{C_m}q_m^{\rm dl}\right) \nonumber \\
= & \sum\limits_{k=1}^K\beta_k\sigma^2+\sum\limits_{k=1}^Kp_k^{{\rm dl}}\bigg(\sum\limits_{j\neq k}\beta_j|\bar{\mv{u}}_k^H\mv{h}_j|^2  +\sum\limits_{m=1}^M\lambda_m|\bar{u}_{k,m}|^2+\sigma^2-\frac{\beta_k|\bar{\mv{u}}_k^H\mv{h}_k|^2}{2^{R_k}-1}\bigg) \nonumber \\
& \quad +\sum\limits_{m=1}^Mq_m^{{\rm dl}}\left(\sum\limits_{k=1}^K\beta_k|h_{k,m}|^2+\lambda_m+\sigma^2-2^{C_m}\lambda_m\right). \label{eqn:Lagrangian}
\end{align}\hrulefill
\setcounter{equation}{62}
\end{figure*}

\subsection{Convex Reformulation of Problem (\ref{eqn:problem3 2}) and Its Dual Problem}\label{sec:Convex Reformulation}

First, we transform the problem (\ref{eqn:problem3 2}) for the broadcast relay channel into the following convex optimization problem:
\begin{align} \mathop{\mathrm{minimize}}_{\{p_k^{\rm dl}\},\{q_m^{{\rm dl}}\}} & ~ \sum\limits_{k=1}^Kp_k^{\rm dl}\sigma^2+\sum\limits_{m=1}^Mq_m^{{\rm dl}}\sigma^2  \label{eqn:problem3 3} \\
	 \mathrm {subject ~ to}  & ~ \frac{p_k^{{\rm dl}}|\bar{\mv{u}}_k^H\mv{h}_k|^2}{2^{R_k}-1}-\sum\limits_{j\neq  k}p_j^{\rm dl}|\bar{\mv{u}}_j^H\mv{h}_k|^2-\sum\limits_{m=1}^Mq_m^{{\rm dl}}|h_{k,m}|^2 \nonumber \\
& \qquad\qquad\qquad\qquad ~ -\sigma^2\geq 0, ~ \forall k\in \mathcal{K}, \label{eqn:downlink rate constraint3 1} \\
& ~ \sum\limits_{k=1}^Kp_k^{\rm dl}|{\bar{u}}_{k,m}|^2+q_m^{\rm dl}-2^{C_m}q_m^{\rm dl}\leq 0, ~ \forall m\in \mathcal{M}, \label{eqn:downlink fronthaul constraint3 2} \\ \hspace{-8pt} & ~ (\ref{eqn:nonnegative power}), ~ (\ref{eqn:positive quantization noise3}). \nonumber
\end{align}Note that in the objective function, the sum power is multiplied by a constant $\sigma^2$ without loss of generality. In fact, problem (\ref{eqn:problem3 3}) is an LP, since the objective function and constraints are all linear in $p_k^{{\rm dl}}$'s and $q_m^{{\rm dl}}$'s.

Take a set of strictly feasible $\{R_k\}$. Since the problem (\ref{eqn:problem3 3}) is a convex problem, 
strong duality holds \cite{Boyd04}, i.e., problem (\ref{eqn:problem3 3}) is equivalent to its dual problem.
In the following, we derive the dual problem of the problem (\ref{eqn:problem3 3}).

\setcounter{equation}{66}
The Lagrangian of the problem (\ref{eqn:problem3 3}) is \eqref{eqn:Lagrangian} shown at the top of the next page,
where $\beta_k\geq 0$'s and $\lambda_m\geq 0$'s are the dual variables associated with constraints (\ref{eqn:downlink rate constraint3 1}) and (\ref{eqn:downlink fronthaul constraint3 2}) in problem (\ref{eqn:problem3 3}), respectively. The dual function is then defined as
\begin{align}\label{eqn:dual function 1}
& g(\{\beta_k\},\{\lambda_m\}) \nonumber \\
& ~ =\min\limits_{p_k^{{\rm dl}}\geq 0, \forall k\in \mathcal{K},q_m^{{\rm dl}}\geq 0, \forall m\in \mathcal{M}} ~ L(\{p_k^{\rm dl},\beta_k\},\{q_m^{{\rm dl}},\lambda_m\})
\end{align}
Finally, the dual problem of problem (\ref{eqn:problem3 3}) is expressed as
\begin{align}
 \mathop{\mathrm{maximize}}_{\{\beta_k\},\{\lambda_m\}} & ~ g(\{\beta_k\},\{\lambda_m\})  \label{eqn:dual problem3 3} \\
 \mathrm {subject ~ to}  & ~ \beta_k\geq 0, ~~~ \forall k\in \mathcal{K}, \label{eqn:broadcast positve beta}\\ & ~ \lambda_m\geq 0, ~~~ \forall m\in \mathcal{M}. \label{eqn:broadcast positive lambda}
\end{align}
Note that according to (\ref{eqn:dual function 1}), $g(\{\beta_k\},\{\lambda_m\})=\sum_{k=1}^K\beta_k\sigma^2$ if and only if
\begin{align}
& \sum\limits_{j\neq k}\beta_j|\bar{\mv{u}}_k^H\mv{h}_j|^2+\sum\limits_{m=1}^M\lambda_m|\bar{u}_{k,m}|^2+\sigma^2 \nonumber \\
& \qquad\qquad\qquad\qquad ~~~ -\frac{\beta_k|\bar{\mv{u}}_k^H\mv{h}_k|^2}{2^{R_k}-1}
\geq 0, ~ \forall k\in \mathcal{K}, \label{eqn:dual constraint 1} \\
& \sum\limits_{k=1}^K\beta_k|h_{k,m}|^2+\lambda_m+\sigma^2-2^{C_m}\lambda_m\geq 0, ~ \forall m\in \mathcal{M}. \label{eqn:dual constraint fronthaul}
\end{align}
Otherwise, $g(\{\beta_k\},\{\lambda_m\})=-\infty$. As a result, problem (\ref{eqn:dual problem3 3}) can be transformed into the following equivalent problem:
\begin{align}
 \mathop{\mathrm{maximize}}_{\{\beta_k\},\{\lambda_m\}} & ~ \sum\limits_{k=1}^K\beta_k\sigma^2  \label{eqn:dual problem3 4} \\
 \mathrm {subject ~ to}  & ~ (\ref{eqn:broadcast positve beta}), ~ (\ref{eqn:broadcast positive lambda}), ~ (\ref{eqn:dual constraint 1}), ~ (\ref{eqn:dual constraint fronthaul}). \nonumber
\end{align}
This problem is now very similar to the multiple-access relay channel problem.
The physical interpretation of the above dual problem is the following. We can view $\beta_k$
as the transmit power of user $k$, $\forall k\in \mathcal{K}$,
and $\lambda_m$ as the quantization noise level of relay $m$ in the multiple-access relay channel,
$\forall m\in \mathcal{M}$. Then, problem (\ref{eqn:dual problem3 4}) aims to maximize the sum-power of all the users, while constraint (\ref{eqn:dual constraint 1}) requires that user $k$'s rate is no larger than $R_k$, $\forall k\in \mathcal{K}$, and constraint (\ref{eqn:dual constraint fronthaul}) requires that relay $m$'s fronthaul rate is no smaller than $C_m$, $\forall m\in \mathcal{M}$. In the following, we show that for the multiple-access relay channel, this power maximization problem (\ref{eqn:dual problem3 4}) is equivalent to the power minimization problem (\ref{eqn:problem3 1}).

\subsection{Equivalence Between Power Maximization Problem (\ref{eqn:dual problem3 4}) and Power Minimization Problem (\ref{eqn:problem3 1})}
To show the equivalence between problem (\ref{eqn:dual problem3 4}) and problem (\ref{eqn:problem3 1}), we first have the following proposition.
\begin{proposition}\label{proposition broadcast 1}
At the optimal solution to problem (\ref{eqn:dual problem3 4}), constraints (\ref{eqn:dual constraint 1}) and (\ref{eqn:dual constraint fronthaul}) should hold with equality, i.e.,
\begin{align}
& \sum\limits_{j\neq k}\beta_j|\bar{\mv{u}}_k^H\mv{h}_j|^2+\sum\limits_{m=1}^M\lambda_m|\bar{u}_{k,m}|^2+\sigma^2 \nonumber \\
& \qquad\qquad\qquad\qquad ~~~ -\frac{\beta_k|\bar{\mv{u}}_k^H\mv{h}_k|^2}{2^{R_k}-1}
= 0, ~ \forall k\in \mathcal{K}, \label{eqn:dual constraint 11} \\
& \sum\limits_{k=1}^K\beta_k|h_{k,m}|^2+\lambda_m+\sigma^2-2^{C_m}\lambda_m= 0, ~ \forall m\in \mathcal{M}. \label{eqn:dual constraint fronthaul 1}
\end{align}
\end{proposition}

\begin{IEEEproof}
Please refer to Appendix \ref{appendix broadcast 1}.
\end{IEEEproof}

To satisfy (\ref{eqn:broadcast positve beta}), (\ref{eqn:broadcast positive lambda}), (\ref{eqn:dual constraint 11}), and (\ref{eqn:dual constraint fronthaul 1}), it can be shown that
\begin{align}
& \beta_k>0, ~~~ \forall k\in \mathcal{K}, \label{eqn:broadcast positive beta 11} \\
& \lambda_m>0, ~~~ \forall m\in \mathcal{M}. \label{eqn:broadcast positive lambda 11}
\end{align}As a result, Proposition \ref{proposition broadcast 1} indicates that problem (\ref{eqn:dual problem3 4}) is equivalent to the following problem
\begin{align}
 \mathop{\mathrm{maximize}}_{\{\beta_k\},\{\lambda_m\}} & ~ \sum\limits_{k=1}^K\beta_k\sigma^2  \label{eqn:dual problem3 5} \\
 \mathrm {subject ~ to}  & ~ (\ref{eqn:dual constraint 11}), ~ (\ref{eqn:dual constraint fronthaul 1}), ~ (\ref{eqn:broadcast positive beta 11}), ~ (\ref{eqn:broadcast positive lambda 11}). \nonumber
\end{align}

\begin{proposition}\label{proposition broadcast 2}
If there exists one set of solutions $\beta_k$'s and $\lambda_m$'s that satisfies (\ref{eqn:dual constraint 11}), (\ref{eqn:dual constraint fronthaul 1}), (\ref{eqn:broadcast positive beta 11}), and (\ref{eqn:broadcast positive lambda 11}) in problem (\ref{eqn:dual problem3 5}), then this solution is unique.
\end{proposition}

\begin{IEEEproof}
Please refer to Appendix \ref{appendix broadcast 2}.
\end{IEEEproof}

Since there is a unique solution that satisfies all the constraints in problem (\ref{eqn:dual problem3 5}), it follows that the maximization problem (\ref{eqn:dual problem3 4}) is equivalent to the following minimization problem:
\begin{align}
 \mathop{\mathrm{minimize}}_{\{\beta_k\},\{\lambda_m\}} & ~ \sum\limits_{k=1}^K\beta_k\sigma^2  \label{eqn:dual problem3 6} \\
\mathrm {subject ~ to}  & ~ (\ref{eqn:dual constraint 11}), ~ (\ref{eqn:dual constraint fronthaul 1}), ~ (\ref{eqn:broadcast positive beta 11}), ~ (\ref{eqn:broadcast positive lambda 11}). \nonumber
\end{align}

At last, we relate problem (\ref{eqn:dual problem3 6}) to the power minimization problem (\ref{eqn:problem3 1}) in the multiple-access relay channel by the following proposition.
\begin{proposition}\label{proposition broadcast 3}
Problem (\ref{eqn:dual problem3 6}) is equivalent to
\begin{align}
 \mathop{\mathrm{minimize}}_{\{\beta_k\},\{\lambda_m\}} & ~ \sum\limits_{k=1}^K\beta_k\sigma^2  \label{eqn:dual problem3 7} \\
 \mathrm {subject ~ to}  & ~  \sum\limits_{j\neq k}\beta_j|\bar{\mv{u}}_k^H\mv{h}_j|^2+\sum\limits_{m=1}^M\lambda_m|\bar{u}_{k,m}|^2+\sigma^2 \nonumber \\
& \qquad\qquad ~ -\frac{\beta_k|\bar{\mv{u}}_k^H\mv{h}_k|^2}{2^{R_k}-1}\leq 0, ~ \forall k\in \mathcal{K}, \label{eqn:dual constraint 2} \\
& ~ \sum\limits_{k=1}^K\beta_k|h_{k,m}|^2+\lambda_m+\sigma^2-2^{C_m}\lambda_m \le 0, \nonumber \\
& \qquad\qquad\qquad\qquad\qquad\qquad ~~ \forall m\in \mathcal{M}, \label{eqn:dual constraint fronthaul 2} \\ & ~ (\ref{eqn:broadcast positive beta 11}), ~ (\ref{eqn:broadcast positive lambda 11}). \nonumber
\end{align}
\end{proposition}

\begin{IEEEproof}
Similar to Proposition \ref{proposition broadcast 1}, it can be shown that the optimal solution to problem (\ref{eqn:dual problem3 7}) must have its constraints (\ref{eqn:dual constraint 2}) and (\ref{eqn:dual constraint fronthaul 2}) satisfied with equality. As a result, problems (\ref{eqn:dual problem3 6}) and (\ref{eqn:dual problem3 7}) are equivalent to each other.
\end{IEEEproof}

The equivalence between the dual problem of the problem (\ref{eqn:problem3 3}), i.e., problem (\ref{eqn:dual problem3 4}), and the problem (\ref{eqn:dual problem3 7}) is therefore established. The key point here is that by viewing the dual variable $\beta_k$ as the transmit power of user $k$, $\forall k\in \mathcal{K}$, and the dual variable $\lambda_m$ as the quantization noise level of relay $m$, $\forall m \in \mathcal{M}$, in the multiple-access relay channel, the problem (\ref{eqn:dual problem3 7}) is exactly the power minimization problem (\ref{eqn:problem3 1}).
As a result, we have shown that the problem (\ref{eqn:problem3 2}) for the broadcast relay channel is equivalent to the problem (\ref{eqn:problem3 1}) for the multiple-access relay channel if the user rate targets $\{R_k\}$ and fronthaul rate constraints $\{C_m\}$ are strictly feasible in both the uplink and downlink.

\section{Proof of Theorem \ref{theorem2}}\label{sec 2}
In this section, we prove the duality between the multiple-access relay channel with successive interference cancellation at the CP as well as independent compression across the relays and the broadcast relay channel with dirty-paper coding at the CP as well as independent compression across the relays.

Similar to Section \ref{sec 1}, we fix the same beamforming vectors $\bar{\mv{u}}_k$'s in the multiple-access relay channel and the broadcast relay channel as shown in (\ref{eqn:beamforming vector}), where $\bar{\mv{u}}_k$'s satisfy (\ref{eqn:assumption}). Next, we assume that the successive interference cancellation order at the CP for the multiple-access relay channel and the dirty-paper encoding order at the CP for the broadcast relay channel are reverse of each other, i.e.,
\begin{align}\label{eqn:encoding decoding order}
\tau^{{\rm ul}}(k)=\tau^{{\rm dl}}(K+1-k)=\bar{\tau}(k), ~~~ \forall k\in \mathcal{K}.
\end{align}For example, if the decoding order in the multiple-access relay channel is $1,\ldots,K$, then the encoding order in the broadcast relay channel is $K,\ldots,1$, i.e., $\bar{\tau}^{{\rm ul}}(k)=k$, $\forall k$.

Let $\{R_k > 0, k \in \mathcal{K} \}$ and $\{C_m> 0, m \in \mathcal{M} \}$ be sets of strictly feasible user rate targets and fronthaul constraints for both the uplink and the downlink.
Given the receive beamforming vectors (\ref{eqn:beamforming vector}) and decoding order (\ref{eqn:encoding decoding order}), for the multiple-access relay channel, the transmit power minimization problem subject to the individual rate constraints as well as the individual fronthaul capacity constraints is formulated as
\begin{align} \mathop{\mathrm{minimize}}_{\{p_k^{\rm ul}\},\{q_m^{\rm ul}\}} & ~ P^{\rm ul}(\{p_k^{\rm ul}\})  \label{eqn:problem4 1} \\
 \mathrm {subject ~ to}  & ~ R_{\bar{\tau}(k)}^{\rm ul,SIC}(\{p_k^{\rm ul},\bar{\mv{u}}_k\},\{q_m^{\rm ul}\},\{\bar{\tau}(k)\}) \nonumber \\
& \qquad\qquad\qquad\qquad ~ \geq R_{\bar{\tau}(k)}, ~ \forall k \in \mathcal{K}, \label{eqn:uplink rate constraint4} \\  & ~ (\ref{eqn:uplink fronthaul constraint3}), ~ (\ref{eqn:uplink nonnegative power}), ~ (\ref{eqn:uplink positive}). \nonumber
\end{align}
Likewise, given the transmit beamforming vectors (\ref{eqn:beamforming vector}) and a reverse encoding order (\ref{eqn:encoding decoding order}), for the broadcast relay channel, the transmit power minimization problem is formulated as
\begin{align} \mathop{\mathrm{minimize}}_{\{p_k^{\rm dl}\},\mv{Q}_{{\rm diag}}} & ~ P^{\rm dl}(\{p_k^{\rm dl}\},\mv{Q}_{{\rm diag}})  \label{eqn:problem4 2} \\
 \mathrm {subject ~ to}  & ~ R_{\bar{\tau}(K+1-k)}^{\rm dl,DPC}(\{p_k^{\rm dl},\bar{\mv{u}}_k\}\},\mv{Q}_{{\rm diag}},\{\bar{\tau}(K+1-k)\}) \nonumber \\
& \qquad\qquad~ \geq R_{\bar{\tau}(K+1-k)}, ~ \forall k \in \mathcal{K}, \label{eqn:downlink rate constraint4} \\  & ~ (\ref{eqn:downlink fronthaul constraint3}), ~ (\ref{eqn:nonnegative power}), ~ (\ref{eqn:positive quantization noise3}). \nonumber
\end{align}

Similar to the equivalence between problem (\ref{eqn:problem3 1}) and problem (\ref{eqn:dual problem3 4}) in Case I, it can be shown that problem (\ref{eqn:problem4 1}) is equivalent to the following convex problem:
\begin{align} \mathop{\mathrm{maximize}}_{\{p_k^{\rm ul}\},\{q_m^{{\rm ul}}\}} & ~ \sum\limits_{k=1}^Kp_k^{\rm ul}\sigma^2  \label{eqn:problem4 4} \\
	 \mathrm {subject ~ to}  & ~  \sum\limits_{j> k}p_{\bar\tau(j)}^{{\rm ul}}|\bar{\mv{u}}_{\bar\tau(k)}^H\mv{h}_{\bar\tau(j)}|^2+\sum\limits_{m=1}^Mq_m^{{\rm ul}}|\bar{u}_{{\bar\tau(k)},m}|^2+\sigma^2 \nonumber \\
& \qquad\qquad ~ \geq \frac{p_{\bar\tau(k)}^{{\rm ul}}|\bar{\mv{u}}_{\bar\tau(k)}^H\mv{h}_{\bar\tau(k)}|^2}{2^{R_{\bar\tau(k)}}-1}, ~ \forall k\in \mathcal{K},  \\
& \sum\limits_{k=1}^Kp_k^{{\rm ul}}|h_{k,m}|^2+q_m^{{\rm ul}}+\sigma^2\geq 2^{C_m}q_m^{{\rm ul}}, ~ \forall m\in \mathcal{M}, \\ & ~ (\ref{eqn:uplink nonnegative power}), ~ (\ref{eqn:uplink positive}). \nonumber
\end{align}
Moreover, we can transform problem (\ref{eqn:problem4 2}) into the following convex problem:
\begin{align} \mathop{\mathrm{minimize}}_{\{p_k^{\rm dl}\},\{q_m^{{\rm dl}}\}} & ~ \sum\limits_{k=1}^Kp_k^{\rm dl}\sigma^2+\sum\limits_{m=1}^Mq_m^{{\rm dl}}\sigma^2  \label{eqn:problem4 3} \\
 \mathrm {subject ~ to}  & ~ \frac{p_{\bar{\tau}(k)}^{{\rm dl}}|\bar{\mv{u}}_{\bar{\tau}(k)}^H\mv{h}_{\bar{\tau}(k)}|^2}{2^{R_{\bar{\tau}(k)}}-1}\geq \sum\limits_{j\leq  k}p_{\bar{\tau}(j)}^{\rm dl}|\mv{v}_{\bar{\tau}(j)}^H\mv{h}_{\bar{\tau}(k)}|^2 \nonumber \\
& \qquad ~ +\sum\limits_{m=1}^Mq_m^{{\rm dl}}|h_{\bar{\tau}(k),m}|^2+\sigma^2, ~ \forall k \in \mathcal{K}, \label{eqn:downlink rate constraint4 1} \\  & ~ (\ref{eqn:downlink fronthaul constraint3 2}), ~ (\ref{eqn:nonnegative power}), ~ (\ref{eqn:positive quantization noise3}). \nonumber
\end{align}

Similar to Case I, we can show that problem (\ref{eqn:problem4 4}) is the Lagrangian dual of problem (\ref{eqn:problem4 3}). Since the method adopted is almost exactly the same as that in Section \ref{sec 1}, we omit the details here. As a result, under the same fronthaul capacity constraints $\{C_m\}$ and total transmit power constraint $P$, any rate-tuple achievable in the multiple-access relay channel can be shown to be achievable also in the broadcast relay channel by setting the transmit beamforming vectors as the receive beamforming vectors in the multiple-access relay channel and the encoding order to be reverse of the decoding order in the multiple-access channel, and vice versa. By trying all the feasible beamforming vectors, it follows that
\begin{align}
\bar{\mathcal{R}}_{{\rm II}}^{\rm ul}(\{C_m\},P,\{\bar{\tau}(k)\})=\mathcal{\bar R}_{{\rm II}}^{\rm dl}(\{C_m\},P,\{\bar{\tau}(K+1-k)\}).
\end{align}Then, by trying all the encoding/decoding orders, we can show that the achievable rate regions of the multiple-access relay channel under successive interference cancellation as well as independent compression and the broadcast relay channel under dirty-paper coding as well as independent compression are identical, i.e., $\mathcal{R}_{{\rm II}}^{\rm ul}(\{C_m\},P)=\mathcal{R}_{{\rm II}}^{\rm dl}(\{C_m\},P)$.

\section{Proof of Theorem \ref{theorem1}}\label{sec 3}
In this section, we prove the duality between the multiple-access relay channel with linear decoding at the CP as well as Wyner-Ziv compression across the relays and the broadcast relay channel with linear encoding at the CP as well as multivariate compression across the relays.

Similar to Sections \ref{sec 1} and \ref{sec 2}, we fix the same beamforming vectors $\bar{\mv{u}}_k$'s in the multiple-access relay channel and the broadcast relay channel as shown in (\ref{eqn:beamforming vector}), where $\bar{\mv{u}}_k$'s satisfy (\ref{eqn:assumption}). Next, for Wyner-Ziv compression across the relays in the multiple-access relay channel and multivariate compression across relays in the broadcast relay channel, we assume that the decompression order is the reverse of the compression order, i.e.,
\begin{align}\label{eqn:compression decompression order}
\rho^{{\rm ul}}(m)=\rho^{{\rm dl}}(M+1-m)=\bar{\rho}(m), ~~~ \forall m\in \mathcal{M}.
\end{align}For example, if the decompression order in the multiple-access relay channel is $1,\ldots,M$, then the compression order in the broadcast relay channel is $M,\ldots,1$, i.e., $\bar{\rho}(m)=m$, $\forall m$.

Let $\{R_k > 0, k \in \mathcal{K} \}$ and $\{C_m> 0, m \in \mathcal{M} \}$ be sets of strictly feasible user rate targets and fronthaul constraints for both the uplink and the downlink.
For the multiple-access relay channel, given the beamforming vectors (\ref{eqn:beamforming vector}) and the decompression order (\ref{eqn:compression decompression order}), the transmit power minimization problem subject to the individual rate constraints as well as the individual fronthaul capacity constraints is formulated as
\begin{align} \mathop{\mathrm{minimize}}_{\{p_k^{\rm ul}\},\{q_m^{\rm ul}\}} & ~ P^{\rm ul}(\{p_k^{\rm ul}\})  \label{eqn:problem 1} \\
 \mathrm {subject ~ to}  & ~ C_{\bar{\rho}(m)}^{{\rm ul,WZ}}(\{p_k^{\rm ul}\},q_{\rho^{{\rm ul}}(1)}^{\rm ul},\ldots,q_{\rho^{{\rm ul}}(m)}^{\rm ul},\{\bar{\rho}(m)\}) \nonumber \\ & \qquad\qquad\qquad\qquad ~ \leq C_{\bar{\rho}(m)}, ~ \forall m \in \mathcal{M}, \label{eqn:uplink fronthaul constraint} \\  & ~ (\ref{eqn:uplink rate constraint}), ~ (\ref{eqn:uplink nonnegative power}), ~ (\ref{eqn:uplink positive}). \nonumber
\end{align}
Likewise, 
for the broadcast relay channel as described in Section \ref{sec:Broadcast Relay Channel}, given the same transmit beamforming vectors (\ref{eqn:beamforming vector}) and a reverse compression order (\ref{eqn:compression decompression order}), the transmit power minimization problem is formulated as
\begin{align} \mathop{\mathrm{minimize}}_{\{p_k^{\rm dl}\},\mv{Q}} & ~ P^{\rm dl}(\{p_k^{\rm dl}\},\mv{Q})  \label{eqn:problem 2} \\
 \mathrm {subject ~ to}  & ~ C_{\bar{\rho}(M+1-m)}^{\rm dl,MV}(\{p_k^{\rm dl},\bar{\mv{u}}_k\},\mv{Q},\{\bar{\rho}(M+1-m)\}) \nonumber \\ & \qquad\qquad\qquad ~ \leq C_{\bar{\rho}(M+1-m)}, ~ \forall m \in \mathcal{M}, \label{eqn:downlink fronthaul constraint} \\ \hspace{-8pt} & ~ \mv{Q}\succeq \mv{0}, \label{eqn:positive} \\  & ~ (\ref{eqn:downlink rate constraint}), ~ (\ref{eqn:nonnegative power}).  \nonumber
\end{align}
If we can show that problem (\ref{eqn:problem 1}) and problem (\ref{eqn:problem 2}) are equivalent, then it implies that under the same fronthaul capacity constraints $\{C_m\}$ and total transmit power constraint $P$, any achievable rate-tuple in the multiple-access relay channel must also be achievable in the broadcast relay channel by setting the transmit beamforming vectors as the receive beamforming vectors in the multiple-access relay channel and the compression order as the reverse of the decompression order in the multiple-access channel, and vice versa. In addition, by trying all the feasible beamforming vectors, it follows that
\begin{align}
\bar{\mathcal{R}}_{{\rm III}}^{\rm ul}(\{C_m\},P,\{\bar{\rho}(m)\})=\mathcal{\bar R}_{{\rm III}}^{\rm dl}(\{C_m\},P,\{\bar{\rho}(M+1-m)\}).
\end{align}
Finally, by trying all the compression/decompression orders, we can show that the achievable rate regions of the multiple-access relay channel under linear decoding as well as Wyner-Ziv compression and the broadcast relay channel under linear encoding as well as multivariate compression are identical, i.e., $\mathcal{R}_{{\rm III}}^{\rm ul}(\{C_m\},P)=\mathcal{R}_{{\rm III}}^{\rm dl}(\{C_m\},P)$.

The key to proving Theorem \ref{theorem1} is therefore to show that the power minimization problems (\ref{eqn:problem 1}) and (\ref{eqn:problem 2}) are equivalent.
However,
different from problems (\ref{eqn:problem3 1}) and (\ref{eqn:problem3 2}) in Section \ref{sec 1} or problems (\ref{eqn:problem4 1}) and (\ref{eqn:problem4 2}) in Section \ref{sec 2}, problems (\ref{eqn:problem 1}) and (\ref{eqn:problem 2}) cannot be transformed into an LP due to the complicated expression of the fronthaul rates given in (\ref{eqn:uplink fronthaul rate}) and (\ref{eqn:downlink fronthaul rate}). In the rest of this section, we validate the equivalence between problems (\ref{eqn:problem 1}) and (\ref{eqn:problem 2}) based on Lagrangian duality of SDP. For convenience, we merely consider the case when the decompression order in the multiple-access relay channel and the compression order in the broadcast relay channel are respectively set as
\begin{align}
& \rho^{{\rm ul}}(m)=m, ~~~ \forall m\in \mathcal{M}, \label{eqn:de order} \\
& \rho^{{\rm dl}}(m)=M+1-m, ~~~ \forall m\in \mathcal{M}.\label{eqn:com order}
\end{align}For the other decompression order and the corresponding reversed compression order, the equivalence between problems (\ref{eqn:problem 1}) and (\ref{eqn:problem 2}) can be proved in a similar way.
The proof involves the following three steps.

\setcounter{equation}{105}
\begin{figure*}[t]
	\normalsize\begin{align}
\mathop{\mathrm{maximize}}_{\{\beta_k\},\{\mv{\Lambda}_m\}} & ~ \sum\limits_{k=1}^K\beta_k \sigma^2 \label{eqn:problem 4} \\
\mathrm {subject ~ to} & ~ \sigma^2+\sum\limits_{j\neq k}\beta_j|\bar{\mv{u}}_k^H\mv{h}_j|^2+\sum\limits_{m=1}^M\mv{\Lambda}_m^{(m,m)}|\bar{u}_{k,m}|^2 -\frac{\beta_k|\bar{\mv{u}}_k^H\mv{h}_k|^2}{2^{R_k}-1}\geq 0, ~ \forall k\in \mathcal{K}, \label{eqn:dual uplink SINR constraint} \\ & ~ \sigma^2\mv{I}+\sum\limits_{k=1}^K\beta_k\mv{h}_k\mv{h}_k^H+\sum\limits_{m=1}^M\mv{E}_m^H\mv{\Lambda}_m\mv{E}_m -  \sum\limits_{m=1}^M2^{C_m}\left[\begin{array}{cc}\mv{0}_{(m-1)\times (m-1)} & \mv{0}_{(m-1)\times (M-m+1)} \\ \mv{0}_{(M-m+1)\times (m-1)} & \mv{\Lambda}_m^{(m:M,m:M)}\end{array}\right] \succeq \mv{0}, \label{eqn:dual uplink fronthaul constraint} \\ & ~ \beta_k\geq 0, ~ \forall k\in \mathcal{K}, \label{eqn:nonegative beta} \\ & ~ \mv{\Lambda}_m\succeq \mv{0}, ~ \forall m\in \mathcal{M}. \label{eqn:nonegative Phi}
\end{align}\hrulefill
\end{figure*}
\setcounter{equation}{110}
\begin{figure*}[!t]
	\normalsize\begin{align}
\mathop{\mathrm{maximize}}_{\{\beta_k\},\{\mv{\Lambda}_m,\mv{A}_m\}} & ~ \sum\limits_{k=1}^K\beta_k \sigma^2 \label{eqn:problem 5} \\
\mathrm {subject ~ to} \ \ & ~ \sigma^2\mv{I}+\sum\limits_{k=1}^K\beta_k\mv{h}_k\mv{h}_k^H+\sum\limits_{m=1}^M\mv{E}_m^H\mv{\Lambda}_m\mv{E}_m \succeq \mv{A}_1, \label{eqn:dual uplink fronthaul constraint 1} \\ & ~ \mv{A}_m=2^{C_m}\mv{\Lambda}_m^{(m:M,m:M)}+\left[\begin{array}{cc}0 & \mv{0}_{1\times (M-m)} \\ \mv{0}_{(M-m)\times 1} & \mv{A}_{m+1} \end{array}\right], ~ m=1,\ldots,M-1, \label{eqn:A1} \\ & ~ \mv{A}_M=2^{C_M}\mv{\Lambda}_M^{(M,M)}, \label{eqn:A2} \\ & ~ (\ref{eqn:dual uplink SINR constraint}), ~ (\ref{eqn:nonegative beta}), ~ (\ref{eqn:nonegative Phi}). \nonumber
\end{align}\hrulefill 
\end{figure*}
\setcounter{equation}{102}

\subsection{Convex Reformulation of Problem (\ref{eqn:problem 2}) and Its Dual Problem}
First, we transform problem (\ref{eqn:problem 2}) into an equivalent convex problem.

\begin{proposition}\label{proposition1}
Power minimization problem (\ref{eqn:problem 2}) in the broadcast relay channel is equivalent to the following problem:
\begin{align} \mathop{\mathrm{minimize}}_{\{p_k^{\rm dl}\},\mv{Q}} & ~ \sum\limits_{k=1}^Kp_k^{\rm dl}\sigma^2+{\rm tr}(\mv{Q})\sigma^2  \label{eqn:problem 3} \\
 \mathrm {subject ~ to}  & ~ \frac{p_k^{\rm dl}|\bar{\mv{u}}_k^H\mv{h}_k|^2}{2^{R_k}-1}\geq \sum\limits_{j\neq k} p_j^{\rm dl}|\bar{\mv{u}}_j^H\mv{h}_k|^2+{\rm tr}(\mv{Q}\mv{h}_k\mv{h}_k^H) \nonumber \\ & \qquad\qquad\qquad\qquad\qquad ~ +\sigma^2, ~~~ \forall k\in \mathcal{K}, \label{eqn:downlink convex rate constraint} \\ & ~ 2^{C_m}\left[\begin{array}{cc}\mv{0}_{(m-1)\times (m-1)} & \mv{0}_{(m-1)\times (M-m+1)} \\ \mv{0}_{(M-m+1)\times (m-1)} & \mv{Q}^{(m:M,m:M)}\end{array}\right] \nonumber \\ & \quad ~ -\mv{E}_m(\mv{Q}+\mv{\Psi})\mv{E}_m^H\succeq \mv{0}, ~~~ \forall m\in \mathcal{M}, \label{eqn:downlink convex fronthaul constraint} \\ & ~ (\ref{eqn:nonnegative power}), ~ (\ref{eqn:positive}), \nonumber
\end{align}where $\mv{E}_m\in \mathbb{C}^{M\times M}$ denotes the matrix where the $m$-th diagonal element is $1$, while the other elements are $0$, and $\mv{\Psi}={\rm diag}\left(\sum_{k=1}^Kp_k^{\rm dl}|\bar{u}_{k,1}|^2,\ldots,\sum_{k=1}^Kp_k^{\rm dl}|\bar{u}_{k,M}|^2\right)$.
\end{proposition}
\begin{IEEEproof}
Please refer to Appendix \ref{appendix1}.
\end{IEEEproof}

It can be seen that problem (\ref{eqn:problem 3}) is an SDP problem and is convex. Further, by our choice of feasible $\{R_k\}$, it satisfies the Slater's condition. As a result, strong duality holds for problem (\ref{eqn:problem 3}). In other words, problem (\ref{eqn:problem 3}) is equivalent to its dual problem. In the following, we derive the dual problem of (\ref{eqn:problem 3}).

\begin{proposition}\label{proposition2}
The dual problem of problem (\ref{eqn:problem 3}) is problem \eqref{eqn:problem 4} at the top of the page,
where $\beta_k$'s and $\mv{\Lambda}_m\in \mathbb{C}^{M\times M}$'s are the dual variables associated with constraints (\ref{eqn:downlink convex rate constraint}) and (\ref{eqn:downlink convex fronthaul constraint}), respectively.
\end{proposition}

\begin{IEEEproof}
Please refer to Appendix \ref{appendix2}.
\end{IEEEproof}

It can be observed that similar to the uplink-downlink duality shown in Section \ref{sec 1}, the dual problem (\ref{eqn:problem 4}) of the power minimization problem (\ref{eqn:problem 3}) in the broadcast relay channel is closely related to the power minimization problem (\ref{eqn:problem 1}) of the multiple-access relay channel.
Specifically, if we interpret the dual variables $\beta_k$'s
as the uplink transmit powers $p_k^{{\rm ul}}$'s and
the dual variables $\mv{\Lambda}_m^{(m,m)}$'s as the
uplink quantization noise levels $q_m^{{\rm ul}}$'s, then the objective of problem (\ref{eqn:problem 4}) is to maximize the total transmit power, and constraint (\ref{eqn:dual uplink SINR constraint}) is to make sure that the rate of each user $k$ in the multiple-access relay channel is no larger than its rate requirement. The remaining challenge is to transform constraint (\ref{eqn:dual uplink fronthaul constraint}) in problem (\ref{eqn:problem 4}) into a set of $M$ fronthaul capacity constraints that are in the same form of constraints (\ref{eqn:uplink fronthaul constraint}) in problem (\ref{eqn:problem 1}).
Note that in contrast to the case with independent compression shown in Section \ref{sec 1}, where constraint (\ref{eqn:dual constraint fronthaul}) of the dual problem (\ref{eqn:dual problem3 4}) in the broadcast relay channel is directly the reverse of constraint (\ref{eqn:uplink fronthaul constraint3}) of problem (\ref{eqn:problem3 1}) in the multiple-access relay channel,
in the case with Wyner-Ziv compression and multivariate compression, 
the validation of the equivalence between problem (\ref{eqn:problem 1}) and problem (\ref{eqn:problem 2}) is considerably more complicated.

\setcounter{equation}{118}
\begin{figure*}[!t]
	\normalsize\begin{align}
\mathop{\mathrm{maximize}}_{\{\beta_k\},\{\mv{\Lambda}_m^{(m,m)}\}} & ~ \sum\limits_{k=1}^K\beta_k \sigma^2 \label{eqn:problem 6} \\
	\mathrm {subject ~ to} \ \ & ~ 2^{C_m}\mv{\Lambda}_m^{(m,m)}\leq \mv{\Omega}^{(m,m)} -\mv{\Omega}^{(m,1:m-1)}(\mv{\Omega}^{(1:m-1,1:m-1)})^{-1}\mv{\Omega}^{(1:m-1,m)}, ~ \forall m \in \mathcal{M}, \label{eqn:eqv dual uplink fronthaul constraint} \\ & ~ \mv{\Lambda}_m^{(m,m)} \ge 0, \label{eqn:positive Phi} \\ & ~ (\ref{eqn:dual uplink SINR constraint}), ~ (\ref{eqn:nonegative beta}). \nonumber
\end{align}\hrulefill 
\end{figure*}
\setcounter{equation}{114}

\subsection{Equivalent Transformation of Constraint (\ref{eqn:dual uplink fronthaul constraint})
in the Dual Problem (\ref{eqn:problem 4})}\label{sec 31}

In the following, we transform constraint (\ref{eqn:dual uplink fronthaul constraint}) in problem (\ref{eqn:problem 4}) into the form of constraint (\ref{eqn:uplink fronthaul constraint}) in problem (\ref{eqn:problem 1}). First, we introduce some auxiliary variables to problem (\ref{eqn:problem 4}).

\begin{proposition}\label{proposition3}
Problem (\ref{eqn:problem 4}) is equivalent to problem \eqref{eqn:problem 5} at the top of the page,
where $\mv{A}_m\in \mathbb{C}^{(M-m+1)\times (M-m+1)}$'s, $m=1,\ldots,M$, are auxiliary variables.
\end{proposition}
\setcounter{equation}{114}

\begin{IEEEproof}
Please refer to Appendix \ref{appendix3}.
\end{IEEEproof}

Next, we show some key properties of the optimal solution to problem (\ref{eqn:problem 5}).
\begin{proposition}\label{lemma1}
The optimal solution to problem (\ref{eqn:problem 5}) must satisfy
\begin{align}
& \sigma^2\mv{I}+\sum\limits_{k=1}^K\beta_k\mv{h}_k\mv{h}_k^H+\sum\limits_{m=1}^M\mv{E}_m^H\mv{\Lambda}_m\mv{E}_m = \mv{A}_1, \label{eqn:optimal condition 4} \\
& \mv{A}_m\succ \mv{0}, ~~~ m=1,\ldots,M, \label{eqn:optimal condition 2} \\
& 2^{C_m}\mv{\Lambda}_m^{(m:M,m:M)}=\frac{\mv{A}_m^{(1:M-m+1,1)}\mv{A}_m^{(1,1:M-m+1)}}{\mv{A}_m^{(1,1)}}, ~ \forall m. \label{eqn:optimal condition}
\end{align}
\end{proposition}

\begin{IEEEproof}
Please refer to Appendix \ref{appendix4}.
\end{IEEEproof}

Proposition \ref{lemma1} indicates that the optimal $\mv{\Lambda}_m$'s to problem (\ref{eqn:problem 5}) are rank-one matrices. This property can be utilized to simplify problem (\ref{eqn:problem 5}) as follows.

\begin{proposition}\label{proposition4}
Define
\begin{align}\label{eqn:Omega}
\mv{\Omega}&=\sigma^2\mv{I}+\sum\limits_{k=1}^K\beta_k\mv{h}_k\mv{h}_k^H+\sum\limits_{m=1}^M\mv{E}_m^H\mv{\Lambda}_m\mv{E}_m \nonumber \\ & =\sigma^2\mv{I}+\sum\limits_{k=1}^K\beta_k\mv{h}_k\mv{h}_k^H+{\rm diag}(\mv{\Lambda}_1^{(1,1)},\ldots,\mv{\Lambda}_M^{(M,M)})\succ \mv{0}.
\end{align}Then, problem (\ref{eqn:problem 5}) is equivalent to problem \eqref{eqn:problem 6} at the top of the page.
\end{proposition}
\setcounter{equation}{121}

\begin{IEEEproof}
Please refer to Appendix \ref{appendix5}.
\end{IEEEproof}

Note that the difference between problem (\ref{eqn:problem 6}) and problem (\ref{eqn:problem 4}) are two-fold. First, the optimization variables reduce from $\mv{\Lambda}_m$'s to their $m$-th diagonal elements, i.e., $\mv{\Lambda}_m^{(m,m)}$'s. Second, constraint (\ref{eqn:dual uplink fronthaul constraint}) in the matrix form reduces to $M$ constraints given in (\ref{eqn:eqv dual uplink fronthaul constraint}) in the scalar form.

If we interpret the dual variables $\beta_k$'s as the transmit powers $p_k^{{\rm ul}}$'s
and the dual variables $\mv{\Lambda}_m^{(m,m)}$'s as the quantization noise levels $q_m^{{\rm ul}}$'s in the multiple-access relay channel, then problem (\ref{eqn:problem 6}) is equivalent to the maximization of the transmit power subject to the constraints that the rate of each user $k$ is no larger than $R_k$, $\forall k$, and the fronthaul rate of each relay $m$ is no smaller than $C_m$, $\forall m$. As a result, problem (\ref{eqn:problem 6}) is a reverse problem to the power minimization problem (\ref{eqn:problem 1}) in the multiple-access relay channel. In the following, we show that this problem is indeed equivalent to the power minimization problem (\ref{eqn:problem 1}).

\subsection{Equivalence Between Power Maximization Problem (\ref{eqn:problem 6}) and Power Minimization Problem (\ref{eqn:problem 1})}\label{sec:step 3}

First, we prove one important property of the optimal solution to problem (\ref{eqn:problem 6}).

\begin{proposition}\label{proposition5}
The optimal solution to problem (\ref{eqn:problem 6}) must have its constraints (\ref{eqn:dual uplink SINR constraint}) and (\ref{eqn:eqv dual uplink fronthaul constraint}) satisfied with equality, i.e.,
\begin{align}
& \sigma^2+\sum\limits_{j\neq k}\beta_j|\bar{\mv{u}}_k^H\mv{h}_j|^2+\sum\limits_{m=1}^M\mv{\Lambda}_m^{(m,m)}|\bar{u}_{k,m}|^2
-\frac{\beta_k|\bar{\mv{u}}_k^H\mv{h}_k|^2}{2^{R_k}-1}
\nonumber \\
& \qquad\qquad\qquad\qquad\qquad\qquad\qquad =  0, ~ \forall k\in \mathcal{K}, \label{eqn:dual uplink SINR constraint 01} \\
& 2^{C_m}\mv{\Lambda}_m^{(m,m)} = \nonumber \\ & \quad \mv{\Omega}^{(m,m)} -\mv{\Omega}^{(m,1:m-1)}(\mv{\Omega}^{(1:m-1,1:m-1)})^{-1}\mv{\Omega}^{(1:m-1,m)}, \nonumber \\
& \qquad\qquad\qquad\qquad\qquad\qquad\qquad\qquad ~ \forall m\in \mathcal{M}. \label{eqn:eqv dual uplink fronthaul constraint 01}
\end{align}
\end{proposition}

\begin{IEEEproof}
Please refer to Appendix \ref{appendix6}.
\end{IEEEproof}

According to Proposition \ref{proposition5}, problem (\ref{eqn:problem 6}) is equivalent to the following problem
\begin{align}\hspace{-8pt}
\mathop{\mathrm{maximize}}_{\{\beta_k\},\{\mv{\Lambda}_m^{(m,m)}\}} & ~ \sum\limits_{k=1}^K\beta_k \sigma^2 \label{eqn:problem 7} \\
\mathrm {subject ~ to} \ \ & ~ (\ref{eqn:nonegative beta}), ~ (\ref{eqn:positive Phi}), ~ (\ref{eqn:dual uplink SINR constraint 01}), ~ (\ref{eqn:eqv dual uplink fronthaul constraint 01}). \nonumber
\end{align}

Next, we show one important property of the optimal solution to problem (\ref{eqn:problem 7}).

\begin{proposition}\label{proposition6}
	If there exists one set of solutions $\{\beta_k, \mv{\Lambda}_m^{(m,m)}\}$ that satisfies the constraints (\ref{eqn:nonegative beta}), (\ref{eqn:positive Phi}), (\ref{eqn:dual uplink SINR constraint 01}), and (\ref{eqn:eqv dual uplink fronthaul constraint 01}) in problem (\ref{eqn:problem 7}), then this solution is unique.
\end{proposition}

\begin{IEEEproof}
Please refer to Appendix \ref{appendix7}.
\end{IEEEproof}

Note that the above proposition is very similar to Proposition \ref{proposition broadcast 2} for the case with independent compression shown in Section \ref{sec 1}. However, the proof of Proposition \ref{proposition6} is much more complicated. This is because for the case with independent compression, equations in (\ref{eqn:dual constraint 11}) and (\ref{eqn:dual constraint fronthaul 1}) are linear in terms of $\beta_k$'s and $\lambda_m$'s, while for the case with Wyner-Ziv compression and multivariate compression, equations in (\ref{eqn:eqv dual uplink fronthaul constraint 01}) are not linear in terms of $\beta_k$'s and $\mv{\Lambda}^{(m,m)}$'s.

Since there is a unique solution that satisfies all the constraints in problem (\ref{eqn:problem 7}), it follows that maximization problem (\ref{eqn:problem 7}) is equivalent to the following minimization problem
\begin{align}\hspace{-8pt}
\mathop{\mathrm{minimize}}_{\{\beta_k\},\{\mv{\Lambda}_m^{(m,m)}\}} & ~ \sum\limits_{k=1}^K\beta_k \sigma^2 \label{eqn:problem 9} \\
\mathrm {subject ~ to} \ \ & ~ (\ref{eqn:nonegative beta}), ~ (\ref{eqn:positive Phi}), ~ (\ref{eqn:dual uplink SINR constraint 01}), ~ (\ref{eqn:eqv dual uplink fronthaul constraint 01}). \nonumber
\end{align}

Last, we relate problem (\ref{eqn:problem 9}) to the power minimization problem (\ref{eqn:problem 1}) in the multiple-access relay channel by the following proposition.

\begin{figure*}[!t]
	\normalsize\begin{align}\hspace{-8pt}
\mathop{\mathrm{minimize}}_{\{\beta_k\},\{\mv{\Lambda}_m^{(m,m)}\}} & ~ \sum\limits_{k=1}^K\beta_k \sigma^2 \label{eqn:problem 10} \\
	\mathrm {subject ~ to} \ \ & ~ \sigma^2+\sum\limits_{j\neq k}\beta_j|\bar{\mv{u}}_k^H\mv{h}_j|^2+\sum\limits_{m=1}^M\mv{\Lambda}_m^{(m,m)}|\bar{u}_{k,m}|^2 -\frac{\beta_k|\bar{\mv{u}}_k^H\mv{h}_k|^2}{2^{R_k}-1}\leq   0, ~ \forall k \in \mathcal{K}, \label{eqn:dual uplink SINR constraint 02} \\  & ~ 2^{C_m}\mv{\Lambda}_m^{(m,m)}\geq \mv{\Omega}^{(m,m)} -\mv{\Omega}^{(m,1:m-1)}(\mv{\Omega}^{(1:m-1,1:m-1)})^{-1}\mv{\Omega}^{(1:m-1,m)}, ~ \forall m \in \mathcal{M}. \label{eqn:eqv dual uplink fronthaul constraint 02} \\ & ~ (\ref{eqn:nonegative beta}), ~ (\ref{eqn:positive Phi}). \nonumber
\end{align}\hrulefill 
\end{figure*}

\begin{proposition}\label{proposition7}
Problem (\ref{eqn:problem 9}) is equivalent to problem \eqref{eqn:problem 10} at the top of the next page.
\end{proposition}

\begin{IEEEproof}
Similar to Proposition \ref{proposition5}, it can be shown that the optimal solution to problem (\ref{eqn:problem 10}) must have its constraints, shown as (\ref{eqn:dual uplink SINR constraint 02}) and (\ref{eqn:eqv dual uplink fronthaul constraint 02}) on top of the next page, satisfied with equality. As a result, problem (\ref{eqn:problem 9}) is equivalent to problem (\ref{eqn:problem 10}). Proposition \ref{proposition7} is thus proved.
\end{IEEEproof}

In problem (\ref{eqn:problem 10}), the dual variables $\beta_k$'s
can be viewed as the transmit powers $p_k^{{\rm ul}}$'s,
and the dual variables $\mv{\Lambda}_m^{(m,m)}$'s can be viewed as the
quantization noise levels $q_m^{{\rm ul}}$'s in the multiple-access relay channel. Moreover, according to (\ref{eqn:Gamma}) and (\ref{eqn:Omega}), $\mv{\Omega}$ can be viewed as the covariance matrix of the received signal in the multiple-access relay channel, i.e., $\mv{\Gamma}$ as defined by (\ref{eqn:Gamma}). As a result, by combining Propositions \ref{proposition1} -- \ref{proposition7}, we can conclude that the power minimization problem (\ref{eqn:problem 2}) for the broadcast relay channel is equivalent to problem (\ref{eqn:problem 10}), and is thus equivalent to power minimization problem (\ref{eqn:problem 1}) for the multiple-access relay channel.
Theorem \ref{theorem1} is thus proved.

To summarize the difference in the methodology for proving Theorem \ref{theorem3} and Theorem \ref{theorem1}, we note that although the proof of Theorem \ref{theorem1} for
the case of Wyner-Ziv compression in the multiple-access relay channel and multivariate compression in the broadcast relay channel follows the same line of reasoning as the proof of Theorem \ref{theorem3} for the case with independent compression (i.e., we first use the Lagrangian duality method to find the dual problem of the power minimization problem in the broadcast relay channel, then show that this dual problem is equivalent to the power minimization problem in the multiple-access relay channel), the validation of the equivalence between the dual problem in the broadcast relay channel and the power minimization problem in the multiple-access channel is much more involved for the case of Wyner-Ziv and multivariate compression. This is because (i) the downlink power minimization problem becomes an SDP, rather than an LP for the case of independent compression, and the duality of SDPs are more complicated than that of LPs; (ii) the step in Section \ref{sec 31} is needed, since constraint (\ref{eqn:dual uplink fronthaul constraint}) in the dual problem (\ref{eqn:problem 4}) is not a direct reverse of constraint (\ref{eqn:uplink fronthaul constraint}) in problem (\ref{eqn:problem 1}); (iii)
Proposition \ref{proposition6} involves nonlinear equations, and it is more much difficult to show that its solution is unique.

\section{Proof of Theorem \ref{theorem4}}\label{sec 4}
In this section, we prove the duality between the multiple-access relay channel with successive interference cancellation at the CP as well as Wyner-Ziv compression across the relays and the broadcast relay channel with dirty-paper coding at the CP as well as multivariate compression across the relays.

Similar to Sections \ref{sec 1}, \ref{sec 2}, and \ref{sec 3}, we fix the same beamforming vectors $\bar{\mv{u}}_k$'s in the multiple-access relay channel and the broadcast relay channel as shown in (\ref{eqn:beamforming vector}), where $\bar{\mv{u}}_k$'s satisfy (\ref{eqn:assumption}). Next, for successive interference cancellation at the CP in the multiple-access relay channel and dirty-paper coding at the CP in the broadcast relay channel, we assume that the decoding order is the reverse of the encoding order, i.e., (\ref{eqn:encoding decoding order}). For example, if the decoding order in the multiple-access relay channel is $1,\ldots,K$, then the encoding order in the broadcast relay channel is $K,\ldots,1$. Moreover, for Wyner-Ziv compression across the relays in the multiple-access relay channel and multivariate compression across the relays in the broadcast relay channel, we assume the decompression order is the reverse of the compression order, i.e., (\ref{eqn:compression decompression order}). For example, if the decompression order in the multiple-access relay channel is $1,\ldots,M$, then the compression order in the broadcast relay channel is $M,\ldots,1$.

\begin{table*}
\begin{center}
\newcommand{\tabincell}[2]{\begin{tabular}{@{}#1@{}}#2\end{tabular}}
\caption{Primal and dual relationships between the sum-power minimization problems for the multiple-access relay channel and the broadcast relay channel}
\label{table:interpretation}
\begin{tabular}{|c|l|l|}
\hline
	\multirow{2}{*}{} & \multicolumn{1}{c|}{Broadcast relay channel} & \multicolumn{1}{c|}{Multiple-access relay channel} \\ \cline{2-3}
	& \multicolumn{2}{c|}{Fixing beamformers $\{\mathbf{w}_k = \mathbf{v}_k$\}} \\ \hline
	Case I & \tabincell{l}{Primal problem: Power minimization (\ref{eqn:problem3 2}) \\ $\bullet$ Rate constraint (\ref{eqn:downlink rate constraint}): Optimal dual variables $\{\beta_k\}$\\$\bullet$ Fronthaul constraint (\ref{eqn:downlink fronthaul constraint3}): Optimal dual variables $\{\lambda_m\}$} & \tabincell{l}{Dual problem: Power minimization (\ref{eqn:problem3 1}) \\ $\bullet$ Optimal transmit powers: $\{p_k^{{\rm ul}}=\beta_k\}$\\$\bullet$ Optimal quantization noises: $\{q_m^{{\rm ul}}=\lambda_m\}$ } \\
       \hline
       Case II & \tabincell{l}{Primal problem: Power minimization (\ref{eqn:problem4 2}) \\ $\bullet$ Rate constraint (\ref{eqn:downlink rate constraint4}): Optimal dual variables $\{\beta_k\}$\\$\bullet$ Fronthaul constraint (\ref{eqn:downlink fronthaul constraint3}): Optimal dual variables $\{\lambda_m\}$} & \tabincell{l}{Dual problem: Power minimization (\ref{eqn:problem4 1}) \\ $\bullet$ Optimal transmit powers: $\{p_k^{{\rm ul}}=\beta_k\}$\\$\bullet$ Optimal quantization noises: $\{q_m^{{\rm ul}}=\lambda_m\}$ } \\
       \hline
       Case III & \tabincell{l}{Primal problem: Power minimization (\ref{eqn:problem 2}) \\ $\bullet$ Rate constraint (\ref{eqn:downlink rate constraint}): Optimal dual variables $\{\beta_k\}$\\$\bullet$ Fronthaul constraint (\ref{eqn:downlink convex fronthaul constraint}): Optimal dual variables $\{\mv{\Lambda}_m\}$} & \tabincell{l}{Dual problem: Power minimization (\ref{eqn:problem 1}) \\ $\bullet$ Optimal transmit powers: $\{p_k^{{\rm ul}}=\beta_k\}$\\$\bullet$ Optimal quantization noises: $\{q_m^{{\rm ul}}=\mv{\Lambda}_m^{(m,m)}\}$ } \\
       \hline
       Case IV & \tabincell{l}{Primal problem: Power minimization (\ref{eqn:problem5 2}) \\ $\bullet$ Rate constraint (\ref{eqn:downlink rate constraint4}): Optimal dual variables $\{\beta_k\}$\\$\bullet$ Fronthaul constraint (\ref{eqn:downlink convex fronthaul constraint}): Optimal dual variables $\{\mv{\Lambda}_m\}$} & \tabincell{l}{Dual problem: Power minimization (\ref{eqn:problem5 1}) \\ $\bullet$ Optimal transmit powers: $\{p_k^{{\rm ul}}=\beta_k\}$\\$\bullet$ Optimal quantization noises: $\{q_m^{{\rm ul}}=\mv{\Lambda}_m^{(m,m)}\}$ } \\
		\hline
	\end{tabular}
\end{center}
\end{table*}

Let $\{R_k>0, k \in \mathcal{K} \}$ and $\{C_m>0, m \in \mathcal{M} \}$ denote sets of strictly feasible user rate requirements and fronthaul rate requirements under the beamforming vectors (\ref{eqn:beamforming vector}), decoding order (\ref{eqn:encoding decoding order}), and decompression order (\ref{eqn:compression decompression order}) for both the uplink and the downlink. For the multiple-access relay channel, 
the transmit power minimization problem subject to the individual rate constraints as well as the individual fronthaul capacity constraints is formulated as
\begin{align}\hspace{-8pt} \mathop{\mathrm{minimize}}_{\{p_k^{\rm ul}\},\{q_m^{\rm ul}\}} & ~ P^{\rm ul}(\{p_k^{\rm ul}\})  \label{eqn:problem5 1} \\
\hspace{-8pt} \mathrm {subject ~ to}  & ~ (\ref{eqn:uplink nonnegative power}), ~ (\ref{eqn:uplink positive}), ~ (\ref{eqn:uplink rate constraint4}), ~ (\ref{eqn:uplink fronthaul constraint}). \nonumber
\end{align}
Likewise, given the transmit beamforming vectors (\ref{eqn:beamforming vector}), a reversed encoding order (\ref{eqn:encoding decoding order}), and a reversed compression order (\ref{eqn:compression decompression order}), for the broadcast relay channel, 
the transmit power minimization problem is formulated as
\begin{align}\hspace{-8pt} \mathop{\mathrm{minimize}}_{\{p_k^{\rm dl}\},\mv{Q}} & ~ P^{\rm dl}(\{p_k^{\rm dl}\},\mv{Q})  \label{eqn:problem5 2} \\
\hspace{-8pt} \mathrm {subject ~ to}  & ~ (\ref{eqn:nonnegative power}), ~ (\ref{eqn:positive quantization noise3}), ~ (\ref{eqn:downlink rate constraint4}), ~ (\ref{eqn:downlink fronthaul constraint}). \nonumber
\end{align}

Similar to Section \ref{sec 3}, we can apply the Lagrangian duality method to show that problem (\ref{eqn:problem5 2}) is equivalent to problem (\ref{eqn:problem5 1}). Since the method adopted is almost the same as that in Section \ref{sec 3}, we omit the proof here. As a result, under the same fronthaul capacity constraints $C_m$'s and total transmit power constraint $P$, any achievable rate-tuple for the multiple-access relay channel is also achievable for the broadcast relay channel by setting the transmit beamforming vectors as the receive beamforming vectors in the multiple-access relay channel and by setting the encoding and compression order to be the reverse of the decoding and decompression order in the multiple-access relay channel, and vice versa. In other words, by trying all the feasible beamforming vectors, it follows that
\begin{multline}
\bar{\mathcal{R}}_{{\rm IV}}^{\rm ul}(\{C_m\},P,\{\bar{\rho}(m)\},\{\bar{\tau}(k)\}) \\ = \mathcal{\bar R}_{{\rm IV}}^{\rm dl}(\{C_m\},P,\{\bar{\rho}(M+1-m)\},\{\bar{\tau}(K+1-k)\}).
\end{multline}
Then, by trying all the encoding/decoding orders and compression/decompression orders, we can show that the achievable rate regions of the multiple-access relay channel under successive interference cancellation as well as Wyner-Ziv compression and the broadcast relay channel under dirty-paper coding as well as multivariate compression are identical, i.e., $\mathcal{R}_{{\rm IV}}^{\rm ul}(\{C_m\},P)=\mathcal{R}_{{\rm IV}}^{\rm dl}(\{C_m\},P)$.

\section{Duality Relationships}\label{sec:Summary}


The main result of this paper is that the Lagrangian duality technique can be used to establish the duality between the sum-power minimization problems for the multiple-access relay channel and the broadacast relay channel. The key observation is that under fixed beamformers, the sum-power minimization problems in both the uplink and the downlink can be transformed into convex optimization problems. In particular, with
independent compression (i.e., Cases I and II), both the uplink and downlink
problems are LPs, while with Wyner-Ziv or multivariate compression (i.e., Cases III and
IV), both the uplink and downlink problems can be transformed into SDPs.
Note that for the uplink problem under the Wyner-Ziv compression, the transformation into SDP also
involves reversing the minimization into the maximization and reversing the
direction of inequalities involving the rate and fronthaul constraints.
We further show that the resulting convex problems in the uplink and the downlink
after the transformations are the Lagrangian duals of each other for Cases I--IV. Moreover, the dual variables have interesting physical interpretations. Specifically, the Lagrangian dual variables corresponding to the downlink achievable rate constraints are the optimal uplink transmit powers; the dual variables corresponding to the downlink fronthaul rate constraints are the optimal uplink quantization noise levels. These interpretations are summarized in Table \ref{table:interpretation}.



In the prior literature, the traditional uplink-downlink duality relationship
is established by showing that any achievable rate-tuple in the uplink is also
achievable in the downlink, and vice versa. However, it is difficult to apply
this approach to verify the duality results of this paper, at least for the case with the
Wyner-Ziv and multivariate compression strategies. This is because the
condition to ensure a feasible solution is easy to characterize only under
linear constraints (i.e., with independent compression). But with Wyner-Ziv
and multivariate compressions, the fronthaul rates are nonlinear
functions of the transmit powers and the quantization noises. For this reason,
this paper takes an alternative approach of fixing the beamforming vectors then transforming the
sum-power minimization problems subject to the individual rate and fronthaul
constraints into suitable convex forms. As a result, we are able to take a
unified approach to establish the uplink-downlink duality using the Lagrangian
duality theory. This approach in fact works for both
independent compression and multivariate compression.


\setcounter{equation}{142}
\begin{figure*}
\begin{align}\label{eqn:interference function case I}
I_k(\mv{p}^{{\rm ul}})= \frac{2^{R_k}-1}{\mv{h}_k^H \left(\sum\limits_{j\neq k}p_j^{{\rm ul}} \mv{h}_j\mv{h}_j^H+{\rm diag}(q_1^{{\rm ul}}(\mv{p}^{{\rm ul}}),\ldots,q_M^{{\rm ul}}(\mv{p}^{{\rm ul}}))+\sigma^2\mv{I}\right)^{-1} \mv{h}_k}.
\end{align}\hrulefill 
\end{figure*}
\setcounter{equation}{131}

\section{Application of Duality}\label{sec:Application}

While the main technical proofs in this paper are for the case of fixed beamformers, the duality relationship also extends to the scenario where the beamformers need to be jointly optimized with the transmit powers and quantization noises. In this section, we consider algorithms for such joint optimization problems and show that the duality relationship gives an efficient way of solving the downlink joint optimization problem via its uplink counterpart.

The joint optimization of the beamforming vectors, transmit
powers, and the quantization noises for the broadcast relay channel is more difficult
than the corresponding optimization for the multiple-access relay channel.
This is because in the multiple-access relay channel, the receive beamforming
vectors $\{\mv{w}_k\}$ affect the user rates only, but do not affect the
fronthaul rates; 
the optimal receive beamformers 
are simply the minimum mean-squared-error (MMSE) receiver.
This is in contrast to the broadcast relay channel, where the transmit beamforming vectors
$\{\mv{v}_k\}$ affect both the user rates and the fronthaul rates, 
which makes the optimization highly nontrivial. Furthermore, the optimization of the
quantization noise solution is also conceptually easier in the multiple-access relay
channel since the quantization noise levels are scalars, while in the broadcast
relay channel, the optimization of the quantization noises is over their
covariance matrix. Due to the above reasons, it would be appealing if we can solve
the broadcast relay channel problem via its dual multiple-access counterpart.
In the following two subsections, we show how this can be done under
independent compression and Wyner-Ziv/multivariate compression, respectively.
The key observation is that in both cases, the sum-power minimization
problem in the multiple-access relay channel can be solved globally via a
fixed-point iteration method.

\subsection{Independent Compression}\label{sec:Independent Compression}
First, consider the sum-power minimization problems in the multiple-access and broadcast relay channels for Cases I and II with independent compression/decompression. For simplicity, we focus on Case I, i.e., linear encoding/decoding, and show how to solve the problem in the broadcast relay channel via solving the dual problem in the multiple-access relay channel. Similar approach can be applied to Case II under nonlinear encoding/decoding. Specifically, in Case I, the sum-power minimization problems in the multiple-access channel and the broadcast relay channel are formulated as
\begin{align}\hspace{-8pt} \mathop{\mathrm{minimize}}_{\{p_k^{\rm ul},\mv{w}_k\},\{q_m^{\rm ul}\}} & ~ P^{\rm ul}(\{p_k^{\rm ul}\})  \label{eqn:problem uplink case I} \\
	\hspace{-8pt} \mathrm {subject ~ to}  & ~ R_k^{\rm ul,TIN}(\{p_k^{\rm ul},\mv{w}_k\},\{q_m^{\rm ul}\})\geq R_k, ~ \forall k\in \mathcal{K}, \label{eqn:Case I 1}\\ & ~ C_m^{\rm ul,IN}(\{p_k^{\rm ul}\},q_m^{\rm ul})\leq C_m, ~ \forall m\in \mathcal{M}, \label{eqn:uplink fronthaul constraint app} \\ \hspace{-8pt} & ~ (\ref{eqn:uplink nonnegative power}), ~ (\ref{eqn:uplink positive}), \nonumber
\end{align}
and
\begin{align}\hspace{-8pt} \mathop{\mathrm{minimize}}_{\{p_k^{\rm dl},\mv{v}_k\},\mv{Q}_{{\rm diag}}} & ~ P^{\rm dl}(\{p_k^{\rm dl}\},\mv{Q}_{{\rm diag}})  \label{eqn:problem downlink case I} \\
\hspace{-8pt} \mathrm {subject ~ to}  & ~ R_k^{\rm dl,LIN}(\{p_k^{\rm dl},\mv{v}_k\},\mv{Q}_{{\rm diag}})\geq R_k, ~ \forall k\in \mathcal{K}, \label{eqn:Case I 2} \\ \hspace{-8pt} & ~ C_m^{\rm dl,IN}(\{p_k^{\rm dl},\mv{v}_k\},\mv{Q}_{{\rm diag}})\leq C_m, ~ \forall m\in \mathcal{M},  \\ \hspace{-8pt} & ~ (\ref{eqn:nonnegative power}), ~ (\ref{eqn:positive quantization noise3}). \nonumber
\end{align}Note that different from problems (\ref{eqn:problem3 1}) and (\ref{eqn:problem3 2}), here the beamforming vectors need to be jointly optimized with the transmit powers and quantization noises. In the following, we propose a fixed-point iteration method that can globally solve problem (\ref{eqn:problem uplink case I}) with low complexity. The main idea is to transform problem (\ref{eqn:problem uplink case I}) that jointly optimizes the transmit powers, receive beamforming vectors, and quantization noise levels into a power control problem. Specifically, according to Proposition \ref{proposition broadcast 3}, at the optimal solution, constraint (\ref{eqn:uplink fronthaul constraint app}) in problem (\ref{eqn:problem uplink case I}) should be satisfied with equality. In this case, the fronthaul capacity constraint (\ref{eqn:uplink fronthaul constraint app}) gives the following relationship between the optimal quantization noise levels and the optimal transmit powers:
\begin{align}\label{eqn:optimal quantization Case I}
q_m^{{\rm ul}}(\mv{p}^{{\rm ul}})=\frac{\sum\limits_{k=1}^Kp_k^{{\rm ul}}|h_{m,k}|^2+\sigma^2}{2^{C_m}-1}, ~~~ \forall m\in \mathcal{M},
\end{align}where $\mv{p}^{{\rm ul}}=[p_1^{{\rm ul}},\ldots,p_K^{{\rm ul}}]^T$. Moreover, it is observed from problem (\ref{eqn:problem uplink case I}) that the receive beamforming vectors affect the user rates only, but do not affect the fronthaul rates. It is well-known that the optimal receive beamforming vectors to maximize the user rates are the MMSE receivers, i.e.,
\begin{align}\label{eqn:MMSE}
\mv{w}_k=\frac{\tilde{\mv{w}}_k}{\|\tilde{\mv{w}}_k\|}, ~~~ \forall k\in \mathcal{K},
\end{align}where
\begin{align}
& \tilde{\mv{w}}_k= \nonumber \\ & \left(\sum\limits_{j\neq k}p_j^{{\rm ul}} \mv{h}_j\mv{h}_j^H+{\rm diag}(q_1^{{\rm ul}}(\mv{p}^{{\rm ul}}),\ldots,q_M^{{\rm ul}}(\mv{p}^{{\rm ul}}))+\sigma^2\mv{I}\right)^{-1}\mv{h}_k,
\end{align}with $q_m^{{\rm ul}}(\mv{p}^{{\rm ul}})$ as given in (\ref{eqn:optimal quantization Case I}). By plugging the above MMSE beamformers into constraint (\ref{eqn:Case I 1}), problem (\ref{eqn:problem uplink case I}) is now equivalent to the following power control problem:
\begin{align}\mathop{\mathrm{minimize}}_{\mv{p}^{{\rm ul}}} & ~ P^{\rm ul}(\{p_k^{\rm ul}\})  \label{eqn:problem uplink case I 1} \\
\mathrm {subject ~ to}  & ~ p_k^{{\rm ul}}\geq I_k(\mv{p}^{{\rm ul}}), ~ \forall k\in \mathcal{K}, \label{eqn:Case I 11} 
\end{align}
where $\forall k\in \mathcal{K}$, and $I_k(\mv{p}^{{\rm ul}})$ is expressed in
(\ref{eqn:interference function case I}) at the top of the page.

\setcounter{equation}{143}

\begin{lemma}\label{lemmacaseI}
Given $\mv{p}^{{\rm ul}}\geq \mv{0}$, the functions $I_k(\mv{p}^{{\rm ul}})$'s, $k=1,\ldots,K$, as defined by (\ref{eqn:interference function case I}) satisfy the following three properties:
\begin{itemize}
\item[1.] $I_k(\mv{p}^{{\rm ul}})>0$, $\forall k$;
\item[2.] Given any $\alpha>1$, it follows that $I_k(\alpha \mv{p}^{{\rm ul}})< \alpha I_k(\mv{p}^{{\rm ul}})$, $\forall k$;
\item[3.] If $\bar{\mv{p}}^{{\rm ul}}\geq \mv{p}^{{\rm ul}}$, then $I_k(\bar{\mv{p}}^{{\rm ul}})\geq I_k(\mv{p}^{{\rm ul}})$, $\forall k$.
\end{itemize}
\end{lemma}

\begin{IEEEproof}
Please refer to Appendix \ref{appendix8}.
\end{IEEEproof}

Lemma \ref{lemmacaseI} shows that $\mv{I}(\mv{p}^{{\rm ul}})=[I_1(\mv{p}^{{\rm ul}}),\ldots,I_K(\mv{p}^{{\rm ul}})]^T$ is a standard interference function \cite{Yates95}. According to \cite[Theorem 2]{Yates95}, as long as the original problem is feasible, given any initial power solution $\mv{p}^{{\rm ul},(0)}=[p_1^{{\rm ul},(0)},\ldots,p_K^{{\rm ul},(0)}]^T$ that satisfies $\mv{p}^{{\rm ul},(0)}\geq \mv{0}$, the following fixed-point iteration must converge to the globally optimal power control solution to problem (\ref{eqn:problem uplink case I 1}):
\begin{align}\label{eqn:fixed-point case I}
p_k^{{\rm ul},(t+1)}=I_k(\mv{p}^{{\rm ul},(t)}), ~~~ \forall k\in \mathcal{K},
\end{align}where $I_k(\mv{p}^{{\rm ul}})$'s are defined in (\ref{eqn:interference function case I}), and $\mv{p}^{{\rm ul},(t)}$ is the power obtained after the $t$-th iteration of the fixed-point update (\ref{eqn:fixed-point case I}). After the optimal power solution denoted by $\mv{p}^{{\rm ul},\ast}$ is obtained via the fixed-point iteration, the optimal quantization noise levels denoted by $\{q_m^{{\rm ul},\ast}\}$ can be obtained via (\ref{eqn:optimal quantization Case I}), and the optimal receive beamforming vectors denoted by $\{\mv{w}_k^\ast\}$ can be obtained via (\ref{eqn:MMSE}). Note that the above algorithm to solve problem (\ref{eqn:problem uplink case I}) is very simple, because all the updates have closed-form expressions.

\begin{figure*}
\setcounter{equation}{154}
\begin{align}\label{eqn:optimal solution Case III 1}
q_m^{{\rm ul}}(\mv{p}^{{\rm ul}})= \frac{\sigma^2+\sum\limits_{k=1}^Kp_k^{{\rm ul}}h_{k,m}h_{k,m}^H-\mv{\Gamma}^{(m,1:m-1)}(\mv{p}^{{\rm ul}})(\mv{\Gamma}^{(1:m-1,1:m-1)}(\mv{p}^{{\rm ul}}))^{-1}\mv{\Gamma}^{(1:m-1,m)}(\mv{p}^{{\rm ul}})}{2^{C_m}-1},~ \forall m \in \mathcal{M}.
\end{align}
\hrulefill
\setcounter{equation}{144}
\end{figure*}

After problem (\ref{eqn:problem uplink case I}) is solved globally, we now show how to obtain the optimal solution to problem (\ref{eqn:problem downlink case I}) using the uplink-downlink duality. It is shown in Section \ref{sec 1} that given the same beamforming vectors, the sum-power minimization problems in the multiple-access channel and the broadcast relay channel are equivalent to each other. As a result, the optimal beamforming vectors for problem (\ref{eqn:problem uplink case I}), i.e., $\{\mv{w}_k^\ast\}$, are also the optimal beamforming vectors for problem (\ref{eqn:problem downlink case I}). After setting $\mv{v}_k^\ast=\mv{w}_k^\ast$, $\forall k$, the problem reduces to an optimization problem over the transmit powers and quantization noise levels. As problem (\ref{eqn:problem downlink case I}) is convex in $p_k^{\rm dl}$ and $\mv{Q}_{{\rm diag}}$ given the fixed beamforming vectors as shown in Section \ref{sec 1}, it can be solved efficiently. 

We contrast the above approach with the direct optimization of the downlink.
It can be shown that if the unit-power beamformers $\mv{v}_k$'s and transmit powers $p_k^{{\rm dl}}$'s are combined together as $\tilde{\mv{v}}_k=p_k^{{\rm dl}}\mv{v}_k=[\tilde{v}_{k,1},\ldots,\tilde{v}_{k,M}]^T$, $\forall k$, then problem (\ref{eqn:problem downlink case I}) can be transformed into the following optimization problem:
\begin{align}\hspace{-8pt} \mathop{\mathrm{minimize}}_{\{\tilde{\mv{v}}_k\},\{q_m^{{\rm dl}}\}} & ~ \sum\limits_{k=1}^K\|\tilde{\mv{v}}_k\|^2\sigma^2+\sum\limits_{m=1}^Mq_m^{{\rm dl}}\sigma^2  \label{eqn:problem downlink case I 1} \\
	\hspace{-8pt} \mathrm {subject ~ to}  & ~ \frac{|\tilde{\mv{v}}_k^H\mv{h}_k|^2}{2^{R_k}-1}-\sum\limits_{j\neq  k}|\tilde{\mv{v}}_j^H\mv{h}_k|^2-\sum\limits_{m=1}^Mq_m^{{\rm dl}}|h_{k,m}|^2 \nonumber \\ & ~  \qquad\qquad\qquad\qquad \quad-\sigma^2\geq 0, ~ \forall k\in \mathcal{K}, \label{eqn:downlink rate case I 1} \\
	\hspace{-8pt} & ~ \sum\limits_{k=1}^K|\tilde{v}_{k,m}|^2+q_m^{\rm dl}-2^{C_m}q_m^{\rm dl}\leq 0, ~ \forall m\in \mathcal{M}, \label{eqn:downlink fronthaul case I 1} \\ \hspace{-8pt} & ~ (\ref{eqn:nonnegative power}), ~ (\ref{eqn:positive quantization noise3}). \nonumber
\end{align}
Note that (\ref{eqn:downlink rate case I 1}) can be transformed into a second-order cone constraint \cite{duality8}. As a result, problem (\ref{eqn:problem downlink case I 1}) for the broadcast relay channel is convex. However, the joint optimization of the beamforming vectors and quantization noise levels in problem (\ref{eqn:problem downlink case I 1}) is over a higher dimension, so it can potentially be more computationally complex than the duality-based approach where the beamforming vectors are first obtained by solving problem (\ref{eqn:problem uplink case I}) using fixed-point iterations, then the powers and quantization noise levels are optimized given the beamforming vectors.

\setcounter{equation}{157}
\begin{figure*}[!t]
	\normalsize\begin{align}\label{eqn:interference function case III}
& I_k(\mv{p}^{{\rm ul}})= \frac{2^{R_k}-1}{\mv{h}_k^H \left(\sum\limits_{j\neq k}p_j^{{\rm ul}} \mv{h}_j\mv{h}_j^H\!+\!{\rm diag}(q_1^{{\rm ul}}(\mv{p}^{{\rm ul}}),\ldots,q_M^{{\rm ul}}(\mv{p}^{{\rm ul}}))\!+\!\sigma^2\mv{I}\right)^{-1} \mv{h}_k}, ~ \forall k\in \mathcal{K}.
\end{align}\hrulefill 
\end{figure*}
\setcounter{equation}{147}

\subsection{Wyner-Ziv and Multivariate Compression}\label{sec:Wyner-Ziv and Multivariate Compression}

The above approach can also be used for the scenarios with Wyner-Ziv and multivariate compressions, i.e., Case III under linear encoding/decoding and Case IV under nonlinear encoding/decoding. For simplicity, in the following, we focus on Case III; similar analysis also applies to Case IV. Suppose that the decompression order for Wyner-Ziv compression and the compression order for multivariate compression are given by (\ref{eqn:compression decompression order}). Then, the sum-power minimization problems under Case III in the multiple-access relay channel and the broadcast relay channel are respectively given by
\begin{align}\hspace{-8pt} \mathop{\mathrm{minimize}}_{\{p_k^{\rm ul},\mv{w}_k\},\{q_m^{\rm ul}\}} & ~ P^{\rm ul}(\{p_k^{\rm ul}\})  \label{eqn:problem uplink Case III} \\
\hspace{-8pt} \mathrm {subject ~ to}
	& ~ R_k^{\rm ul,TIN}(\{p_k^{\rm ul},\mv{w}_k\},\{q_m^{\rm ul}\})\geq R_k, ~ \forall k\in \mathcal{K}, \label{eqn:uplink rate constraint app 2} \\
	& ~ C_{\bar\rho(m)}^{{\rm ul,WZ}}(\{p_k^{\rm ul}\},q_{\bar\rho(1)}^{\rm ul},\ldots,q_{\bar\rho(m)}^{\rm ul},\{\bar{\rho}(m)\}) \nonumber \\ &  \qquad\qquad\qquad\qquad~ \leq C_{\bar\rho(m)}, ~ \forall m \in \mathcal{M}, \label{eqn:uplink fronthaul constraint app 2} \\
	& ~ (\ref{eqn:uplink nonnegative power}), ~ (\ref{eqn:uplink positive}),  \nonumber
\end{align}
and
\begin{align}\hspace{-8pt} \mathop{\mathrm{minimize}}_{\{p_k^{\rm dl},\mv{v}_k\},\mv{Q}} & ~ P^{\rm dl}(\{p_k^{\rm dl}\},\mv{Q})  \label{eqn:problem downlink Case III} \\
\hspace{-8pt} \mathrm {subject ~ to}  &
~ R_k^{\rm dl,LIN}(\{p_k^{\rm dl},\mv{v}_k\}\},\mv{Q}) \geq R_k, ~ \forall k\in \mathcal{K}, \label{eqn:Case I 2 app} \\
	\hspace{-8pt} &
	~ C_{\bar\rho(M+1-m)}^{\rm dl,MV}(\{p_k^{\rm dl},\mv{v}_k\},\mv{Q},\{\bar{\rho}(M+1-m)\}) \nonumber \\ &  \qquad\qquad\qquad~~~ \leq C_{\bar\rho(M+1-m)}, ~ \forall m \in \mathcal{M},  \\
	& ~ (\ref{eqn:positive}), ~ (\ref{eqn:nonnegative power}).  \nonumber
\end{align}
Next, we show how to solve problem (\ref{eqn:problem downlink Case III}) via problem (\ref{eqn:problem uplink Case III}). Similar to problem (\ref{eqn:problem uplink case I}), in the following, we transform problem (\ref{eqn:problem uplink Case III}) into a power control problem, then propose a fixed-point iteration method to solve this power control problem globally. First, according to Proposition \ref{proposition7}, with the optimal solution to problem (\ref{eqn:problem uplink Case III}), constraint (\ref{eqn:uplink fronthaul constraint app 2}) in problem (\ref{eqn:problem uplink Case III}) is satisfied with equality. In this case, the quantization noise level of relay $1$ can be expressed as a function of the transmit powers as follows:
\begin{align}\label{eqn:optimal solution Case III}
q_1^{{\rm ul}}(\mv{p}^{{\rm ul}})= \frac{\sigma^2+\sum\limits_{k=1}^Kp_k^{{\rm ul}}h_{k,1}h_{k,1}^H}{2^{C_1}-1}.
\end{align}Moreover, if $q_1^{{\rm ul}}(\mv{p}^{{\rm ul}}),\ldots,q_{m-1}^{{\rm ul}}(\mv{p}^{{\rm ul}})$ are already expressed as functions of $\mv{p}^{{\rm ul}}$, then $\mv{\Gamma}^{(1:m-1,1:m-1)}$, $\mv{\Gamma}^{(m,1:m-1)}$, and $\mv{\Gamma}^{(1:m-1,m)}$ can also be expressed as functions of $\mv{p}^{{\rm ul}}$ according to (\ref{eqn:Gamma}), given the specific decompression order (\ref{eqn:compression decompression order}). For convenience, we use $\mv{\Gamma}^{(1:m-1,1:m-1)}(\mv{p}^{{\rm ul}})$, $\mv{\Gamma}^{(m,1:m-1)}(\mv{p}^{{\rm ul}})$, and $\mv{\Gamma}^{(1:m-1,m)}(\mv{p}^{{\rm ul}})$ to indicate that they are functions of $\mv{p}^{{\rm ul}}$. In this case, $q_m^{{\rm ul}}(\mv{p}^{{\rm ul}})$ can be uniquely expressed as a function of $\mv{p}^{{\rm ul}}$ as given in (\ref{eqn:optimal solution Case III 1}) on top of the page. In other words, given $\mv{p}^{{\rm ul}}$, we can first characterize $q_1^{{\rm ul}}(\mv{p}^{{\rm ul}})$ according to (\ref{eqn:optimal solution Case III}), then characterize $q_2^{{\rm ul}}(\mv{p}^{{\rm ul}})$ given $q_1^{{\rm ul}}(\mv{p}^{{\rm ul}})$ according to (\ref{eqn:optimal solution Case III 1}), then characterize $q_3^{{\rm ul}}(\mv{p}^{{\rm ul}})$ given $q_1^{{\rm ul}}(\mv{p}^{{\rm ul}})$ and $q_2^{{\rm ul}}(\mv{p}^{{\rm ul}})$ according to (\ref{eqn:optimal solution Case III 1}), and so on.

Next, similar to problem (\ref{eqn:problem uplink case I}), given the transmit power solution $\mv{p}^{{\rm ul}}$ and the quantization noise levels $\{q_m^{{\rm ul}}(\mv{p}^{{\rm ul}})\}$ as in (\ref{eqn:optimal solution Case III}) and (\ref{eqn:optimal solution Case III 1}), the optimal MMSE beamforming vectors are given in (\ref{eqn:MMSE}). By plugging them into problem (\ref{eqn:problem uplink Case III}), we have the following power control problem:
\setcounter{equation}{155}
\begin{align}
\mathop{\mathrm{minimize}}_{\mv{p}^{{\rm ul}}} & ~ P^{\rm ul}(\{p_k^{\rm ul}\})  \label{eqn:problem uplink case III 1} \\
\hspace{-8pt} \mathrm {subject ~ to}  & ~ p_k^{{\rm ul}}\geq I_k(\mv{p}^{{\rm ul}}), ~ \forall k\in \mathcal{K}, \label{eqn:Case III 11} 
\end{align}
where $\left\{I_k(\mv{p}^{{\rm ul}})\right\}$ are given in \eqref{eqn:interference function case III} at the top of the page and
$\{q_m^{{\rm ul}}(\mv{p}^{{\rm ul}})\}$ are given in (\ref{eqn:optimal solution Case III}) and (\ref{eqn:optimal solution Case III 1}).
\setcounter{equation}{158}

\begin{lemma}\label{lemmacaseIII}
Given $\mv{p}^{{\rm ul}}\geq \mv{0}$, the functions $I_k(\mv{p}^{{\rm ul}})$'s, $k=1,\ldots,K$, defined by (\ref{eqn:interference function case III}) satisfy the following three properties:
\begin{itemize}
\item[1.] $I_k(\mv{p}^{{\rm ul}})>0$, $\forall k$;
\item[2.] Given any $\alpha>1$, it follows that $I_k(\alpha \mv{p}^{{\rm ul}})< \alpha I_k(\mv{p}^{{\rm ul}})$, $\forall k$;
\item[3.] If $\bar{\mv{p}}^{{\rm ul}}\geq \mv{p}^{{\rm ul}}$, then $I_k(\bar{\mv{p}}^{{\rm ul}})\geq I_k(\mv{p}^{{\rm ul}})$, $\forall k$.
\end{itemize}
\end{lemma}

\begin{IEEEproof}
Please refer to Appendix \ref{appendix9}.
\end{IEEEproof}

Similar to Lemma \ref{lemmacaseI}, Lemma \ref{lemmacaseIII} shows that $\mv{I}(\mv{p}^{{\rm ul}})=[I_1(\mv{p}^{{\rm ul}}),\ldots,I_K(\mv{p}^{{\rm ul}})]^T$ is a standard interference function \cite{Yates95}. According to \cite[Theorem 2]{Yates95}, as long as the original problem is feasible, given any initial power solution $\mv{p}^{{\rm ul},(0)}=[p_1^{{\rm ul},(0)},\ldots,p_K^{{\rm ul},(0)}]^T$ that satisfies $\mv{p}^{{\rm ul},(0)}\geq \mv{0}$, the following fixed-point iteration must converge to the globally optimal power control solution to problem (\ref{eqn:problem uplink case III 1}):
\begin{align}\label{eqn:fixed-point case III}
p_k^{{\rm ul},(t+1)}=I_k(\mv{p}^{{\rm ul},(t)}), ~~~ \forall k\in \mathcal{K},
\end{align}where $I_k(\mv{p}^{{\rm ul}})$'s are defined in (\ref{eqn:interference function case III}), and $\mv{p}^{{\rm ul},(t)}$ denotes the power solution obtained at the $t$-th iteration of the fixed-point update (\ref{eqn:fixed-point case III}). To implement the above fixed-point iteration, given $\mv{p}^{{\rm ul},(t)}$, we can first get $q_1^{{\rm ul}}(\mv{p}^{{\rm ul},(t)})$ according to (\ref{eqn:optimal solution Case III}), then $q_2^{{\rm ul}}(\mv{p}^{{\rm ul},(t)})$ given $q_1^{{\rm ul}}(\mv{p}^{{\rm ul},(t)})$ according to (\ref{eqn:optimal solution Case III 1}), and so on. Thus, given $\mv{p}^{{\rm ul},(t)}$, we get $\{q_m^{{\rm ul}}(\mv{p}^{{\rm ul},(t)})\}$, then $\{I_k(\mv{p}^{{\rm ul},(t)})\}$ in (\ref{eqn:fixed-point case III}) can be computed by using (\ref{eqn:interference function case III}).

After the optimal power $\mv{p}^{{\rm ul},\ast}$ is obtained by the fixed-point method (\ref{eqn:fixed-point case III}), the optimal quantization noise levels $\{q_m^{{\rm ul},\ast}\}$ can then be obtained via (\ref{eqn:optimal solution Case III}) and (\ref{eqn:optimal solution Case III 1}), and the optimal receive beamforming vectors $\{\mv{w}_k^\ast\}$ can be obtained via (\ref{eqn:MMSE}). Note that the above algorithm to solve problem (\ref{eqn:problem uplink Case III}) is simple, because there are closed-form expressions for the update of all the variables.

Finally, we solve problem (\ref{eqn:problem downlink Case III}) via duality. 
According to the duality result in Section \ref{sec 3}, we can set the transmit beamforming vectors in problem (\ref{eqn:problem downlink Case III}) as $\mv{v}_k^\ast=\mv{w}_k^\ast$, $\forall k\in \mathcal{K}$. Then, we can obtain the optimal transmit power and quantization covariance matrix given this beamforming solution by solving the convex problem (\ref{eqn:problem 3}).
Note that it is not yet known whether problem (\ref{eqn:problem downlink Case III}) has a convex reformulation. Nevertheless, the above algorithm gives a globally optimal solution for (\ref{eqn:problem downlink Case III}). As a final remark, it is worth emphasizing that a condition for the convergence of the fixed-point iteration algorithm is that the original problem is feasible to start with. However, determining whether a set of user target rates and fronthaul capacities is feasible is by itself not necessarily always easy to do computationally, unless the problem can be reformulated as a convex optimization.


\subsection{Numerical Example}

\begin{figure}[t]
  \centering
  \includegraphics[width=9cm]{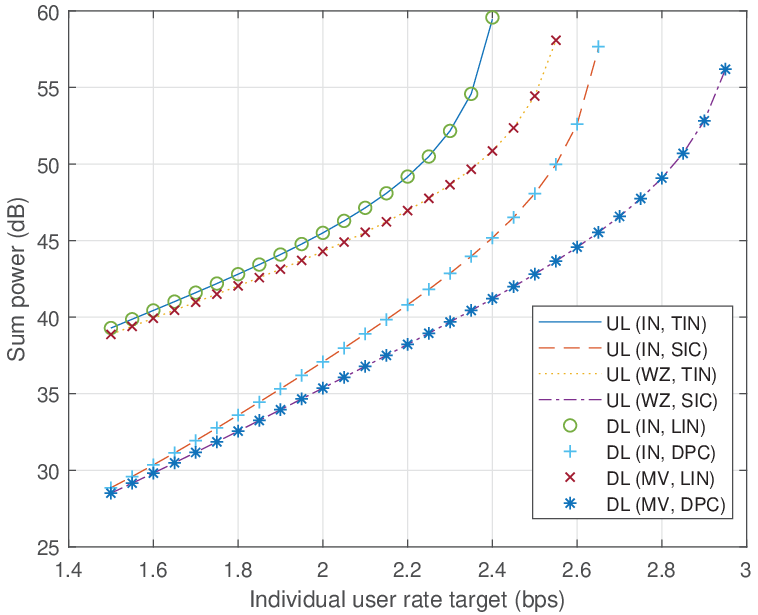}
  \caption{Illustration of the numerical algorithm for using duality to solve the downlink sum-power minimization problem via the dual uplink problem.}
\label{simulation}
\end{figure}

As a numerical example demonstrating the algorithm of using uplink-downlink
duality to solve the broadcast relay channel problem, we consider a
network with $M=3$ relays and $K=3$ users, where the wireless channels between
these relays and users are generated based on the independent and identically
distributed Rayleigh fading model with zero mean and unit variance, and the
fronthaul capacities between all the relays and the CP are set to be $3$ bps.
Moreover, the noise powers at the relays in the uplink and at the users in the
downlink are set to be $\sigma^2=1$. The rate targets for all the users are
assumed to be identical.
Under the above setup, Fig.~\ref{simulation}
shows the optimal values of the uplink (UL) problem
(\ref{eqn:problem uplink case I})
and the downlink (DL) problem
(\ref{eqn:problem downlink case I})
under independent compression (IN), and respectively the
UL problem
(\ref{eqn:problem uplink Case III})
and the DL problem
(\ref{eqn:problem downlink Case III})
under the Wyner-Ziv (WZ) and multivariate (MV) compression, as functions of the user rate target,
for both the cases of treating interference as noise (TIN) and linear encoding (LIN), respectively.
In addition, the corresponding curves for the cases of successive interference
cancellation (SIC) in the uplink or dirty-paper coding (DPC) in the downlink are also plotted.
The numerical values are obtained under the optimal beamforming vectors
computed based on the multiple-access relay channel; the same beamformers are
also optimal for the broadcast relay channel. Under these fixed beamformers,
the numerical results show that for both the cases of independent compression
and Wyner-Ziv/multivariate compression, at the same feasible rate targets, the minimum
sum power in the broadcast relay channel is the same as that in the
multiple-access relay channel. Moreover, as a consequence of this duality relationship, the ranges of the feasible rate targets are also the same for the broadcast relay channel and the multiple-access relay channel, under various encoding/decoding and compression/decompression strategies. For example, under the case of linear encoding/decoding and independent compression/decompression, if the common rate target is beyond about $2.4$ bps, the sum power minimization problem becomes infeasible in both the uplink and the downlink. It is worth noting that the minimum sum power
under the Wyner-Ziv and multivariate compression is smaller than that under the
independent compression, especially when the user rate target is high. This is
because the Wyner-Ziv compression can utilize the fronthaul more efficiently by
using the decompressed signals as side information, while the multivariate
compression can reduce the interference caused by compression as seen by the
users. Finally, successive decoding and dirty-paper coding have significant
benefit as compared to treating interference as noise and linear precoding,
respectively.

\section{Concluding Remarks}\label{sec:Conclusion}
This paper reveals an interesting duality relationship between the Gaussian
multiple-access relay channel and the Gaussian broadcast relay channel.
Specifically, we first show that if independent compression is applied in both
the uplink and the downlink, then the achievable rate regions of the
multiple-access relay channel and the broadcast relay channel are identical
under the same total transmit power constraint and individual fronthaul
capacity constraints. Furthermore, this duality continues to hold when the
Wyner-Ziv compression strategy is applied in the multiple-access relay channel
and the multivariate compression strategy is applied in the broadcast relay
channel. This duality relationship has an intimate connection to the Lagrangian
duality theory in convex optimization. Under fixed beamformers, the power
minimization problems for the multiple-access and broadcast relay channels are
the Lagrangian duals of each other. The optimal dual variables corresponding to
the rate constraints in the broadcast relay channel problem are the optimal
transmit powers in the dual multiple-access relay channel problem.  The optimal
dual variables corresponding to the fronthaul constraints in the broadcast
relay channel problem are the optimal quantization noise levels in the dual
multiple-access relay channel problem. Furthermore, this paper shows that
for jointly optimizing over the transmit powers, the receive beamforming
vectors, and the quantization noise levels, the uplink sum-power minimization
problem can be globally solved using a low-complexity fixed-point iteration
method. Thus, the duality also gives a computationally simple way of finding
the optimal beamformers in the downlink problem via its dual uplink.

We remark that the duality relationship established in this paper is specific
to the compression based relaying strategies. The compression strategies
considered in this paper are special cases of the more general noisy network
coding \cite{Kim_NNC} and distributed decode-forward \cite{Kim_DDF} schemes for
general multihop relay networks. An \emph{operational} duality between
noisy network coding and distributed decode-forward has already been observed
in \cite{Kim_DDF}, thus the main contribution of this paper is in establishing
a \emph{computational} duality for the specific compression schemes in the
setting of the specific two-hop relay network.  Is there a \emph{computational}
duality for the more general network? Is there a \emph{capacity region} duality
for either the two-hop or more general relay networks?  These remain
interesting open questions.

\begin{appendix}
\subsection{Proof of Proposition \ref{proposition broadcast 1}}\label{appendix broadcast 1}
First, suppose that at the optimal solution to problem (\ref{eqn:dual problem3 4}), which is denoted by $\beta_k^\ast$'s and $\lambda_m^\ast$'s, there exists a $\bar{k}$ such that
\begin{align}
\sum\limits_{j\neq \bar{k}}\beta_j^\ast|\bar{\mv{u}}_{\bar{k}}^H\mv{h}_j|^2+\sum\limits_{m=1}^M\lambda_m^\ast|\bar{u}_{\bar{k},m}|^2+\sigma^2-\frac{\beta_{\bar{k}}^\ast|\bar{\mv{u}}_{\bar{k}}^H\mv{h}_{\bar{k}}|^2}{2^{R_{\bar{k}}}-1}>0.
\end{align}Then, consider another solution where $\lambda_m=\lambda_m^\ast$, $\forall m$,  $\beta_k=\beta_k^\ast$, $\forall k\neq \bar{k}$, and
\begin{align}\label{eqn:new transmit power}
\beta_{\bar{k}} &= \frac{\left(\sum\limits_{j\neq \bar{k}}\beta_j^\ast|\bar{\mv{u}}_{\bar{k}}^H\mv{h}_j|^2+\sum\limits_{m=1}^M\lambda_m^\ast|\bar{u}_{\bar{k},m}|^2+\sigma^2\right)(2^{R_{\bar{k}}}-1)}{|\bar{\mv{u}}_{\bar{k}}^H\mv{h}_{\bar{k}}|^2} \nonumber \\
&> ~ \beta_{\bar{k}}^\ast.
\end{align}
	It can be shown that the new solution is a feasible solution to problem (\ref{eqn:dual problem3 4}). Moreover, due to (\ref{eqn:new transmit power}), at the new solution, the objective value of problem (\ref{eqn:dual problem3 4}) is increased. This contradicts to the fact that the optimal solution to problem (\ref{eqn:dual problem3 4}) is $\{\beta_k^\ast, \lambda_m^\ast\}$. As a result, at the optimal solution to problem (\ref{eqn:dual problem3 4}), constraint (\ref{eqn:dual constraint 1}) should hold with equality.

Next, suppose that at the optimal solution to problem (\ref{eqn:dual problem3 4}), which is denoted by $\beta_k^\ast$'s and $\lambda_m^\ast$'s, there exists an $\bar{m}$ such that
\begin{align}
\sum\limits_{k=1}^K\beta_k^\ast|h_{k,\bar{m}}|^2+\lambda_{\bar{m}}^\ast+\sigma^2-2^{C_{\bar{m}}}\lambda_{\bar{m}}^\ast> 0.
\end{align}Then, consider another solution where $\beta_k=\beta_k^\ast$, $\forall k$, $\lambda_m=\lambda_m^\ast$, $\forall m\neq \bar{m}$, and
\begin{align}\label{eqn:new quantization noise level}
\lambda_{\bar{m}}=\frac{\sum\limits_{k=1}^K\beta_k^\ast|h_{k,\bar{m}}|^2+\sigma^2}{2^{C_{\bar{m}}}-1}> \lambda_{\bar{m}}^\ast.
\end{align}
	It can be shown that the new solution is a feasible solution to problem (\ref{eqn:dual problem3 4}). Moreover, due to (\ref{eqn:new quantization noise level}), at the new solution, the objective value of problem (\ref{eqn:dual problem3 4}) is increased. This contradicts to the fact that the optimal solution to problem (\ref{eqn:dual problem3 4}) is $\{\beta_k^\ast,\lambda_m^\ast\}$. As a result, at the optimal solution to problem (\ref{eqn:dual problem3 4}), constraint (\ref{eqn:dual constraint fronthaul}) should hold with equality.

Proposition \ref{proposition broadcast 1} is thus proved.
\setcounter{equation}{165}
\begin{figure*}[!t]
	\normalsize\begin{equation}\label{eqn:standard function}
I_k(\mv{\beta})=\frac{\!\left(\!\sum\limits_{j\neq k}\beta_j|\bar{\mv{u}}_k^H\mv{h}_j|^2\!+\!\sum\limits_{m=1}^M\frac{\left(\sum\limits_{k=1}^K\beta_k|h_{k,m}|^2\!+\!\sigma^2\!\right)|\bar{u}_{k,m}|^2}{2^{C_m}\!-\!1}\!+\!\sigma^2\right)(2^{R_k}\!-\!1)}{|\bar{\mv{u}}_k^H\mv{h}_k|^2}.
\end{equation}\hrulefill 
\end{figure*}
\setcounter{equation}{167}
\begin{figure*}[!t]
	\normalsize\begin{align}
I_k(\alpha\mv{\beta}) &= \frac{\left(\sum\limits_{j\neq k}\alpha\beta_j|\bar{\mv{u}}_k^H\mv{h}_j|^2\!+\!\sum\limits_{m=1}^M\frac{\left(\sum\limits_{k=1}^K\alpha\beta_k|h_{k,m}|^2+\sigma^2\right)|\bar{u}_{k,m}|^2}{2^{C_m}-1}\!+\!\sigma^2\right)(2^{R_k}\!-\!1)}{\left|\bar{\mv{u}}_k^H\mv{h}_k\right|^2}  \label{167}\\ &<  \frac{\alpha\left(\sum\limits_{j\neq k}\beta_j|\bar{\mv{u}}_k^H\mv{h}_j|^2+\sum\limits_{m=1}^M\frac{\left(\sum\limits_{k=1}^K\beta_k|h_{k,m}|^2+\sigma^2\right)|\bar{u}_{k,m}|^2}{2^{C_m}-1}+\sigma^2\right)(2^{R_k}\!-\!1)}{\left|\bar{\mv{u}}_k^H\mv{h}_k\right|^2} \\ &= \alpha I_k(\mv{\beta}), ~~~ \forall k\in \mathcal{K}.\label{169}
\end{align}\hrulefill 
\end{figure*}
\begin{figure*}[!t]
	\normalsize\begin{align}
I_k(\bar{\mv{\beta}}) &= \frac{\left(\sum\limits_{j\neq k}\bar{\beta}_j|\bar{\mv{u}}_k^H\mv{h}_j|^2+\sum\limits_{m=1}^M\frac{\left(\sum\limits_{k=1}^K\bar{\beta}_k|h_{k,m}|^2+\sigma^2\right)|\bar{u}_{k,m}|^2}{2^{C_m}-1}+\sigma^2\right)(2^{R_k}-1)}{\left|\bar{\mv{u}}_k^H\mv{h}_k\right|^2}\label{170} \\  &\geq \frac{\left(\sum\limits_{j\neq k}\beta_j|\bar{\mv{u}}_k^H\mv{h}_j|^2+\sum\limits_{m=1}^M\frac{\left(\sum\limits_{k=1}^K\beta_k|h_{k,m}|^2+\sigma^2\right)|\bar{u}_{k,m}|^2}{2^{C_m}-1}+\sigma^2\right)(2^{R_k}-1)}{\left|\bar{\mv{u}}_k^H\mv{h}_k\right|^2}  \\
&= I_k(\mv{\beta}), ~~~ \forall k\in \mathcal{K}.\label{172}
\end{align}\hrulefill 
\end{figure*}
\begin{figure*}[!t]
	\normalsize\begin{align}\label{eqn:downlink new fronthaul constraint}
	2^{C_m}\mv{Q}^{(m,m)}-\mv{Q}^{(m,m)}-\sum\limits_{k=1}^{K} p_k^{\rm dl} \left|\bar{u}_{k,m}\right|^2  -2^{C_m}\mv{Q}^{(m,m+1:M)}(\mv{Q}^{(m+1:M,m+1:M)})^{-1}\mv{Q}^{(m+1:M,m)}\geq 0,	~ \forall m.
	\end{align}\hrulefill 
\end{figure*}

\setcounter{equation}{163}

\subsection{Proof of Proposition \ref{proposition broadcast 2}}\label{appendix broadcast 2}
First, according to (\ref{eqn:dual constraint fronthaul 1}), it follows that
\begin{align}\label{eqn:equal quantization noise level}
\lambda_m=\frac{\sum\limits_{k=1}^K\beta_k|h_{k,m}|^2+\sigma^2}{2^{C_m}-1}, ~~~ \forall m\in \mathcal{M}.
\end{align}By plugging (\ref{eqn:equal quantization noise level}) into (\ref{eqn:dual constraint 11}), it follows that
\begin{align}\label{eqn:standard function 1}
\beta_k=I_k(\mv{\beta}), ~~~ \forall k\in \mathcal{K},
\end{align}where $\mv{\beta}=[\beta_1,\ldots,\beta_K]^T$, and $I_k(\mv{\beta})$ is given in \eqref{eqn:standard function} on top of the page. It can be shown that with $\beta_k>0$, $\forall k$, we have
\begin{align}\setcounter{equation}{166}
I_k(\mv{\beta})>0, ~~~ \forall k\in \mathcal{K}.
\end{align}Next, given $\alpha>1$, it can be shown that \eqref{167}--\eqref{169} on top of the page are true. Last, given $\bar{\mv{\beta}}=[\bar{\beta}_1,\ldots,\bar{\beta}_K]^T\geq \mv{\beta}$, it follows that \eqref{170}--\eqref{172} on top of the page are true. As a result, the function $\mv{I}(\mv{\beta})=[I_1(\mv{\beta}),\ldots,I_K(\mv{\beta})]^T$ defined by (\ref{eqn:standard function}) is a standard interference function \cite{Yates95}. It then follows from \cite[Theorem 1]{Yates95} that there exists a unique solution $\mv{\beta}>\mv{0}$ to equation (\ref{eqn:standard function 1}). Moreover, with this solution of $\mv{\beta}$, there is a unique solution of $\lambda_m>0$'s to equation (\ref{eqn:equal quantization noise level}). As a result, if problem (\ref{eqn:dual problem3 5}) is feasible, there exists only one solution to the constraints (\ref{eqn:dual constraint 11}), (\ref{eqn:dual constraint fronthaul 1}), (\ref{eqn:broadcast positive beta 11}), and (\ref{eqn:broadcast positive lambda 11}) in problem (\ref{eqn:dual problem3 5}).

Proposition \ref{proposition broadcast 2} is thus proved.

\begin{figure*}[!t]
	\setcounter{equation}{174}
	\begin{align}\label{eqn:D matrix}
	\mv{D}_m=\left[\begin{array}{cc} 2^{C_m}\mv{Q}^{(m,m)}-\mv{Q}^{(m,m)}-\sum\limits_{k=1}^{K} p_k^{\rm dl} \left|\bar{u}_{k,m}\right|^2 & 2^{C_m} \mv{Q}^{(m,m+1:M)} \\ 2^{C_m} \mv{Q}^{(m+1:M,m)} & 2^{C_m} \mv{Q}^{(m+1:M,m+1:M)} \end{array} \right]\in \mathbb{C}^{(M-m+1)\times (M-m+1)}.
	\end{align}
	\hrulefill
\end{figure*}
\begin{figure*}[!t]
	\normalsize\begin{align}
	2^{C_m}\left[\begin{array}{cc}\mv{0}_{(m-1)\times (m-1)} & \mv{0}_{(m-1)\times (M-m+1)} \\ \mv{0}_{(M-m)\times (m-1)} & \mv{Q}^{(m:M,m:M)}\end{array}\right]-\mv{E}_m(\mv{Q}+\mv{\Psi})\mv{E}_m^H =\left[\begin{array}{cc}\mv{0}_{(m-1)\times (m-1)} & \mv{0}_{(m-1)\times (M-m+1)} \\ \mv{0}_{(M-m+1)\times (m-1)} & \mv{D}_m \end{array} \right], ~~~ \forall m.\label{178}
	\end{align}	\hrulefill
\end{figure*}
\setcounter{equation}{176}
\begin{figure*}[!t]
	\normalsize\begin{align}
	&\mathcal{L}(\{p_k^{\rm dl}\},\mv{Q},\{\beta_k\},\{\mv{\Lambda}_m\}) \nonumber\\
	& = \sum\limits_{k=1}^Kp_k^{\rm dl}\sigma^2+{\rm tr}(\mv{Q})\sigma^2  -\sum\limits_{k=1}^K\beta_k\bigg(\frac{p_k^{\rm dl}|\bar{\mv{u}}_k^H\mv{h}_k|^2}{2^{R_k}-1} - \sum\limits_{j\neq k} p_j^{\rm dl}|\bar{\mv{u}}_j^H\mv{h}_k|^2-{\rm tr}(\mv{Q}\mv{h}_k\mv{h}_k^H)-\sigma^2\bigg) \nonumber \\
	& \quad - \sum\limits_{m=1}^M{\rm tr}\bigg(\mv{\Lambda}_m\bigg(2^{C_m}\bigg[\begin{array}{cc}\mv{0}_{(m-1)\times (m-1)} & \mv{0}_{(m-1)\times (M-m+1)} \\ \mv{0}_{(M-m+1)\times (m-1)} & \mv{Q}^{(m:M,m:M)}\end{array}\bigg]  - \mv{E}_m(\mv{Q}+\mv{\Psi})\mv{E}_m^H\bigg)\bigg) \nonumber \\
	& = \sum\limits_{k=1}^Kp_k^{\rm dl}\bigg(\sigma^2+\sum\limits_{j\neq k}\beta_j|\bar{\mv{u}}_k^H\mv{h}_j|^2+\sum\limits_{m=1}^M\mv{\Lambda}_m^{(m,m)}|\bar{u}_{k,m}|^2
	- \frac{\beta_k|\bar{\mv{u}}_k^H\mv{h}_k|^2}{2^{R_k}-1} \bigg) + \sum\limits_{k=1}^K\beta_k \sigma^2  \nonumber\\
	& \quad + {\rm tr}\bigg(\mv{Q}\bigg(\sigma^2\mv{I}\!+\!\sum\limits_{k=1}^K\beta_k\mv{h}_k\mv{h}_k^H\!+\!\sum\limits_{m=1}^M\mv{E}_m^H\mv{\Lambda}_m\mv{E}_m -  \sum\limits_{m=1}^M2^{C_m}\left[\begin{array}{cc}\mv{0}_{(m-1)\times (m-1)} & \mv{0}_{(m-1)\times (M-m+1)} \\ \mv{0}_{(M-m+1)\times (m-1)} & \mv{\Lambda}_m^{(m:M,m:M)}\end{array}\right] \bigg)\bigg).  \label{eqn:downlink Lagrange}
	\end{align}\hrulefill 
\end{figure*}

\subsection{Proof of Proposition \ref{proposition1}}\label{appendix1}
First, it can be easily verified that the SINR constraints (\ref{eqn:downlink convex rate constraint}) are equivalent to the original SINR constraints (\ref{eqn:downlink rate constraint}). Next, we consider the fronthaul constraints (\ref{eqn:downlink fronthaul constraint}), which can be re-formulated as \eqref{eqn:downlink new fronthaul constraint} on top of the page.

Given $m=1,\ldots,M$, define $\mv{D}_m$ as in (\ref{eqn:D matrix}) on top of the next page. Note that $\mv{Q}^{(m,m+1:M)}=(\mv{Q}^{(m+1:M,m)})^H$ and thus $\mv{D}_m$ is a Hermitian matrix, $m=1,\ldots,M$. Note that we must have $\mv{Q}\succ\mv{0}$ in problem (\ref{eqn:problem 2}) since otherwise, the fronthaul rates for relays become infinite based on (\ref{eqn:downlink fronthaul rate}) and (\ref{eqn:assumption}). According to \cite[Theorem 7.7.9]{Horn12}, given $\mv{Q}\succ\mv{0}$, (\ref{eqn:downlink new fronthaul constraint}) is equivalent to $\mv{D}_m\succeq \mv{0}$, $\forall m$.

Moreover, since \eqref{178} on top of the page is true, the fronthaul constraint (\ref{eqn:downlink new fronthaul constraint}) in the broadcast relay channel is equivalent to (\ref{eqn:downlink convex fronthaul constraint}) given (\ref{eqn:positive}).

To summarize, the new SINR constraints (\ref{eqn:downlink convex rate constraint}) and fronthaul constraints (\ref{eqn:downlink convex fronthaul constraint}) are equivalent to the original constraints. Moreover, we multiply the objective function in problem (\ref{eqn:problem 2}) by a constant $\sigma^2$ for convenience. As a result, problem (\ref{eqn:problem 3}) and problem (\ref{eqn:problem 2}) have the same optimal solution. Proposition \ref{proposition1} is proved.

\setcounter{equation}{177}
\subsection{Proof of Proposition \ref{proposition2}}\label{appendix2}
We can write down the Lagrangian of problem (\ref{eqn:problem 3}) as in \eqref{eqn:downlink Lagrange} on top of the page,~where $\beta_k\geq 0$, $k=1,\ldots,K$, and $\mv{\Lambda}_m\in \mathbb{C}^{M\times M}\succeq \mv{0}$, $m=1,\ldots,M$, are the dual variables associated with constraints (\ref{eqn:downlink convex rate constraint}) and (\ref{eqn:downlink convex fronthaul constraint}), respectively. The dual function of problem (\ref{eqn:problem 2}) is thus formulated as
\begin{align}\label{eqn:dual function problem 2}
& g(\{\beta_k\},\{\mv{\Lambda}_m\}) \nonumber \\ & = ~ \min\limits_{p_k^{\rm dl}\geq 0, \forall k\in \mathcal{K}, \mv{Q}\succeq \mv{0}} ~ \mathcal{L}(\{p_k^{\rm dl}\},\mv{Q},\{\beta_k\},\{\mv{\Lambda}_m\}).
\end{align}At last, the dual problem of problem (\ref{eqn:problem 3}) is
\begin{align}\hspace{-8pt}
\mathop{\mathrm{maximize}}_{\{\beta_k\},\{\mv{\Lambda}_m\}} & ~ g(\{\beta_k\},\{\mv{\Lambda}_m\}) \label{eqn:dual problem of problem 2} \\
\mathrm {subject ~ to}  & ~ \beta_k\geq 0, ~ \forall k, \\
& ~ \mv{\Lambda}_m\succeq \mv{0}, ~ \forall m.
\end{align}
Note that according to (\ref{eqn:downlink Lagrange}), the dual function is $g(\{\beta_k\},\{\mv{\Lambda}_m\})=\sum\limits_{k=1}^K\beta_k \sigma^2$ if
and only if
\begin{multline}
\sigma^2+\sum\limits_{j\neq k}\beta_j|\bar{\mv{u}}_k^H\mv{h}_j|^2+\sum\limits_{m=1}^M\mv{\Lambda}_m^{(m,m)}|\bar{u}_{k,m}|^2 \\
 -\frac{\beta_k|\bar{\mv{u}}_k^H\mv{h}_k|^2}{2^{R_k}-1}\geq 0, ~ \forall k \label{eqn:opt con 1}
\end{multline}and\begin{multline} \!\!\!\!\!\sigma^2\mv{I}+\sum\limits_{k=1}^K\beta_k\mv{h}_k\mv{h}_k^H+\sum\limits_{m=1}^M\mv{E}_m^H\mv{\Lambda}_m\mv{E}_m\\-\sum\limits_{m=1}^M2^{C_m}\left[\begin{array}{cc}\mv{0}_{(m-1)\times (m-1)} & \!\!\!\mv{0}_{(m-1)\times (M-m+1)} \\ \mv{0}_{(M-m+1)\times (m-1)} & \!\!\! \mv{\Lambda}_m^{(M-m+1:M-m+1)}\end{array}\right] \succeq \mv{0}. \label{eqn:opt con 2}
\end{multline}Otherwise, the dual function is unbounded, i.e., $g(\{\beta_k\},\{\mv{\Lambda}_m\})=-\infty$. As a result, the optimal solution to problem (\ref{eqn:dual function problem 2}) must satisfy constraints (\ref{eqn:opt con 1}) and (\ref{eqn:opt con 2}). In this case, $g(\{\beta_k\},\{\mv{\Lambda}_m\})=\sum\limits_{k=1}^K\beta_k \sigma^2$. This indicates that the dual problem of problem (\ref{eqn:problem 3}), i.e., problem (\ref{eqn:dual problem of problem 2}), is equivalent to problem (\ref{eqn:problem 4}).

Proposition \ref{proposition1} is thus proved.

\subsection{Proof of Proposition \ref{proposition3}}\label{appendix3}
According to (\ref{eqn:A1}) and (\ref{eqn:A2}), we have
\begin{align}
\mv{A}_1&=2^{C_1}\mv{\Lambda}_1^{(1:M,1:M)}+\left[\begin{array}{cc}0 & \mv{0}_{1\times (M-1)} \\ \mv{0}_{(M-1)\times 1} & \mv{A}_2 \end{array}\right] \\ & = 2^{C_1}\mv{\Lambda}_1^{(1:M,1:M)}+2^{C_2}\left[\begin{array}{cc}0 & \mv{0}_{1\times (M-1)} \\ \mv{0}_{(M-1)\times 1} & \mv{\Lambda}_m^{(M-1:M-1)}\end{array}\right] \nonumber \\ &\quad + \left[\begin{array}{cc}\mv{0}_{2\times 2} & \mv{0}_{2\times (M-2)} \\ \mv{0}_{(M-2)\times 2} & \mv{A}_3 \end{array}\right] \\ & \vdots  \nonumber \\ & = \sum\limits_{m=1}^M2^{C_m}\left[\begin{array}{cc}\mv{0}_{(m-1)\times (m-1)} & \!\!\!\!\!\mv{0}_{(m-1)\times (M-m+1)} \\ \mv{0}_{(M-m+1)\times (m-1)} & \!\!\!\mv{\Lambda}_m^{(M-m+1:M-m+1)}\end{array}\right]. \label{eqn:A3}
\end{align}By combining (\ref{eqn:dual uplink fronthaul constraint 1}) and (\ref{eqn:A3}), we have (\ref{eqn:dual uplink fronthaul constraint}). As a result, problem (\ref{eqn:problem 4}) is equivalent to problem (\ref{eqn:problem 5}). Proposition \ref{proposition3} is thus proved.

\subsection{Proof of Proposition \ref{lemma1}}\label{appendix4}
We first show that at the optimal solution to problem (\ref{eqn:problem 5}), condition (\ref{eqn:optimal condition 4}) is true, i.e.,
\begin{align}
\mv{B}=\sigma^2\mv{I}+\sum\limits_{k=1}^K\beta_k\mv{h}_k\mv{h}_k^H+\sum\limits_{m=1}^M\mv{E}_m^H\mv{\Lambda}_m\mv{E}_m-\mv{A}_1=\mv{0}.
\end{align}

Suppose that at the optimal solution $\{\beta_k, \mv{A}_m, \mv{\Lambda}_m\}$ to problem (\ref{eqn:problem 5}), $\mv{B}\succeq \mv{0}$ but $\mv{B}\neq \mv{0}$. Then, let $\bar{m}$ denote the index of the first positive diagonal element of $\mv{B}$. Since $\mv{B}\succeq \mv{0}$, we have
\begin{align}
\mv{B}=\left[\begin{array}{cc} \mv{0}_{(\bar{m}-1)\times (\bar{m}-1)} & \mv{0}_{(\bar{m}-1)\times (M-\bar{m}+1)} \\ \mv{0}_{(M-\bar{m}+1)\times (\bar{m}-1)} & \mv{B}^{(\bar{m}:M,\bar{m}:M)} \end{array}\right].
\end{align}

Consider another solution given by
\begin{align}
& \hat{\beta}_k=\beta_k, ~~~ \forall k, \label{eqn:opt1}\\
& \hat{\mv{\Lambda}}_m=\mv{\Lambda}_m, ~~~ \forall m\neq \bar{m}, \label{eqn:opt2}\\
& \hat{\mv{\Lambda}}_{\bar{m}}=\mv{\Lambda}_{\bar{m}}+\left[\begin{array}{cc} \mv{0}_{(\bar{m}-1)\times (\bar{m}-1)} & \mv{0}_{(\bar{m}-1)\times (M-\bar{m}+1)} \\ \mv{0}_{(M-\bar{m}+1)\times (\bar{m}-1)} & \mv{C}_{\bar{m}}\end{array}\right], \label{eqn:opt3}\\
& \hat{\mv{A}}_m=2^{C_m}\hat{\mv{\Lambda}}_m^{(m:M,m:M)}+\left[\begin{array}{cc}0 & \mv{0}_{1\times (M-m)} \\ \mv{0}_{(M-m)\times 1} & \hat{\mv{A}}_{m+1} \end{array}\right], \nonumber \\ & \qquad\qquad\qquad\qquad\qquad\qquad\quad ~ m=1,\ldots,M-1, \label{eqn:opt4}\\
& \hat{\mv{A}}_M=2^{C_M}\hat{\mv{\Lambda}}_M^{(M,M)}, \label{eqn:opt5}
\end{align}where $\mv{C}_{\bar{m}}\in \mathbb{C}^{(M-\bar{m}+1)\times (M-\bar{m}+1)}$ in (\ref{eqn:opt3}) satisfies
\begin{multline}\label{eqn:lemma81}
2^{C_{\bar{m}}}\mv{C}_{\bar{m}}-\left[\begin{array}{cc} \mv{C}_{\bar{m}}^{(1,1)} & \mv{0}_{1\times (M-\bar{m})} \\ \mv{0}_{(M-\bar{m})\times 1} & \mv{0}_{(M-\bar{m})\times (M-\bar{m})}\end{array}\right]\\=\mv{B}^{(\bar{m}:M,\bar{m}:M)}.
\end{multline}

Since $\mv{B}^{(\bar{m},\bar{m})}>0$, it follows from (\ref{eqn:lemma81}) that $\mv{C}_{\bar{m}}^{(1,1)}>0$. Moreover, since $\mv{B}^{(\bar{m}:M,\bar{m}:M)} \succeq \mv{0}$ and $\mv{C}_{\bar{m}}^{(1,1)}>0$, according to (\ref{eqn:lemma81}) we have $\mv{C}_{\bar{m}}\succeq \mv{0}$. As a result, (\ref{eqn:opt3}) indicates that $\hat{\mv{\Lambda}}_{\bar{m}}\succeq \mv{0}$, i.e., constraint (\ref{eqn:nonegative Phi}) is satisfied. Further, it follows from (\ref{eqn:opt4}) and (\ref{eqn:opt5}) that
\begin{align}
\hat{\mv{A}}_1&=\mv{A}_1+2^{C_{\bar{m}}}\left[\begin{array}{cc} \mv{0}_{(\bar{m}-1)\times (\bar{m}-1)} & \mv{0}_{(\bar{m}-1)\times (M-\bar{m}+1)} \nonumber\\ \mv{0}_{(M-\bar{m}+1)\times (\bar{m}-1)} & \mv{C}_{\bar{m}}\end{array}\right] \\
&=\mv{A}_1+\mv{B}+{\rm diag}([\mv{0}_{1\times (\bar{m}-1)},\mv{C}_{\bar{m}}^{(1,1)},\mv{0}_{1\times (M-\bar{m})}]) \nonumber\\
&=\sigma^2\mv{I}+\sum\limits_{k=1}^K\beta_k\mv{h}_k\mv{h}_k^H+\sum\limits_{m=1}^M\mv{E}_m^H\mv{\Lambda}_m\mv{E}_m \nonumber \\ & \qquad\qquad~ +{\rm diag}([\mv{0}_{1\times (\bar{m}-1)},\mv{C}_{\bar{m}}^{(1,1)},\mv{0}_{1\times (M-\bar{m})}]) \nonumber\\
&=\sigma^2\mv{I}+\sum\limits_{k=1}^K\hat{\beta}_k\mv{h}_k\mv{h}_k^H+\sum\limits_{m=1}^M\mv{E}_m^H\hat{\mv{\Lambda}}_m\mv{E}_m.
\end{align}
As a result, at the new solution, constraint (\ref{eqn:dual uplink fronthaul constraint 1}) is satisfied with equality. Last, since $\mv{C}_{\bar{m}}^{(1,1)}>0$, it follows from (\ref{eqn:opt2}) and (\ref{eqn:opt3}) that $\hat{\mv{\Lambda}}_m^{(m,m)}=\mv{\Lambda}_m^{(m,m)}$, $\forall m\neq \bar{m}$, and $\hat{\mv{\Lambda}}_{\bar{m}}^{(\bar{m},\bar{m})}>\mv{\Lambda}_{\bar{m}}^{(\bar{m},\bar{m})}$. Moreover, according to condition (\ref{eqn:assumption}), we have $\sum_{k=1}^K|\bar{u}_{k,\bar{m}}|^2>0$. As a result, at the new solution, constraint (\ref{eqn:dual uplink SINR constraint}) is satisfied with strict inequality at least for one $k$.

To summarize, at the new solution, the same objective value of problem (\ref{eqn:problem 5}) is achieved due to (\ref{eqn:opt1}), and all the constraints are also satisfied. Moreover, since constraint (\ref{eqn:dual uplink SINR constraint}) is satisfied with inequality, we can further find a better solution of $\beta_k$'s such that the objective value of problem (\ref{eqn:problem 5}) is enhanced while constraint (\ref{eqn:dual uplink SINR constraint}) is satisfied with equality. This contradicts the assumption that the optimal solution to problem (\ref{eqn:problem 5}) is $\{\beta_k, \mv{A}_m, \mv{\Lambda}_m\}$. As a result, at the optimal solution to problem (\ref{eqn:problem 5}), constraint (\ref{eqn:dual uplink fronthaul constraint 1}) must hold with equality.

Next, we show that at the optimal solution to problem (\ref{eqn:problem 5}), conditions (\ref{eqn:optimal condition 2}) and (\ref{eqn:optimal condition}) are true.

First, according to (\ref{eqn:optimal condition 4}), at the optimal solution, we have $\mv{A}_1\succ \mv{0}$, and thus
\begin{align}\label{eqn:lemma82}
\mv{A}_1^{(1,1)}>0.
\end{align}Then, we show that at the optimal solution, condition (\ref{eqn:optimal condition}) is true for $m=1$, i.e.,
\begin{align}\label{eqn:optimal condition 1}
2^{C_1}\mv{\Lambda}_1=\frac{\mv{A}_1^{(1:M,1)}\mv{A}_1^{(1,1:M)}}{\mv{A}_1^{(1,1)}}.
\end{align}

On the right-hand side (RHS) of (\ref{eqn:A1}), given each $m$, the first row and the first column of the overall matrix is merely contributed by the first row and the first column of $\mv{\Lambda}_m^{(m:M,m:M)}$. With $\mv{A}_1^{(1,1)}>0$, to make (\ref{eqn:A1}) hold, the optimal solution $2^{C_1}\mv{\Lambda}_1$ must be in the following form
\begin{align}\label{eqn:A4}
2^{C_1}\mv{\Lambda}_1=\frac{\mv{A}_1^{(1:M,1)}\mv{A}_1^{(1,1:M)}}{\mv{A}_1^{(1,1)}}+\left[\begin{array}{cc}0 & \mv{0}_{1\times (M-1)} \\ \mv{0}_{(M-1)\times 1} & \mv{T}_1\end{array}\right],
\end{align}where $\mv{T}_1$ satisfies
\begin{align}\label{eqn:T}
\left[\begin{array}{cc}0 & \mv{0}_{1\times (M-1)} \\ \mv{0}_{(M-1)\times 1} & \mv{T}_1+\mv{A}_{2}\end{array}\right]=\mv{A}_1-\frac{\mv{A}_1^{(1:M,1)}\mv{A}_1^{(1,1:M)}}{\mv{A}_1^{(1,1)}}.
\end{align}It can be shown from (\ref{eqn:T}) that $\mv{T}_1+\mv{A}_{2}\succeq \mv{0}$.

\begin{figure*}
	\setcounter{equation}{206}
	\begin{align}\label{eqn:A matrix}
	\tilde{\mv{A}}_m=\left\{\begin{array}{ll}\mv{A}_m, & {\rm if} ~ m=1 ~ {\rm or} ~ 1+\tilde{m}<m\leq M, \\ \mv{A}_m+\left[\begin{array}{cc}\mv{0}_{(1+\tilde{m}-m)\times (1+\tilde{m}-m)} & \mv{0}_{(1+\tilde{m}-m)\times (M-\tilde{m})} \\ \mv{0}_{(M-\tilde{m})\times (1+\tilde{m}-m)} & \mv{T}_1^{(\tilde{m}:M-1,\tilde{m}:M-1)}/2^{C_1}\end{array}\right], & {\rm if} ~ 1<m<1+\tilde{m}, \\
	\mv{A}_m+\mv{T}_1^{(\tilde{m}:M-1,\tilde{m}:M-1)}/2^{C_1}, & {\rm if} ~ m=1+\tilde{m}.\end{array} \right.
	\end{align}
	\setcounter{equation}{199}
	\hrulefill
\end{figure*}

\begin{lemma}\label{lemma7}
Suppose that $\mv{a}\in \mathbb{C}^{M\times 1}$ with $\mv{a}^{(1)}>0$ and $\mv{B}\in \mathbb{C}^{(M-1)\times (M-1)}$. Then,
\begin{align}\label{eqn:rank1}
\mv{a}\mv{a}^H+\left[\begin{array}{cc}0 & \mv{0}_{1\times M} \\ \mv{0}_{M\times 1} & \mv{B}\end{array}\right]\succeq \mv{0},
\end{align}implies $\mv{B}\succeq \mv{0}$.
\end{lemma}

\begin{IEEEproof}
Suppose that $\mv{B}$ is not a positive semidefinite matrix. Then, there exists some $\mv{x}\in \mathbb{C}^{(M-1)\times 1}$ such that $\mv{x}^H\mv{B}\mv{x}<0$. Next, with $\mv{a}^{(1)}>0$, define $\mv{y}=[-\mv{x}^H\mv{a}^{(2:M)}/\mv{a}^{(1)},\mv{x}^H]^H\in \mathbb{C}^{M\times 1}$. It can then be shown that
\begin{align}
\mv{y}^H \left(\mv{a}\mv{a}^H+\left[\begin{array}{cc}0 & \mv{0}_{1\times M} \\ \mv{0}_{M\times 1} & \mv{B}\end{array}\right]\right)\mv{y}=\mv{x}^H\mv{B}\mv{x}<0.
\end{align}As a result, as long as $\mv{B}$ is not a positive semidefinite matrix, (\ref{eqn:rank1}) does not hold. Lemma \ref{lemma7} is thus proved.
\end{IEEEproof}

According to Lemma \ref{lemma7} and (\ref{eqn:lemma82}), if $2^{C_1}\mv{\Lambda}_1$ defined in (\ref{eqn:A4}) is a positive semidefinite matrix, we must have
\begin{align}
\mv{T}_1\succeq \mv{0}.
\end{align}

Now, suppose that the optimal solution $\{\beta_k, \mv{\Lambda}_m, \mv{A}_m\}$ to problem (\ref{eqn:problem 5}) satisfies  $\mv{T}_1\neq \mv{0}$. Define $\tilde{m}$ as the index of the first positive diagonal element in $\mv{T}_1$. Since $\mv{T}_1\succeq \mv{0}$, $\mv{T}_1$ must be in the form of
\begin{align}
\mv{T}_1=\left[\begin{array}{cc}\mv{0}_{(\tilde{m}-1)\times (\tilde{m}-1)} & \mv{0}_{(\tilde{m}-1)\times (M-\tilde{m})} \\ \mv{0}_{(M-\tilde{m})\times (\tilde{m}-1)} & \mv{T}_1^{(\tilde{m}:M-1,\tilde{m}:M-1)}\end{array}\right],
\end{align}with $\mv{T}_1^{(\tilde{m}:M-1,\tilde{m}:M-1)}\succeq \mv{0}$.

Then, consider a new solution in which $\tilde{\beta}_k=\beta_k$, $\forall k$, $\tilde{\mv{\Lambda}}_m$'s are given as follows
\begin{align}
\tilde{\mv{\Lambda}}_1&=\mv{\Lambda}_1-\left[\begin{array}{cc}0 & \mv{0}_{1\times (M-1)} \\ \mv{0}_{(M-1)\times 1} & \mv{T}_1/2^{C_1}\end{array}\right]\nonumber\\
&=\frac{\mv{A}_1^{(1:M,1)}\mv{A}_1^{(1,1:M)}}{\mv{A}_1^{(1,1)}} \nonumber \\ & \succeq  \mv{0}, \label{eqn:new solution 1}\\
\tilde{\mv{\Lambda}}_{1+\tilde{m}} &=\mv{\Lambda}_{1+\tilde{m}}+\left[\begin{array}{cc}0 & \mv{0}_{1\times (M-1)} \\ \mv{0}_{(M-1)\times 1} & \mv{T}_1/2^{C_1}\end{array}\right] \nonumber \\ & =\mv{\Lambda}_{1+\tilde{m}}+\left[\begin{array}{cc}\mv{0}_{\tilde{m}\times \tilde{m}} & \mv{0}_{\tilde{m}\times (M-\tilde{m})} \\ \mv{0}_{(M-\tilde{m})\times \tilde{m}} & \mv{T}_1^{(\tilde{m}:M-1,\tilde{m}:M-1)}/2^{C_1}\end{array}\right] \nonumber \\ & \succeq \mv{0}, \label{eqn:new solution 2} \\
\tilde{\mv{\Lambda}}_m & =\mv{\Lambda}_m, ~ \forall m\neq 1 ~ {\rm and} ~ m\neq 1+\tilde{m}, \label{eqn:new solution 3}
\end{align}and $\tilde{\mv{A}}_m$'s are given as in (\ref{eqn:A matrix}) on top of the page.

It can be shown that the above solution satisfies constraints (\ref{eqn:A1}), (\ref{eqn:A2}), (\ref{eqn:nonegative beta}), and (\ref{eqn:nonegative Phi}) in problem (\ref{eqn:problem 5}). Moreover, with this new solution, we have
\setcounter{equation}{207}
\begin{align}
& \tilde{\mv{\Lambda}}_1^{(1,1)}=\mv{\Lambda}_1^{(1,1)}-0=\mv{\Lambda}_1^{(1,1)},
\\
& \tilde{\mv{\Lambda}}_{1+\tilde{m}}^{(1+\tilde{m},1+\tilde{m})}=\mv{\Lambda}_{1+\tilde{m}}^{(1+\tilde{m},1+\tilde{m})}+\mv{T}_1^{(\tilde{m},\tilde{m})}/2^{C_1},\\
& \tilde{\mv{\Lambda}}_m^{(m,m)}=\mv{\Lambda}_m^{(m,m)}, ~ \forall m\neq 1 ~ {\rm and} ~ m\neq 1+\tilde{m}.
\end{align}Since $\mv{T}_1(\tilde{m},\tilde{m})>0$, it follows that $\tilde{\mv{\Lambda}}_m^{(m,m)}=\mv{\Lambda}_m^{(m,m)}$ if $m\neq 1+\tilde{m}$, and $\tilde{\mv{\Lambda}}_m^{(m,m)} > \mv{\Lambda}_m^{(m,m)}$ if $m=1+\tilde{m}$. Moreover, according to (\ref{eqn:assumption}), we have $\sum_{k=1}^K|\bar{u}_{k,1+\tilde{m}}|^2>0$. As a result, at the new solution, constraint (\ref{eqn:dual uplink SINR constraint}) in problem (\ref{eqn:problem 5}) is satisfied with strict inequality at least for one $k$. Last, since
\begin{align}
\sum\limits_{m=1}^M\mv{E}_m^H\tilde{\mv{\Lambda}}_m\mv{E}_m \succeq \sum\limits_{m=1}^M\mv{E}_m^H\mv{\Lambda}_m\mv{E}_m,
\end{align}constraint (\ref{eqn:dual uplink fronthaul constraint 1}) in problem (\ref{eqn:problem 5}) is satisfied.

To summarize, with the new solution $\{\tilde{\beta}_k, \tilde{\mv{\Lambda}}_m, \tilde{\mv{A}}_m\}$, all the constraints in problem (\ref{eqn:problem 5}) are satisfied, while the optimal objective value is also achieved. Moreover, since constraint (\ref{eqn:dual uplink SINR constraint}) is satisfied with strict inequality for some $k$'s, we can further find a better solution of $\beta_k$'s such that the objective value is enhanced while constraint (\ref{eqn:dual uplink SINR constraint}) is satisfied with equality. This contradicts the assumption that the optimal solution to problem (\ref{eqn:problem 5}) is $\{\beta_k, \mv{\Lambda}_m, \mv{A}_m\}$. As a result, at the optimal solution to problem (\ref{eqn:problem 5}), condition (\ref{eqn:optimal condition 1}) must hold.

Given (\ref{eqn:optimal condition 1}), it follows from (\ref{eqn:A1}) that
\begin{align}
\left[\begin{array}{cc}0 & \mv{0}_{1\times (M-1)} \\ \mv{0}_{(M-1)\times 1} & \mv{A}_2\end{array}\right]=\mv{A}_1-\frac{\mv{A}_1^{(1:M,1)}\mv{A}_1^{(1,1:M)}}{\mv{A}_1^{(1,1)}}.
\end{align}

Since ${\rm rank}(\mv{A}_1)=M$ and ${\rm rank}(\mv{A}_1^{(1:M,1)}\mv{A}_1^{(1,1:M)})=1$, we have ${\rm rank}(\mv{A}_2)=M-1$. As a result, we have $\mv{A}_2\succ \mv{0}$, and thus $\mv{A}_2^{(1,1)}>0$. Similar to the way to prove (\ref{eqn:optimal condition 1}), we can show that with $\mv{A}_2^{(1,1)}>0$, condition (\ref{eqn:optimal condition}) must hold for $m=2$. Then, similar to the way to prove $\mv{A}_2\succ \mv{0}$, we can show that if condition (\ref{eqn:optimal condition}) holds for $m=2$, then $\mv{A}_3\succ \mv{0}$. We can keep applying the above method to show that with the optimal solution to problem (\ref{eqn:problem 5}), conditions (\ref{eqn:optimal condition 2}) and (\ref{eqn:optimal condition}) are true.

Proposition \ref{lemma1} is thus proved.
\setcounter{equation}{216}
\begin{figure*}[!t]
	\normalsize\begin{multline}
	\mv{A}_n^{(m-n+1,m-n+1)}- \mv{A}_n^{(m-n+1,1:m-n)}(\mv{A}_n^{(1:m-n,1:m-n)})^{-1}\mv{A}_n^{(1:m-n,m-n+1)} \\ = \mv{A}_{n-1}^{(m-n+2,m-n+2)}- \mv{A}_{n-1}^{(m-n+2,1:m-n+1)}(\mv{A}_{n-1}^{(1:m-n+1,1:m-n+1)})^{-1}\mv{A}_{n-1}^{(1:m-n+1,m-n+2)}, ~ n=2,\ldots,m. \label{eqn:optimal solution 4}
	\end{multline}\hrulefill 
\end{figure*}
\begin{figure*}[!t]
	\normalsize\begin{align}
	& \mv{A}_n^{(m-n+1,m-n+1)}- \mv{A}_n^{(m-n+1,1:m-n)}(\mv{A}_n^{(1:m-n,1:m-n)})^{-1}\mv{A}_n^{(1:m-n,m-n+1)} \nonumber \\
	& =  \mv{A}_{n-1}^{(m-n+2,m-n+2)}-\frac{\mv{A}_{n-1}^{(m-n+2,1)}\mv{A}_{n-1}^{(1,m-n+2)}}{\mv{A}_{n-1}^{(1,1)}}-\left(\mv{A}_{n-1}^{(m-n+2,2:m-n+1)}-\frac{\mv{A}_{n-1}^{(m-n+2,1)}\mv{A}_{n-1}^{(1,2:m-n+1)}}{\mv{A}_{n-1}^{(1,1)}}\right) \nonumber \\
	&\quad \times \mv{B}_{n-1,m}^{-1}\left(\mv{A}_{n-1}^{(2:m-n+1,m-n+2)}-\frac{\mv{A}_{n-1}^{(1,m-n+2)}\mv{A}_{n-1}^{(2:m-n+1,1)}}{\mv{A}_{n-1}^{(1,1)}}\right), ~ n=2,\ldots,m.
	\label{eqn:optimal solution 21}
	\end{align}\hrulefill 
\end{figure*}\setcounter{equation}{219}
\begin{figure*}[!t]
	\normalsize\begin{align}
	& \mv{A}_{n-1}^{(m-n+2,m-n+2)}-\mv{A}_{n-1}^{(m-n+2,1:m-n+1)} \times (\mv{A}_{n-1}^{(1:m-n+1,1:m-n+1)})^{-1}\mv{A}_{n-1}^{(1:m-n+1,m-n+2)} \nonumber \\
	& = \mv{A}_{n-1}^{(m-n+2,m-n+2)}-[\mv{A}_{n-1}^{(m-n+2,1)},\mv{A}_{n-1}^{(m-n+2,2:m-n+1)}]\nonumber \\
	& \quad \times \left(\begin{array}{cc}\mv{A}_{n-1}^{(1,1)} & \mv{A}_{n-1}^{(1,2:m-n+1)} \\ \mv{A}_{n-1}^{(2:m-n+1,1)} & \mv{A}_{n-1}^{(2:m-n+1,2:m-n+1)}\end{array}\right)^{-1} \times \left[\begin{array}{c} \mv{A}_{n-1}^{(1,m-n+2)} \\ \mv{A}_{n-1}^{(2:m-n+1,m-n+2)}\end{array}\right], ~ n=2,\ldots,m. \label{eqn:optimal solution 23}
	\end{align}\hrulefill 
\end{figure*}

\setcounter{equation}{212}
\subsection{Proof of Proposition \ref{proposition4}}\label{appendix5}
According to (\ref{eqn:A1}), (\ref{eqn:A2}) and (\ref{eqn:optimal condition}), the optimal solution to problem (\ref{eqn:problem 5}) must satisfy
\begin{align}
&\left[\begin{array}{cc}0 & \mv{0}_{1\times (M-m+1)} \\ \mv{0}_{(M-m+1)\times 1} & \mv{A}_{m} \end{array}\right] \nonumber \\ &= \mv{A}_{m-1}-2^{C_{m-1}}\mv{\Lambda}_{m-1}^{(m-1:M,m-1:M)} \\ &= \mv{A}_{m-1} - \frac{\mv{A}_{m-1}^{(1:M-m+2,1)}\mv{A}_{m-1}^{(1,1:M-m+2)}}{\mv{A}_{m-1}^{(1,1)}}, ~ m=2,\ldots,M. \label{eqn:optimal solution 1}
\end{align}In other words, we have
\begin{align}\label{eqn:optimal solution 2}
\!\!\!\mv{A}_m& =\mv{A}_{m-1}^{(2:M-m+2,2:M-m+2)} \nonumber \\ \!\!\!&\quad - \frac{\mv{A}_{m-1}^{(2:M-m+2,1)}\mv{A}_{m-1}^{(1,2:M-m+2)}}{\mv{A}_{m-1}^{(1,1)}}, ~ m=2,\ldots,M.
\end{align}It then follows from (\ref{eqn:optimal solution 2}) that
\begin{align}\label{eqn:optimal solution 3}
\mv{A}_m^{(1,1)}=\mv{A}_{m-1}^{(2,2)}-\frac{\mv{A}_{m-1}^{(2,1)}\mv{A}_{m-1}^{(1,2)}}{\mv{A}_{m-1}^{(1,1)}}, ~ m=2,\ldots,M.
\end{align}

\begin{lemma}\label{lemma2}
If (\ref{eqn:optimal solution 2}) holds and $\mv{A}_m \succ \mv{0}$, $\forall m$, then for any $m=2,\ldots,M$, we have \eqref{eqn:optimal solution 4} on top of the page.
\end{lemma}

\begin{IEEEproof}
First, if (\ref{eqn:optimal solution 2}) holds, \eqref{eqn:optimal solution 21} on top of the page then follows, where\setcounter{equation}{218}
\begin{multline}\label{eqn:optimal solution 22}
\mv{B}_{n-1,m}=
\mv{A}_{n-1}^{(2:m-n+1,2:m-n+1)}\\-\frac{\mv{A}_{n-1}^{(2:m-n+1,1)}\mv{A}_{n-1}^{(1,2:m-n+1)}}{\mv{A}_{n-1}^{(1,1)}}.
\end{multline}

On the other hand, we have \eqref{eqn:optimal solution 23} on top of the page.
\setcounter{equation}{220}

\begin{lemma}\label{lemma6}
\cite[Section 0.7.3]{Horn12} Consider an invertible matrix
\begin{align}
\mv{X}=\left[\begin{array}{cc}\mv{C} & \mv{D} \\ \mv{E} & \mv{F}\end{array}\right].
\end{align}Then the inverse of $\mv{X}$ is given in (\ref{eqn:X inverse}) on top of the next page.
\end{lemma}
\begin{figure*}[!t]
	\setcounter{equation}{221}
	\begin{align}\label{eqn:X inverse}
	\mv{X}^{-1}=& \left[\begin{array}{cc} \mv{C}^{-1}+\mv{C}^{-1}\mv{D}(\mv{F}-\mv{E}\mv{C}^{-1}\mv{D})^{-1}\mv{E}\mv{C}^{-1} & -\mv{C}^{-1}\mv{D}(\mv{F}-\mv{E}\mv{C}^{-1}\mv{D})^{-1} \\ -(\mv{F}-\mv{E}\mv{C}^{-1}\mv{D})^{-1} \mv{E}\mv{C}^{-1} & (\mv{F}-\mv{E}\mv{C}^{-1}\mv{D})^{-1} \end{array}\right].
	\end{align}\hrulefill
\end{figure*}

According to Lemma \ref{lemma6}, we have (\ref{eqn:optimal solution 24}) on the next page. \begin{figure*}[!t]
	\normalsize
	\begin{align}
	&[\mv{A}_{n-1}^{(m-n+2,1)},\mv{A}_{n-1}^{(m-n+2,2:m-n+1)}]\left(\begin{array}{cc}\mv{A}_{n-1}^{(1,1)} & \mv{A}_{n-1}^{(1,2:m-n+1)} \\ \mv{A}_{n-1}^{(2:m-n+1,1)} & \mv{A}_{n-1}^{(2:m-n+1,2:m-n+1)}\end{array}\right)^{-1} \left[\begin{array}{c} \mv{A}_{n-1}^{(1,m-n+2)} \\ \mv{A}_{n-1}^{(2:m-n+1,m-n+2)}\end{array}\right] \nonumber \\
	& = [\mv{A}_{n-1}^{(m-n+2,1)},\mv{A}_{n-1}^{(m-n+2,2:m-n+1)}] \nonumber \\
	& \quad  \times \left[\begin{array}{cc}\frac{1}{\mv{A}_{n-1}^{(1,1)}}-\frac{\mv{A}_{n-1}^{(1,2:m-n+1)}\mv{B}_{n-1,m}^{-1}\mv{A}_{n-1}^{(2:m-n+1,1)}}{(\mv{A}_{n-1}^{(1,1)})^2} & -\frac{\mv{A}_{n-1}^{(1,2:m-n+1)}\mv{B}_{n-1,m}^{-1}}{\mv{A}_{n-1}^{(1,1)}} \\ -\frac{\mv{B}_{n-1,m}^{-1}\mv{A}_{n-1}^{(2:m-n+1,1)}}{\mv{A}_{n-1}^{(1,1)}} & \mv{B}_{n-1,m}^{-1}\end{array}\right] \left[\begin{array}{c} \mv{A}_{n-1}^{(1,m-n+2)} \\ \mv{A}_{n-1}^{(2:m-n+1,m-n+2)}\end{array}\right] \nonumber \\
	& = \frac{\mv{A}_{n-1}^{(m-n+2,1)}\mv{A}_{n-1}^{(1,m-n+2)}}{\mv{A}_{n-1}^{(1,1)}}+\frac{\mv{A}_{n-1}^{(m-n+2,1)}\mv{A}_{n-1}^{(1,m-n+2)}\mv{A}_{n-1}^{(1,2:m-n+1)}\mv{B}_{n-1,m}^{-1}\mv{A}_{n-1}^{(2:m-n+1,1)}}{(\mv{A}_{n-1}^{(1,1)})^2} \nonumber \\
	& \quad - \frac{\mv{A}_{n-1}^{(1,m-n+2)}\mv{A}_{n-1}^{(m-n+2,2:m-n+1)}\mv{B}_{n-1,m}^{-1}\mv{A}_{n-1}^{(2:m-n+1,1)}}{\mv{A}_{n-1}^{(1,1)}} -\frac{\mv{A}_{n-1}^{(m-n+2,1)}\mv{A}_{n-1}^{(1,2:m-n+1)}\mv{B}_{n-1,m}^{-1}\mv{A}_{n-1}^{(2:m-n+1,m-n+2)}}{\mv{A}_{n-1}^{(1,1)}}\nonumber \\
	& \quad  + \mv{A}_{n-1}^{(m-n+2,2:m-n+1)}\mv{B}_{n-1,m}^{-1}\mv{A}_{n-1}^{(2:m-n+1,m-n+2)}, ~ n=2,\ldots,m-1. \label{eqn:optimal solution 24}
	\end{align}
	\setcounter{equation}{218}
	\hrulefill
\end{figure*}
By taking (\ref{eqn:optimal solution 24}) into (\ref{eqn:optimal solution 23}), \eqref{eqn:optimal solution 25} on the next page follows.\begin{figure*}[!t]
	\normalsize
\setcounter{equation}{223}
\begin{align}
&\mv{A}_{n-1}^{(m-n+2,m-n+2)}-\mv{A}_{n-1}^{(m-n+2,1:m-n+1)}\nonumber \times (\mv{A}_{n-1}^{(1:m-n+1,1:m-n+1)})^{-1}\mv{A}_{n-1}^{(1:m-n+1,m-n+2)} \nonumber \\
&=  \mv{A}_{n-1}^{(m-n+2,m-n+2)}-\frac{\mv{A}_{n-1}^{(m-n+2,1)}\mv{A}_{n-1}^{(1,m-n+2)}}{\mv{A}_{n-1}^{(1,1)}}-\left(\mv{A}_{n-1}^{(m-n+2,2:m-n+1)}-\frac{\mv{A}_{n-1}^{(m-n+2,1)}\mv{A}_{n-1}^{(1,2:m-n+1)}}{\mv{A}_{n-1}^{(1,1)}}\right) \nonumber \\
& \quad \times \mv{B}_{n-1,m}^{-1}\bigg(\mv{A}_{n-1}^{(2:m-n+1,m-n+2)}- \frac{\mv{A}_{n-1}^{(1,m-n+2)}\mv{A}_{n-1}^{(2:m-n+1,1)}}{\mv{A}_{n-1}^{(1,1)}}\bigg), ~ n=2,\ldots,m. \label{eqn:optimal solution 25}
\end{align}\hrulefill 
\end{figure*}~According to (\ref{eqn:optimal solution 21}) and (\ref{eqn:optimal solution 25}), it is concluded that if (\ref{eqn:optimal solution 2}) is true, then (\ref{eqn:optimal solution 4}) holds. Lemma \ref{lemma2} is thus proved.
\end{IEEEproof}

\setcounter{equation}{228}
\begin{figure*}[!t]
	\begin{align}\label{eqn:optimal solution 7}
	2^{C_m}\mv{\Lambda}_m^{(m,m)}=\left\{\begin{array}{ll}\mv{A}_1^{(1,1)} & {\rm if} ~ m=1, \\ \mv{A}_1^{(m,m)}-\mv{A}_1^{(m,1:m-1)}(\mv{A}_1^{(1:m-1,1:m-1)})^{-1}\mv{A}_1^{(1:m-1,m)}, & {\rm if} ~ m=2,\ldots,M.\end{array}\right.
	\end{align}\hrulefill
\end{figure*}
By combining (\ref{eqn:optimal solution 3}), Proposition \ref{lemma1}, and Lemma \ref{lemma2}, the optimal solution to problem (\ref{eqn:problem 5}) satisfies
\begin{align}\setcounter{equation}{224}
&\mv{A}_m^{(1,1)} \nonumber \\ &= \mv{A}_{m-1}^{(2,2)}-\frac{\mv{A}_{m-1}^{(2,1)}\mv{A}_{m-1}^{(1,2)}}{\mv{A}_{m-1}^{(1,1)}} \\ &= \mv{A}_{m-2}^{(3,3)}-\mv{A}_{m-2}^{(3,1:2)}(\mv{A}_{m-2}^{(1:2,1:2)})^{-1}\mv{A}_{m-2}^{(1:2,3)} \\ & ~ \vdots \nonumber \\ &=
\mv{A}_1^{(m,m)}-\mv{A}_1^{(m,1:m-1)}(\mv{A}_1^{(1:m-1,1:m-1)})^{-1}\mv{A}_1^{(1:m-1,m)}, \nonumber \\ & \qquad\qquad\qquad\qquad\qquad\qquad\qquad\quad m=2,\ldots,M. \label{eqn:optimal solution 5}
\end{align}Moreover, according to (\ref{eqn:optimal condition}), it follows that
\begin{align}\label{eqn:optimal solution 6}
2^{C_m}\mv{\Lambda}_m^{(m,m)}=\mv{A}_m^{(1,1)}, ~ \forall m.
\end{align}By combining (\ref{eqn:optimal solution 5}) and (\ref{eqn:optimal solution 6}), it can be concluded that the optimal solution to problem (\ref{eqn:problem 5}) must satisfy (\ref{eqn:optimal solution 7}) on the next page.
\setcounter{equation}{229}
To summarize, according to constraints (\ref{eqn:A1}) and (\ref{eqn:A2}) in problem (\ref{eqn:problem 5}), the optimal $\mv{\Lambda}_m^{(m,m)}$'s are just functions of $\mv{A}_1$ as shown in (\ref{eqn:optimal solution 7}). Moreover, constraints (\ref{eqn:dual uplink fronthaul constraint 1}) and (\ref{eqn:dual uplink SINR constraint}) are only dependent of $\mv{\Lambda}_m^{(m,m)}$'s, but are independent of the other elements of $\mv{\Lambda}_m$'s. As a result, problem (\ref{eqn:problem 5}) is equivalent to the following problem
\begin{align}\hspace{-8pt}
\mathop{\mathrm{maximize}}_{\{\beta_k\},\{\mv{\Lambda}_m^{(m,m)}\},\mv{A}_1} & ~~~ \sum\limits_{k=1}^K\beta_k \sigma^2 \label{eqn:problem 8} \\
\mathrm {subject ~ to} \ \ \  & ~ \mv{A}_1\succ \mv{0}, \\ \ \ \ & ~ (\ref{eqn:dual uplink SINR constraint}), ~ (\ref{eqn:nonegative beta}), ~ (\ref{eqn:dual uplink fronthaul constraint 1}), ~  (\ref{eqn:positive Phi}), ~ (\ref{eqn:optimal solution 7}). \nonumber
\end{align}As compared to problem (\ref{eqn:problem 5}), the optimization variables $\mv{A}_2,\ldots,\mv{A}_M$ disappear in problem (\ref{eqn:problem 8}). Moreover, constraints (\ref{eqn:A1}) and (\ref{eqn:A2}) reduce to constraint (\ref{eqn:optimal solution 7}).

\begin{lemma}\label{lemma3}
Consider $\mv{X}\in \mathbb{C}^{M\times M}$ and $\mv{Y}\in \mathbb{C}^{M\times M}$. If $\mv{X}\succ \mv{0}$, $\mv{Y}\succ \mv{0}$, and $\mv{X}\succeq \mv{Y}$, it then follows that
\begin{equation}
\mv{X}^{(1,1)}\geq \mv{Y}^{(1,1)}, \label{eqn:lemma 31}
\end{equation}
and
\begin{align}
&\mv{X}^{(m,m)}-\mv{X}^{(m,1:m-1)}(\mv{X}^{(1:m-1,1:m-1)})^{-1}\mv{X}^{(1:m-1,m)} \nonumber \\
&\geq \mv{Y}^{(m,m)}-\mv{Y}^{(m,1:m-1)}(\mv{Y}^{(1:m-1,1:m-1)})^{-1}\mv{Y}^{(1:m-1,m)}, \nonumber \\ & \qquad\qquad\qquad\qquad\qquad\qquad\qquad\quad ~ m=2,\ldots,M. \label{eqn:optimal solution 8}
\end{align}
\end{lemma}

\begin{IEEEproof}
(\ref{eqn:lemma 31}) directly follows from $\mv{X}\succeq \mv{Y}$. In the following, we prove (\ref{eqn:optimal solution 8}). Since $\mv{X}\succ \mv{0}$, $\mv{Y}\succ \mv{0}$, and $\mv{Y}\succeq \mv{X}$, we have $\mv{X}^{(1:m,1:m)}\succ \mv{0}$, $\mv{Y}^{(1:m,1:m)}\succ \mv{0}$, and $\mv{X}^{(1:m,1:m)}\succeq \mv{Y}^{(1:m,1:m)}$, $\forall m$. As a result, the inverses of $\mv{X}^{(1:m-1,1:m-1)}$'s and $\mv{Y}^{(1:m-1,1:m-1)}$'s exist, $\forall m\geq 2$, and \eqref{largeeq241}--\eqref{largeeq244} on top of the next page are true.

Lemma \ref{lemma3} is thus proved.
\end{IEEEproof}

\setcounter{equation}{233}
\begin{figure*}[!t]
	\normalsize \begin{align}
	&\mv{X}^{(m,m)}-\mv{X}^{(m,1:m-1)}(\mv{X}^{(1:m-1,1:m-1)})^{-1}\mv{X}^{(1:m-1,m)}\nonumber \\
	&= [-\mv{X}^{(m,1:m-1)}(\mv{X}^{(1:m-1,1:m-1)})^{-1}, 1]\times\mv{X}^{(1:m,1:m)} \times [-\mv{X}^{(m,1:m-1)}(\mv{X}^{(1:m-1,1:m-1)})^{-1}, 1]^H \label{largeeq241}\\
	&\geq [-\mv{X}^{(m,1:m-1)}(\mv{X}^{(1:m-1,1:m-1)})^{-1}, 1]\times\mv{Y}^{(1:m,1:m)} \times [-\mv{X}^{(m,1:m-1)}(\mv{X}^{(1:m-1,1:m-1)})^{-1}, 1]^H \\
	&= \mv{Y}^{(m,m)}-\mv{Y}^{(m,1:m-1)}(\mv{Y}^{(1:m-1,1:m-1)})^{-1}\mv{Y}^{(1:m-1,m)}+ [-\mv{X}^{(m,1:m-1)}(\mv{X}^{(1:m-1,1:m-1)})^{-1}\nonumber \\ & \quad + \mv{Y}^{(m,1:m-1)}(\mv{Y}^{(1:m-1,1:m-1)})^{-1}]\times \mv{Y}^{(1:m-1,1:m-1)} \times[-\mv{X}^{(m,1:m-1)}(\mv{X}^{(1:m-1,1:m-1)})^{-1}\nonumber \\ & \quad + \mv{Y}^{(m,1:m-1)}(\mv{Y}^{(1:m-1,1:m-1)})^{-1}]^H \\
	&\geq \mv{Y}^{(m,m)}-\mv{Y}^{(m,1:m-1)}(\mv{Y}^{(1:m-1,1:m-1)})^{-1}\mv{Y}^{(1:m-1,m)}, ~ m=2,\ldots,M.\label{largeeq244}
	\end{align}\hrulefill 
\end{figure*}

\begin{figure*}\setcounter{equation}{240}\begin{align}
	\left[\sigma^2\mv{I}+\sum\limits_{i=1}^K\beta_k\mv{h}_k\mv{h}_k^H\right]^{(m,m)}-\mv{\Omega}^{(m,1:m-1)}(\mv{\Omega}^{(1:m-1,1:m-1)})^{-1}\mv{\Omega}^{(1:m-1,m)} \geq (2^{C_m}-1)\mv{\Lambda}_m^{(m,m)}, ~ \forall m.\label{248}
	\end{align}\hrulefill 
\end{figure*}
\begin{figure*}\begin{align}
	& \left[\sigma^2\mv{I}+\sum\limits_{k=1}^K\beta_k\mv{h}_k\mv{h}_k^H\right]^{(\bar{m},\bar{m})}-\mv{\Omega}^{(\bar{m},1:\bar{m}-1)}(\mv{\Omega}^{(1:\bar{m}-1,1:\bar{m}-1)})^{-1}\mv{\Omega}^{(1:\bar{m}-1,\bar{m})} >  (2^{C_{\bar{m}}}-1)\mv{\Lambda}_{\bar{m}}^{(\bar{m},\bar{m})}.\label{249}
	\end{align}\hrulefill 
\end{figure*}
\begin{figure*}[!t]
	\setcounter{equation}{244}
	\begin{align}\label{eqn:Omega et}
	\tilde{\mv{\Omega}}^{(m_1,m_2)}=\left\{\begin{array}{ll}\mv{\Omega}^{(m_1,m_2)}, & {\rm if} ~ (m_1,m_2)\neq (\bar{m},\bar{m}), \\ \mv{\Omega}^{(\bar{m},\bar{m})}-\mv{\Lambda}_{\bar{m}}^{(\bar{m},\bar{m})}+\tilde{\mv{\Lambda}}_{\bar{m}}^{(\bar{m},\bar{m})}>\mv{\Omega}^{(\bar{m},\bar{m})}, & {\rm otherwise}.\end{array}\right.
	\end{align}
	\hrulefill
\end{figure*}

\setcounter{equation}{237}

According to Lemma \ref{lemma3}, (\ref{eqn:dual uplink fronthaul constraint 1}) and (\ref{eqn:optimal solution 7}) indicate that
\begin{align}
&\mv{\Omega}^{(m,m)}-\mv{\Omega}^{(m,1:m-1)}(\mv{\Omega}^{(1:m-1,1:m-1)})^{-1}\mv{\Omega}^{(1:m-1,m)} \nonumber \\
& \geq \mv{A}_1^{(m,m)}-\mv{A}_1^{(m,1:m-1)}(\mv{A}_1^{(1:m-1,1:m-1)})^{-1}\mv{A}_1^{(1:m-1,m)}\nonumber \\
& =2^{C_m}\mv{\Lambda}_m^{(m,m)}, ~ m=1,\ldots,M. \label{eqn:optimal solution 9}
\end{align}This corresponds to constraint (\ref{eqn:eqv dual uplink fronthaul constraint}). As a result, any feasible solution to problem (\ref{eqn:problem 8}) is feasible to problem (\ref{eqn:problem 6}). In other words, the optimal value of problem (\ref{eqn:problem 6}) is no smaller than that of problem (\ref{eqn:problem 8}).

Next, we show that the optimal value of problem (\ref{eqn:problem 6}) is no larger than that of problem (\ref{eqn:problem 8}).
\begin{lemma}\label{lemma4}
The optimal solution to problem (\ref{eqn:problem 6}) must satisfy
\begin{multline}\label{eqn:optimal solution 10}
 \mv{\Omega}^{(m,m)}-\mv{\Omega}^{(m,1:m-1)}(\mv{\Omega}^{(1:m-1,1:m-1)})^{-1}\mv{\Omega}^{(1:m-1,m)} \\ = 2^{C_m}\mv{\Lambda}_m^{(m,m)}, ~ m=1,\ldots,M.
\end{multline}
\end{lemma}

\begin{IEEEproof}
According to (\ref{eqn:Omega}), we have
\begin{equation}
\mv{\Omega}^{(m,m)}=\left[\sigma^2\mv{I}+\sum\limits_{i=1}^K\beta_k\mv{h}_k\mv{h}_k^H\right]^{(m,m)}+\mv{\Lambda}_m^{(m,m)}, ~ \forall m.
\end{equation}
Therefore, we can re-express constraint (\ref{eqn:eqv dual uplink fronthaul constraint}) in problem (\ref{eqn:problem 6}) as \eqref{248} on top of the page.

Note that $[\sigma^2\mv{I}+\sum_{i=1}^K\beta_k\mv{h}_k\mv{h}_k^H]^{(m,m)}$ and $\mv{\Omega}^{(m,1:m-1)}(\mv{\Omega}^{(1:m-1,1:m-1)})^{-1}\mv{\Omega}^{(1:m-1,m)}$ are not functions of $\mv{\Lambda}_m^{(m,m)}$, $\forall m$.

Suppose that for a given feasible solution to problem (\ref{eqn:problem 6}), denoted by $\beta_k$'s and $\mv{\Lambda}_m^{(m,m)}$'s, there exists an $1\leq \bar{m} \leq M$ such that \eqref{249} on top of the page is true. Then, we consider another solution where $\tilde{\beta}_k=\beta_k$, $\forall k$, and $\tilde{\mv{\Lambda}}_m^{(m,m)}=\mv{\Lambda}_m^{(m,m)}$, $\forall m\neq \bar{m}$, while\setcounter{equation}{242}
\begin{multline}
 \tilde{\mv{\Lambda}}_{\bar{m}}^{(\bar{m},\bar{m})}=\frac{\left[\sigma^2\mv{I}+\sum\limits_{k=1}^K\beta_k\mv{h}_k\mv{h}_k^H\right]^{(\bar{m},\bar{m})}}{2^{C_{\bar{m}}}-1} \\  -\frac{\tilde{\mv{\Omega}}^{(\bar{m},1:\bar{m}-1)}(\tilde{\mv{\Omega}}^{(1:\bar{m}-1,1:\bar{m}-1)})^{-1}\tilde{\mv{\Omega}}^{(1:\bar{m}-1,\bar{m})}}{2^{C_{\bar{m}}}-1}, \end{multline}\begin{equation} \tilde{\mv{\Omega}}=\sigma^2\mv{I}+\sum\limits_{k=1}^K\tilde{\beta}_k\mv{h}_k\mv{h}_k^H+{\rm diag}(\tilde{\mv{\Lambda}}_1^{(1,1)},\ldots,\tilde{\mv{\Lambda}}_M^{(M,M)}).\end{equation}
As a result, $\tilde{\mv{\Lambda}}_{\bar{m}}^{(\bar{m},\bar{m})}>\mv{\Lambda}_{\bar{m}}^{(\bar{m},\bar{m})}$.

Consider the above new solution. First, it can be shown that (\ref{eqn:Omega et}) on top of the page is true. As a result, at the new solution, constraint (\ref{eqn:eqv dual uplink fronthaul constraint}) still holds $\forall m\leq \bar{m}$. For $m>\bar{m}$, we have
\setcounter{equation}{245}
\begin{align}
&2^{C_m}\tilde{\mv{\Lambda}}_m^{(m,m)}\nonumber \\ &= 2^{C_m}\mv{\Lambda}_m^{(m,m)} \nonumber\\ &\leq \mv{\Omega}^{(m,m)}-\mv{\Omega}^{(m,1:m-1)}(\mv{\Omega}^{(1:m-1,1:m-1)})^{-1}\mv{\Omega}^{(1:m-1,m)} \nonumber\\
&= \tilde{\mv{\Omega}}^{(m,m)}-\tilde{\mv{\Omega}}^{(m,1:m-1)}(\mv{\Omega}^{(1:m-1,1:m-1)})^{-1}\tilde{\mv{\Omega}}^{(1:m-1,m)} \nonumber\\
&\leq \tilde{\mv{\Omega}}^{(m,m)}-\tilde{\mv{\Omega}}^{(m,1:m-1)}(\tilde{\mv{\Omega}}^{(1:m-1,1:m-1)})^{-1}\tilde{\mv{\Omega}}^{(1:m-1,m)}, \label{eqn:lemma41}
\end{align}where (\ref{eqn:lemma41}) is because $\tilde{\mv{\Omega}} \succeq \mv{\Omega}$.

Next, consider constraint (\ref{eqn:dual uplink SINR constraint}) in problem (\ref{eqn:problem 6}). Since $\tilde{\mv{\Lambda}}_m^{(m,m)}=\mv{\Lambda}^{(m,m)}$, $\forall m\neq \bar{m}$,  $\tilde{\mv{\Lambda}}_{\bar{m}}^{(\bar{m},\bar{m})}>\mv{\Lambda}_{\bar{m}}^{(\bar{m},\bar{m})}$, and $\sum_{k=1}^K|\bar{u}_{k,\bar{m}}|^2>0$ according to (\ref{eqn:assumption}), constraint (\ref{eqn:dual uplink SINR constraint}) in problem (\ref{eqn:problem 6}) is satisfied with inequality at least for one $k$.

To summarize, the new solution is a feasible solution to problem (\ref{eqn:problem 6}), with constraint (\ref{eqn:dual uplink SINR constraint}) satisfied with inequality. Thus, given $\tilde{\mv{\Lambda}}_m^{(m,m)}$'s, we can further find a better solution of $\beta_k$'s such that the objective value of problem (\ref{eqn:problem 6}) is enhanced while constraint (\ref{eqn:dual uplink SINR constraint}) is satisfied with equality. This indicates that if (\ref{eqn:optimal solution 10}) does not hold for problem (\ref{eqn:problem 6}), we can always find a better solution. Lemma \ref{lemma4} is thus proved.
\end{IEEEproof}

Given any feasible solution $\{\beta_k, \mv{\Lambda}_m^{(m,m)}\}$ to problem (\ref{eqn:problem 6}) that satisfies the optimal condition (\ref{eqn:optimal solution 10}), we can construct a solution to problem (\ref{eqn:problem 8}) as follows. First, $\beta_k$'s and $\mv{\Lambda}_m^{(m,m)}$'s are unchanged in problem (\ref{eqn:problem 8}). Second, $\mv{A}_1$ is set as
\begin{align}\label{eqn:optimal solution 11}
\mv{A}_1=\mv{\Omega}\succ \mv{0}.
\end{align}Since $\beta_k$'s and $\mv{\Lambda}_m^{(m,m)}$'s is a feasible solution to problem (\ref{eqn:problem 6}), it must satisfy constraints (\ref{eqn:dual uplink SINR constraint}) and (\ref{eqn:nonegative beta}). Further, it can be observed from (\ref{eqn:Omega}) that $\mv{\Omega}\succ \mv{0}$. As a result, it follows that
\begin{multline}
\mv{\Omega}^{(m,m)}-\mv{\Omega}^{(m,1:m-1)}(\mv{\Omega}^{(1:m-1,1:m-1)})^{-1}\mv{\Omega}^{(1:m-1,m)}\\ >0, ~ \forall m.
\end{multline}
According to (\ref{eqn:optimal solution 10}), we have $\mv{\Lambda}_m^{(m,m)}>0$, $\forall m$. In other words, constraint (\ref{eqn:positive Phi}) holds. Moreover, (\ref{eqn:optimal solution 11}) guarantees that constraint (\ref{eqn:dual uplink fronthaul constraint 1}) holds in problem (\ref{eqn:problem 8}). At last, (\ref{eqn:optimal solution 9}) and (\ref{eqn:optimal solution 10}) guarantee that
\begin{align}
&\mv{\Omega}^{(m,m)}-\mv{\Omega}^{(m,1:m-1)}(\mv{\Omega}^{(1:m-1,1:m-1)})^{-1}\mv{\Omega}^{(1:m-1,m)}\nonumber \\
&= \mv{A}_1^{(m,m)}-\mv{A}_1^{(m,1:m-1)}(\mv{A}_1^{(1:m-1,1:m-1)})^{-1}\mv{A}_1^{(1:m-1,m)} \nonumber\\
&= 2^{C_m}\mv{\Lambda}_m^{(m,m)}, ~ \forall m.
\end{align}As a result, constraint (\ref{eqn:optimal solution 7}) also holds. To summarize, given any feasible solution to problem (\ref{eqn:problem 6}) that satisfies (\ref{eqn:optimal solution 10}), we can also generate a feasible solution to problem (\ref{eqn:problem 8}). Moreover, according to Lemma \ref{lemma4}, the optimal solution to problem (\ref{eqn:problem 6}) must satisfy condition (\ref{eqn:optimal solution 10}). As a result, the optimal value of problem (\ref{eqn:problem 6}) is achievable for problem (\ref{eqn:problem 8}). In other words, the optimal value of problem (\ref{eqn:problem 6}) is no larger than that of problem (\ref{eqn:problem 8}).

To summarize, we have shown that the optimal value of problem (\ref{eqn:problem 6}) is not smaller than that of problem (\ref{eqn:problem 8}), and at the same time, the optimal value of problem (\ref{eqn:problem 6}) is no larger than that of problem (\ref{eqn:problem 8}). This indicates that problem (\ref{eqn:problem 6}) is equivalent to problem (\ref{eqn:problem 8}), which is further equivalent to problem (\ref{eqn:problem 5}). In other words, problem (\ref{eqn:problem 6}) is equivalent to (\ref{eqn:problem 5}). This completes the proof of Proposition \ref{proposition4}.

\subsection{Proof of Proposition \ref{proposition5}}\label{appendix6}
According to Lemma \ref{lemma4} in Appendix \ref{appendix5}, with the optimal solution to problem (\ref{eqn:problem 6}), all the constraints shown in (\ref{eqn:eqv dual uplink fronthaul constraint}) should hold with equality. In the following, we show that with the optimal solution, to problem (\ref{eqn:problem 6}), all the constraints shown in (\ref{eqn:dual uplink SINR constraint}) should hold with equality, i.e.,
\begin{multline}\label{eqn:optimal solution 51}
 \sigma^2+\sum\limits_{j\neq k}\beta_j|\bar{\mv{u}}_k^H\mv{h}_j|^2+\sum\limits_{m=1}^M\mv{\Lambda}_m^{(m,m)}|\bar{u}_{k,m}|^2 \\ -\frac{\beta_k|\bar{\mv{u}}_k^H\mv{h}_k|^2}{2^{R_k}-1}= 0, ~ \forall k.
\end{multline}

Given any feasible solution to problem (\ref{eqn:problem 6}), suppose that there exists a $\bar{k}$ such that
\begin{multline}
\sigma^2+\sum\limits_{j\neq \bar{k}}\beta_j|\bar{\mv{u}}_{\bar{k}}^H\mv{h}_j|^2+\sum\limits_{m=1}^M\mv{\Lambda}_m^{(m,m)}|\bar{u}_{\bar{k},m}|^2\\-\frac{\beta_{\bar{k}}|\bar{\mv{u}}_k^H\mv{h}_{\bar{k}}|^2}{2^{R_{\bar{k}}}-1}> 0.
\end{multline}
Then, consider another solution of $\tilde{\beta}_{\bar{k}}$ such that
\begin{multline}\label{eqn:optimal solution 52}
\sigma^2+\sum\limits_{j\neq \bar{k}}\beta_j|\bar{\mv{u}}_{\bar{k}}^H\mv{h}_j|^2+\sum\limits_{m=1}^M\mv{\Lambda}_m^{(m,m)}|\bar{u}_{\bar{k},m}|^2\\-\frac{\tilde{\beta}_{\bar{k}}|\bar{\mv{u}}_k^H\mv{h}_{\bar{k}}|^2}{2^{R_{\bar{k}}}-1}= 0.
\end{multline}
As a result, we have
\begin{align}
\tilde{\beta}_{\bar{k}}>\beta_{\bar{k}}.
\end{align}

First, it can be shown that with $\tilde{\beta}_{\bar{k}}$, constraint (\ref{eqn:dual uplink SINR constraint}) holds for all $k$. Moreover, according to (\ref{eqn:Omega}), it follows that
\begin{align}
\mv{\Omega}& =\sigma^2\mv{I}+\sum\limits_{k=1}^K\beta_k\mv{h}_k\mv{h}_k^H+\sum\limits_{m=1}^M\mv{E}_m^H\mv{\Lambda}_m\mv{E}_m \nonumber \\
& \preceq \sigma^2\mv{I}+\sum\limits_{j\neq \bar{k}}\beta_k\mv{h}_k\mv{h}_k^H+\tilde{\beta}_{\bar{k}}\mv{h}_{\bar{k}}\mv{h}_{\bar{k}}^H+\sum\limits_{m=1}^M\mv{E}_m^H\mv{\Lambda}_m\mv{E}_m. \label{eqn:optimal solution 53}
\end{align}According to Lemma \ref{lemma3}, it follows that the new solution also satisfies constraint (\ref{eqn:eqv dual uplink fronthaul constraint}). As a result, the new solution is also a feasible solution to problem (\ref{eqn:problem 6}). However, with the new solution, the objective value of problem (\ref{eqn:problem 6}) is increased. This indicates that with the optimal solution to problem (\ref{eqn:problem 6}), all the constraints shown in (\ref{eqn:dual uplink SINR constraint}) should hold with equality.

Proposition \ref{proposition5} is thus proved.
\setcounter{equation}{266}
\begin{figure*}
	\begin{align}
	\lambda_{\bar{m}}(\alpha\mv{\beta})&=\frac{\sigma^2+\alpha\sum\limits_{k=1}^K\beta_kh_{k,{\bar{m}}}h_{k,{\bar{m}}}^H-\bar{\mv{\Omega}}^{({\bar{m}},1:{\bar{m}}-1)}(\bar{\mv{\Omega}}^{(1:{\bar{m}}-1,1:{\bar{m}}-1)})^{-1}\bar{\mv{\Omega}}^{(1:{\bar{m}}-1,{\bar{m}})}}{2^{C_{\bar{m}}}-1}\label{eqn:optimal solution 712} \\
	&= \frac{\sigma^2+\alpha\sum\limits_{k=1}^K\beta_kh_{k,{\bar{m}}}h_{k,{\bar{m}}}^H-\alpha^2\hat{\mv{\Omega}}^{({\bar{m}},1:{\bar{m}}-1)}(\bar{\mv{\Omega}}^{(1:{\bar{m}}-1,1:{\bar{m}}-1)})^{-1}\hat{\mv{\Omega}}^{(1:{\bar{m}}-1,{\bar{m}})}}{2^{C_{\bar{m}}}-1}\label{eqn:optimal solution 713} \\
	&<\frac{\sigma^2+\alpha\sum\limits_{k=1}^K\beta_kh_{k,{\bar{m}}}h_{k,{\bar{m}}}^H-\alpha^2\hat{\mv{\Omega}}^{({\bar{m}},1:{\bar{m}}-1)}(\alpha\hat{\mv{\Omega}}^{(1:{\bar{m}}-1,1:{\bar{m}}-1)})^{-1}\hat{\mv{\Omega}}^{(1:{\bar{m}}-1,{\bar{m}})}}{2^{C_{\bar{m}}}-1}\label{eqn:optimal solution 714} \\
	&<\frac{\alpha\left(\sigma^2+\sum\limits_{k=1}^K\beta_kh_{k,{\bar{m}}}h_{k,{\bar{m}}}^H-\hat{\mv{\Omega}}^{({\bar{m}},1:{\bar{m}}-1)}(\hat{\mv{\Omega}}^{(1:{\bar{m}}-1,1:{\bar{m}}-1)})^{-1}\hat{\mv{\Omega}}^{(1:{\bar{m}}-1,{\bar{m}})}\right)}{2^{C_{\bar{m}}}-1}\label{eqn:optimal solution 715} \\
	& =  \alpha\lambda_{\bar{m}}(\mv{\beta}). \label{eqn:optimal solution 716}
	\end{align}
	\hrulefill
\end{figure*}
\setcounter{equation}{254}

\subsection{Proof of Proposition \ref{proposition6}}\label{appendix7}
Given any $\mv{\beta}=[\beta_1,\ldots,\beta_K]>\mv{0}$, let $\mv{\lambda}(\mv{\beta})=[\lambda_1(\mv{\beta}),\ldots,\lambda_M(\mv{\beta})]$ denote the solution of $\mv{\Lambda}_m^{(m,m)}$'s (i.e.,$\mv{\Lambda}_m^{(m,m)}=\lambda_m(\mv{\beta})$, $\forall m$) that satisfies constraint (\ref{eqn:eqv dual uplink fronthaul constraint 01}) in problem (\ref{eqn:problem 7}). The uniqueness of $\mv{\lambda}(\mv{\beta})$ given $\mv{\beta}> \mv{0}$ can be shown as follows. First, it follows from (\ref{eqn:eqv dual uplink fronthaul constraint 01}) that
\begin{align}\label{eqn:optimal solution 61}
\lambda_1(\mv{\beta})=\frac{\sigma^2+\sum\limits_{k=1}^K\beta_kh_{k,1}h_{k,1}^H}{2^{C_1}-1}.
\end{align}Next, it can be shown that if $\lambda_1(\mv{\beta}),\ldots,\lambda_{m-1}(\mv{\beta})$, then $\lambda_m(\mv{\beta})$ can be uniquely determined as
\begin{multline}\label{eqn:optimal solution 62}
\lambda_m(\mv{\beta})=\frac{\sigma^2+\sum\limits_{k=1}^K\beta_kh_{k,m}h_{k,m}^H}{2^{C_m}-1} \\ -\frac{\mv{\Omega}^{(m,1:m-1)}(\mv{\Omega}^{(1:m-1,1:m-1)})^{-1}\mv{\Omega}^{(1:m-1,m)}}{2^{C_m}-1},
\end{multline}since $\mv{\Omega}^{(m,1:m-1)}$, $\mv{\Omega}^{(1:m-1,1:m-1)}$, and $\mv{\Omega}^{(1:m-1,m)}$ only depend on $\lambda_1(\mv{\beta}),\ldots,\lambda_{m-1}(\mv{\beta})$. As a result, given any $\mv{\beta}> \mv{0}$, there always exists a unique solution $\mv{\lambda}(\mv{\beta})$ such that constraint (\ref{eqn:eqv dual uplink fronthaul constraint 01}) is satisfied in problem (\ref{eqn:problem 7}). Moreover, the following lemma shows some important properties of $\mv{\lambda}(\mv{\beta})$.

\begin{lemma}\label{lemma5}
Given $\mv{\beta}\geq \mv{0}$, the function $\mv{\lambda}(\mv{\beta})$ defined by (\ref{eqn:optimal solution 61}) and (\ref{eqn:optimal solution 62}) has the following properties:
\begin{itemize}
\item[1.] $\mv{\lambda}(\mv{\beta})>\mv{0}$;
\item[2.] Given any $\alpha>1$, it follows that $\mv{\lambda}(\alpha \mv{\beta})< \alpha \mv{\lambda}(\mv{\beta})$;
\item[3.] If $\bar{\mv{\beta}}\geq \mv{\beta}$, then $\mv{\lambda}(\bar{\mv{\beta}})\geq \mv{\lambda}(\mv{\beta})$.
\end{itemize}
\end{lemma}

\begin{IEEEproof}
First, since $\lambda_m(\mv{\beta})$'s in (\ref{eqn:optimal solution 61}) and (\ref{eqn:optimal solution 62}) satisfy (\ref{eqn:eqv dual uplink fronthaul constraint 01}) and $\mv{\Omega}\succ \mv{0}$, it then follows that $\lambda_m(\mv{\beta})>0$, $\forall m$.

Next, we show that given any $\alpha>1$, we have $\mv{\lambda}(\alpha\mv{\beta})<\alpha\mv{\lambda}(\mv{\beta})$. For convenience, we define
\begin{align}
& \hat{\mv{\Omega}}=\sigma^2\mv{I}+\sum\limits_{k=1}^K\beta_k\mv{h}_k\mv{h}_k^H+{\rm diag}(\lambda_1(\mv{\beta}),\ldots,\lambda_M(\mv{\beta})), \label{eqn:Omega 1} \\
& \bar{\mv{\Omega}}=\sigma^2\mv{I}+\sum\limits_{k=1}^K\alpha\beta_k\mv{h}_k\mv{h}_k^H+{\rm diag}(\lambda_1(\alpha\mv{\beta}),\ldots,\lambda_M(\alpha\mv{\beta})), \label{eqn:Omega 2}
\end{align}as the corresponding $\mv{\Omega}$'s shown in (\ref{eqn:Omega}) with $\mv{\beta}$ and $\mv{\Lambda}_m^{(m,m)}$'s replaced by $\mv{\beta}$ and $\mv{\lambda}(\mv{\beta})$ as well as $\alpha\mv{\beta}$ and $\mv{\lambda}(\alpha\mv{\beta})$, respectively. It is observed from (\ref{eqn:Omega 1}) and (\ref{eqn:Omega 2}) that the non-diagonal elements of $\hat{\mv{\Omega}}$ and $\hat{\mv{\Omega}}$ satisfy
\begin{align}
\alpha\hat{\mv{\Omega}}^{(m1,m2)}=\bar{\mv{\Omega}}^{(m1,m2)}, ~~~ \forall m1\neq m2. \label{eqn:optimal solution 71}
\end{align}
As a result, we have
\begin{align}\label{eqn:optimal solution 77}
\bar{\mv{\Omega}}^{(m,1:m-1)}=\alpha\hat{\mv{\Omega}}^{(m,1:m-1)}, ~ \forall m.
\end{align}

In the following, we shown by induction that $\alpha\lambda_m(\mv{\beta})>\lambda_m(\alpha\mv{\beta})$, $\forall m$.

First, when $m=1$, it follows from (\ref{eqn:optimal solution 61}) that
\begin{align}
\alpha\lambda_1(\mv{\beta})&= \frac{\alpha\sigma^2+\alpha\sum\limits_{k=1}^K\beta_kh_{k,1}h_{k,1}^H}{2^{C{1}}-1} \nonumber\\
 &>\frac{\sigma^2+\alpha\sum\limits_{k=1}^K\beta_kh_{k,1}h_{k,1}^H}{2^{C{1}}-1} \nonumber\\ & = \lambda_1(\alpha\mv{\beta}). \label{eqn:optimal solution 72}
\end{align}

Next, we shown that given any $\bar{m}\geq 2$, if
\begin{align}\label{eqn:optimal solution 75}
\alpha\lambda_m(\mv{\beta}) > \lambda_m(\alpha \mv{\beta}), ~ \forall m=1,\ldots,\bar{m}-1,
\end{align}then
\begin{align}\label{eqn:optimal solution 78}
\alpha\lambda_{\bar{m}}(\mv{\beta})>\lambda_{\bar{m}}(\alpha\mv{\beta}).
\end{align}

Given (\ref{eqn:optimal solution 75}), the diagonal elements of $\hat{\mv{\Omega}}$ and $\bar{\mv{\Omega}}$ defined in (\ref{eqn:Omega 1}) and (\ref{eqn:Omega 2}) satisfy
\begin{align}\label{eqn:optimal solution 79}
\bar{\mv{\Omega}}^{(m,m)}&=\sigma^2+\alpha\sum\limits_{k=1}^K\beta_kh_{k,m}h_{k,m}^H+\lambda_{m}(\alpha\mv{\beta}) \nonumber \\ &< \alpha\left(\sigma^2+\sum\limits_{k=1}^K\beta_kh_{k,m}h_{k,m}^H+\lambda_{m}(\mv{\beta})\right) \nonumber \\
&=\alpha\hat{\mv{\Omega}}^{(m,m)}, ~ \forall m=1,\ldots,\bar{m}-1.
\end{align}Based on (\ref{eqn:optimal solution 77}) and (\ref{eqn:optimal solution 79}), it then follows that
\begin{align}\label{eqn:optimal solution 710}
\bar{\mv{\Omega}}^{(1:\bar{m}-1,1:\bar{m}-1)}\prec\alpha\hat{\mv{\Omega}}^{(1:\bar{m},1:\bar{m}-1)},
\end{align}or
\begin{align}\label{eqn:optimal solution 711}
\left(\bar{\mv{\Omega}}^{(1:\bar{m}-1,1:\bar{m}-1)}\right)^{-1}\succ\left(\alpha\hat{\mv{\Omega}}^{(1:\bar{m},1:\bar{m}-1)}\right)^{-1}.
\end{align}

Last, it follows that (\ref{eqn:optimal solution 712}) -- (\ref{eqn:optimal solution 716}) on top of the page are true, where (\ref{eqn:optimal solution 713}) is due to (\ref{eqn:optimal solution 77}) and (\ref{eqn:optimal solution 714}) is due to (\ref{eqn:optimal solution 711}). As a result, given any $\bar{m}\geq 2$, if (\ref{eqn:optimal solution 75}) is true, then (\ref{eqn:optimal solution 78}) is true. By combining the above with (\ref{eqn:optimal solution 72}), it follows that $\alpha\lambda_m(\mv{\beta})>\lambda_m(\alpha\mv{\beta})$, $\forall m$.
\setcounter{equation}{276}
\begin{figure*}[!t]
	\normalsize
	\begin{align}\label{eqn:interference function}
	& I_k(\mv{\beta})=\frac{(2^{R_k}-1)(\sigma^2+\sum_{j\neq k}\beta_j|\bar{\mv{u}}_k^H\mv{h}_j|^2+\sum_{m=1}^M\lambda_m(\mv{\beta})|\bar{u}_{k,m}|^2)}{|\bar{\mv{u}}_k^H\mv{h}_k|^2}, ~ \forall k.
	\end{align}\hrulefill 
\end{figure*}
\setcounter{equation}{271}
Next, we shown that if $\bar{\mv{\beta}}=[\bar{\beta}_1,\ldots,\bar{\beta}_K]^T$ satisfies $\bar{\beta}_k\geq \beta_k$, $\forall k$, then $\mv{\lambda}(\bar{\mv{\beta}})\geq \mv{\lambda}(\mv{\beta})$ by induction. First, it can be shown from (\ref{eqn:optimal solution 61}) that if $\bar{\mv{\beta}}\geq \mv{\beta}$, then
\begin{align}
\lambda_1(\bar{\mv{\beta}})\geq \lambda_1(\mv{\beta}).
\end{align}In the following, we prove that given any $\bar{m}\geq 2$, if $\lambda_m(\bar{\mv{\beta}})\geq \lambda_m(\mv{\beta})$, $\forall m\leq \bar{m}-1$, then $\lambda_{\bar{m}}(\bar{\mv{\beta}})\geq \lambda_{\bar{m}}(\mv{\beta})$. To prove this, given any $\bar{m}\geq 2$, define
\begin{align}
\mv{X}_{\bar{m}}&=\sigma^2I+\sum\limits_{k=1}^K\beta_k\mv{h}_k^{(1:\bar{m})}[\mv{h}_k^{(1:\bar{m})}]^H\nonumber \\ & ~ +{\rm diag}([\lambda_1(\mv{\beta}),\ldots,\lambda_{\bar{m}-1}(\mv{\beta}),0])\in \mathbb{C}^{\bar{m}\times \bar{m}}, \\
\bar{\mv{X}}_{\bar{m}}&=\sigma^2I+\sum\limits_{k=1}^K\bar{\beta}_k\mv{h}_k^{(1:\bar{m})}[\mv{h}_k^{(1:\bar{m})}]^H \nonumber \\ & ~ +{\rm diag}([\lambda_1(\bar{\mv{\beta}}),\ldots,\lambda_{\bar{m}-1}(\bar{\mv{\beta}}),0])\in \mathbb{C}^{\bar{m}\times \bar{m}}.
\end{align}With $\bar{\mv{\beta}}\geq \mv{\beta}$ and $\lambda_m(\bar{\mv{\beta}})\geq \lambda_m(\mv{\beta})$, $\forall m\leq \bar{m}-1$, it follows that $\bar{\mv{X}}_{\bar{m}} \succeq \mv{X}_{\bar{m}}$. Based on Lemma \ref{lemma3}, it can be shown from (\ref{eqn:optimal solution 62}) that
\begin{align}
&\lambda_{\bar{m}}(\bar{\mv{\beta}})\nonumber \\ &= \frac{\bar{\mv{X}}_{\bar{m}}^{(\bar{m},\bar{m})}-\bar{\mv{X}}_{\bar{m}}^{(1:\bar{m}-1,\bar{m})}(\bar{\mv{X}}_{\bar{m}}^{(1:\bar{m}-1,1:\bar{m}-1)})^{-1}\bar{\mv{X}}_{\bar{m}}^{(\bar{m},1:\bar{m}-1)}}{2^{\bar{m}}-1} \nonumber\\
&\geq \frac{\mv{X}_{\bar{m}}^{(\bar{m},\bar{m})}-\mv{X}_{\bar{m}}^{(1:\bar{m}-1,\bar{m})}(\mv{X}_{\bar{m}}^{(1:\bar{m}-1,1:\bar{m}-1)})^{-1}\mv{X}_{\bar{m}}^{(\bar{m},1:\bar{m}-1)}}{2^{\bar{m}}-1} \nonumber\\
&= \lambda_{\bar{m}}(\mv{\beta}).
\end{align}

To summarize, we have shown that $\lambda_1(\bar{\mv{\beta}})\geq \lambda_1(\mv{\beta})$ and given any $\bar{m}\geq 2$, $\lambda_{\bar{m}}(\bar{\mv{\beta}})\geq \lambda_{\bar{m}}(\mv{\beta})$ if $\lambda_m(\bar{\mv{\beta}})\geq \lambda_m(\mv{\beta})$, $\forall m\leq \bar{m}-1$. Therefore, by deduction, it can be shown that $\mv{\lambda}(\bar{\mv{\beta}})\geq \mv{\lambda}(\mv{\beta})$ if $\bar{\mv{\beta}}\geq \mv{\beta}$.

Lemma \ref{lemma5} is thus proved.
\end{IEEEproof}

Then, the remaining job is to show that there is only a unique solution $\mv{\beta}>\mv{0}$ such that $\mv{\beta}>\mv{0}$ and $\mv{\lambda}(\mv{\beta})$ satisfy constraint (\ref{eqn:dual uplink SINR constraint 01}) in problem (\ref{eqn:problem 7}). In the following, we prove this.

Note that constraint (\ref{eqn:dual uplink SINR constraint 01}) can be re-expressed as
\begin{align}\label{eqn:update}
\mv{\beta}=\mv{I}(\mv{\beta}),
\end{align}where $\mv{I}(\mv{\beta})=[I_1(\mv{\beta}),\ldots,I_K(\mv{\beta})]$ with $I_k(\mv{\beta})$ defined in \eqref{eqn:interference function} on top of the page.

In the following, we show three important properties of the function $\mv{I}(\mv{\beta})$.

\begin{lemma}\label{lemma11}
Given $\mv{\beta}\geq \mv{0}$, the function $\mv{I}(\mv{\beta})$ defined by (\ref{eqn:interference function}) satisfies the following three properties:
\begin{itemize}
\item[1.] $\mv{I}(\mv{\beta})>\mv{0}$;
\item[2.] Given any $\alpha>1$, it follows that $\mv{I}(\alpha \mv{\beta})< \alpha \mv{I}(\mv{\beta})$;
\item[3.] If $\bar{\mv{\beta}}\geq \mv{\beta}$, then $\mv{I}(\bar{\mv{\beta}})\geq \mv{I}(\mv{\beta})$.
\end{itemize}
\end{lemma}
\setcounter{equation}{277}
\begin{IEEEproof}
First, given any $\mv{\beta}\geq \mv{0}$, it can be shown from (\ref{eqn:interference function}) that
\begin{align}\label{eqn:standard interference function 1}
\mv{I}(\mv{\beta})>\mv{0}.
\end{align}

Next, according to Lemma \ref{lemma5}, given any $\alpha>1$, it can be shown that
\begin{align}
&I_k(\alpha \mv{\beta})\nonumber \\ = & \frac{(2^{R_k}-1)\left(\sigma^2+\sum\limits_{j\neq k}\alpha \beta_j|\bar{\mv{u}}_k^H\mv{h}_j|^2+\sum\limits_{m=1}^M\lambda_m(\alpha\mv{\beta})|\bar{u}_{k,m}|^2\right)}{|\bar{\mv{u}}_k^H\mv{h}_k|^2} \nonumber\\
\leq & \frac{\alpha(2^{R_k}-1)\left(\sigma^2+\sum\limits_{j\neq k}\beta_j|\bar{\mv{u}}_k^H\mv{h}_j|^2+\sum\limits_{m=1}^M\lambda_m(\mv{\beta})|\bar{u}_{k,m}|^2\right)}{|\bar{\mv{u}}_k^H\mv{h}_k|^2} \nonumber \\
= & \alpha I_k(\mv{\beta}), ~ \forall k, \label{eqn:standard interference function 3}
\end{align}where the inequality is due to Lemma \ref{lemma5}.

Last, if $\bar{\mv{\beta}}\geq \mv{\beta}$, it then follows that
\begin{align}
&I_k(\bar{\mv{\beta}})\nonumber \\ = &\frac{(2^{R_k}-1)\left(\sigma^2+\sum\limits_{j\neq k}\bar{\beta}_j|\bar{\mv{u}}_k^H\mv{h}_j|^2+\sum\limits_{m=1}^M\lambda_m(\bar{\mv{\beta}})|\bar{u}_{k,m}|^2\right)}{|\bar{\mv{u}}_k^H\mv{h}_k|^2} \nonumber\\
\geq & \frac{(2^{R_k}-1)\left(\sigma^2+\sum\limits_{j\neq k}\beta_j|\bar{\mv{u}}_k^H\mv{h}_j|^2+\sum\limits_{m=1}^M\lambda_m(\mv{\beta})|\bar{u}_{k,m}|^2\right)}{|\bar{\mv{u}}_k^H\mv{h}_k|^2} \nonumber \\
= & I_k(\mv{\beta}), ~ \forall k, \label{eqn:standard interference function 2}
\end{align}where the inequality is due to Lemma \ref{lemma5}.

Lemma \ref{lemma11} is thus proved.
\end{IEEEproof}

Lemma \ref{lemma11} shows that the function $\mv{I}(\mv{\beta})$ defined by (\ref{eqn:interference function}) is a standard interference function \cite{Yates95}. It then follows from \cite[Theorem 1]{Yates95} that there exists a unique solution $\mv{\beta}>\mv{0}$ to equation (\ref{eqn:update}). Note that we have shown in the above that given $\mv{\beta}>\mv{0}$, there exists a unique solution $\mv{\lambda}(\mv{\beta})$ that satisfies constraint (\ref{eqn:eqv dual uplink fronthaul constraint 01}). Therefore, if problem (\ref{eqn:problem 7}) is feasible, there exists only one solution to problem (\ref{eqn:problem 7}).

Proposition \ref{proposition6} is thus proved.

\subsection{Proof of Lemma \ref{lemmacaseI}}\label{appendix8}
First, it can be easily shown that if $\mv{p}^{{\rm ul}}\geq \mv{0}$, then $I_k(\mv{p}^{{\rm ul}})>0$, $\forall k$. Next, given $\alpha>1$, it follows from (\ref{eqn:optimal quantization Case I}) that $q_m^{{\rm ul}}(\alpha \mv{p}^{{\rm ul}})<\alpha q_m^{{\rm ul}}(\mv{p}^{{\rm ul}})$, $\forall m$. As a result, $\forall k \in \mathcal{K}$, we have
\begin{multline}
	\!\!\!\!\!\!\sum\limits_{j\neq k}\alpha p_j^{{\rm ul}}\mv{h}_j\mv{h}_j^H+{\rm diag}(q_1^{{\rm ul}}(\alpha \mv{p}^{{\rm ul}}),\ldots,q_M^{{\rm ul}}(\alpha \mv{p}^{{\rm ul}}))+\sigma^2\mv{I} \\ \prec \alpha\left(\sum\limits_{j\neq k} p_j^{{\rm ul}}\mv{h}_j\mv{h}_j^H+{\rm diag}(q_1^{{\rm ul}}( \mv{p}^{{\rm ul}}),\ldots,q_M^{{\rm ul}}( \mv{p}^{{\rm ul}}))+\sigma^2\mv{I}\right).
\end{multline}
Based on (\ref{eqn:interference function case I}), it follows that $I_k(\alpha \mv{p}^{{\rm ul}})<\alpha I_k(\mv{p}^{{\rm ul}})$, $\forall k$, given $\alpha>1$.

Last, if $\bar{\mv{p}}^{{\rm ul}}\geq \mv{p}^{{\rm ul}}$, then based on (\ref{eqn:optimal quantization Case I}), it follows that $q_m^{{\rm ul}}( \bar{\mv{p}}^{{\rm ul}})\geq q_m^{{\rm ul}}(\mv{p}^{{\rm ul}})$, $\forall m$. As a result, $\forall k \in \mathcal{K}$, we have
\begin{multline}
	\!\!\!\sum\limits_{j\neq k} \bar{p}_j^{{\rm ul}}\mv{h}_j\mv{h}_j^H+{\rm diag}(q_1^{{\rm ul}}(\bar{\mv{p}}^{{\rm ul}}),\ldots,q_M^{{\rm ul}}(\bar{\mv{p}}^{{\rm ul}}))+\sigma^2\mv{I} \\ \succeq \sum\limits_{j\neq k} p_j^{{\rm ul}}\mv{h}_j\mv{h}_j^H+{\rm diag}(q_1^{{\rm ul}}( \mv{p}^{{\rm ul}}),\ldots,q_M^{{\rm ul}}( \mv{p}^{{\rm ul}}))+\sigma^2\mv{I}.
\end{multline}
Based on (\ref{eqn:interference function case I}), it follows that $I_k(\bar{\mv{p}}^{{\rm ul}})> I_k(\mv{p}^{{\rm ul}})$, $\forall k$.

Lemma \ref{lemmacaseI} is thus proved.

\subsection{Proof of Lemma \ref{lemmacaseIII}}\label{appendix9}
First, it can be easily shown that if $\mv{p}^{{\rm ul}}\geq \mv{0}$, then $I_k(\mv{p}^{{\rm ul}})>0$, $\forall k \in \mathcal{K}$. Next, given $\alpha>1$, it follows from Lemma \ref{lemma5} that $\{q_m^{{\rm ul}}(\mv{p}^{{\rm ul}})\}$ defined in (\ref{eqn:optimal solution Case III}) and (\ref{eqn:optimal solution Case III 1}) satisfies $q_m^{{\rm ul}}(\alpha \mv{p}^{{\rm ul}})<\alpha q_m^{{\rm ul}}(\mv{p}^{{\rm ul}})$, $\forall m \in \mathcal{M}$. Then, similar to Appendix \ref{appendix8}, it can be shown that $\{I_k(\mv{p}^{{\rm ul}})\}$ defined in (\ref{eqn:interference function case III}) satisfies $I_k(\alpha \mv{p}^{{\rm ul}})<\alpha I_k(\mv{p}^{{\rm ul}})$, $\forall k \in \mathcal{K}$, given $\alpha>1$. Last, according to Lemma \ref{lemma5}, if $\bar{\mv{p}}^{{\rm ul}}\geq \mv{p}^{{\rm ul}}$, then $q_m^{{\rm ul}}(\bar{\mv{p}}^{{\rm ul}})\geq q_m^{{\rm ul}}(\mv{p}^{{\rm ul}})$, $\forall m \in \mathcal{M}$. Similar to Appendix \ref{appendix8}, it can be shown that $I_k(\bar{\mv{p}}^{{\rm ul}})\geq I_k(\mv{p}^{{\rm ul}})$, $\forall k \in \mathcal{K}$.

Lemma \ref{lemmacaseIII} is thus proved.

\end{appendix}
\bibliographystyle{IEEEtran}
\bibliography{CIC}

\begin{thebibliography}{10}
\providecommand{\url}[1]{#1}
\csname url@samestyle\endcsname
\providecommand{\newblock}{\relax}
\providecommand{\bibinfo}[2]{#2}
\providecommand{\BIBentrySTDinterwordspacing}{\spaceskip=0pt\relax}
\providecommand{\BIBentryALTinterwordstretchfactor}{4}
\providecommand{\BIBentryALTinterwordspacing}{\spaceskip=\fontdimen2\font plus
\BIBentryALTinterwordstretchfactor\fontdimen3\font minus
  \fontdimen4\font\relax}
\providecommand{\BIBforeignlanguage}[2]{{%
\expandafter\ifx\csname l@#1\endcsname\relax
\typeout{** WARNING: IEEEtran.bst: No hyphenation pattern has been}%
\typeout{** loaded for the language `#1'. Using the pattern for}%
\typeout{** the default language instead.}%
\else
\language=\csname l@#1\endcsname
\fi
#2}}
\providecommand{\BIBdecl}{\relax}
\BIBdecl

\bibitem{duality1}
S.~Vishwanath, N.~Jinda, and A.~Goldsmith, ``{Duality, achievable rates, and
  sum-rate capacity of Gaussian MIMO broadcast channels},'' \emph{IEEE Trans.
  Inf. Theory}, vol.~49, no.~10, pp. 2658--2668, Oct. 2003.

\bibitem{Irmer11}
R.~{Irmer}, H.~{Droste}, P.~{Marsch}, M.~{Grieger}, G.~{Fettweis}, S.~{Brueck},
  H.~{Mayer}, L.~{Thiele}, and V.~{Jungnickel}, ``{Coordinated multipoint:
  Concepts, performance, and field trial results},'' \emph{IEEE Commun. Mag.},
  vol.~49, no.~2, pp. 102--111, Feb. 2011.

\bibitem{Kerpez96}
K.~J. Kerpez, ``{A radio access system with distributed antennas},'' \emph{IEEE
  Trans. Veh. Technol.}, vol.~45, no.~2, pp. 265--275, May 1996.

\bibitem{Simeone16}
O.~Simeone, A.~Maeder, M.~Peng, O.~Sahin, and W.~Yu, ``{Cloud radio access
  network: Virtualizing wireless access for dense heterogeneous systems},''
  \emph{J. Commun. Netw.}, vol.~18, no.~2, pp. 135--149, Apr. 2016.

\bibitem{Ngo17}
H.~Q. Ngo, A.~Ashikhmin, H.~Yang, E.~G. Larsson, and T.~L. Marzetta,
  ``{Cell-free massive MIMO versus small cells},'' \emph{IEEE Trans. Wireless
  Commun.}, vol.~16, no.~3, pp. 1834--1850, Mar. 2017.

\bibitem{duality6}
F.~Rashid-Farrokhi, K.~J.~R. Liu, and L.~Tassiulas, ``{Transmit beamforming and
  power control for cellular wireless systems},'' \emph{IEEE J. Sel. Areas
  Commun.}, vol.~16, no.~8, pp. 1437--1450, Oct. 1998.

\bibitem{duality7}
H.~Boche and M.~Schubert, ``{A general duality theory for uplink and downlink
  beamforming},'' in \emph{Proc. IEEE Veh. Tech. Conf. (VTC)}, Sep. 2002.

\bibitem{duality8}
A.~Wiesel, Y.~C. Eldar, and S.~Shamai, ``{Linear precoding via conic
  optimization for fixed MIMO receivers},'' \emph{IEEE Trans. Signal Process.},
  vol.~54, no.~1, pp. 161--176, Jan. 2006.

\bibitem{duality11}
M.~Schubert and H.~Boche, ``{Solution of the multiuser downlink beamforming
  problem with individual SINR constraints},'' \emph{IEEE Trans. Veh.
  Technol.}, vol.~53, no.~1, pp. 18--28, Jan. 2004.

\bibitem{duality9}
B.~Song, R.~L. Cruz, and B.~D. Rao, ``{Network duality for multiuser MIMO
  beamforming networks and applications},'' \emph{IEEE Trans. Commun.},
  vol.~55, no.~3, pp. 618--630, Mar. 2007.

\bibitem{duality3}
P.~Viswanath and D.~N.~C. Tse, ``{Sum capacity of the vector Gaussian broadcast
  channel and uplink-downlink duality},'' \emph{IEEE Trans. Inf. Theory},
  vol.~49, no.~8, pp. 1912--1921, Aug. 2003.

\bibitem{duality4}
W.~Yu, ``{Uplink-downlink duality via minimax duality},'' \emph{IEEE Trans.
  Inf. Theory}, vol.~52, no.~2, pp. 361--374, Feb. 2006.

\bibitem{Marsch11}
P.~Marsch and G.~Fettweis, ``{Uplink CoMP under a constrained backhaul and
  imperfect channel knowledge},'' \emph{IEEE Trans. Wireless Commun.}, vol.~10,
  no.~6, pp. 1730--1742, June 2011.

\bibitem{Yu13}
L.~Zhou and W.~Yu, ``{Uplink multicell processing with limited backhaul via
  per-base-station successive interference cancellation},'' \emph{IEEE J. Sel.
  Areas Commun.}, vol.~31, no.~10, pp. 1981--1993, Oct. 2013.

\bibitem{Gerhard07}
P.~Marsch and G.~Fettweis, ``{A framework for optimizing the uplink performance
  of distributed antenna systems under a constrained backhaul},'' in
  \emph{Proc. IEEE Int. Conf. Commun. (ICC)}, Jun. 2007.

\bibitem{Zhou14}
Y.~Zhou and W.~Yu, ``{Optimized backhaul compression for uplink cloud radio
  access network},'' \emph{IEEE J. Sel. Areas Commun.}, vol.~32, no.~6, pp.
  1295--1307, June 2014.

\bibitem{Yu16}
Y.~Zhou, Y.~Xu, W.~Yu, and J.~Chen, ``{On the optimal fronthaul compression and
  decoding strategies for uplink cloud radio access networks},'' \emph{IEEE
  Trans. Inf. Theory}, vol.~62, no.~2, pp. 7402--7418, Dec. 2016.

\bibitem{Zhou16}
Y.~Zhou and W.~Yu, ``{Fronthaul compression and transmit beamforming
  optimization for multi-antenna uplink C-RAN},'' \emph{IEEE Trans. Signal
  Process.}, vol.~64, no.~16, pp. 4138--4151, Aug. 2016.

\bibitem{Liu15}
L.~Liu, S.~Bi, and R.~Zhang, ``{Joint power control and fronthaul rate
  allocation for throughput maximization in OFDMA-based cloud radio access
  network},'' \emph{IEEE Trans. Commun.}, vol.~63, no.~11, pp. 4097--4110, Nov.
  2015.

\bibitem{Liang15}
L.~Liu and R.~Zhang, ``{Optimized uplink transmission in multi-antenna C-RAN
  with spatial compression and forward},'' \emph{IEEE Trans. Signal Process.},
  vol.~63, no.~19, pp. 5083--5095, Oct. 2015.

\bibitem{Simeone13}
S.~H. Park, O.~Simeone, O.~Sahin, and S.~Shamai, ``{Joint precoding and
  multivariate backhaul compression for the downlink of cloud radio access
  networks},'' \emph{IEEE Trans. Signal Process.}, vol.~61, no.~22, pp.
  5646--5658, Nov. 2013.

\bibitem{Bashar18}
M.~Bashar, K.~Cumanan, A.~G. Burr, H.~Q. Ngo, and M.~Debbah, ``{Cell-free
  massive MIMO with limited backhaul},'' in \emph{Proc. IEEE Int. Conf. Commun.
  (ICC)}, May 2018.

\bibitem{Marzetta17a}
E.~Nayebi, A.~Ashikhmin, T.~L. Marzetta, H.~Yang, and B.~D. Rao, ``{Precoding
  and power optimization in cell-free massive MIMO systems},'' \emph{IEEE
  Trans. Wireless Commun.}, vol.~16, no.~7, pp. 4445--4459, Jul. 2017.

\bibitem{Marzetta17b}
M.~Bashar, K.~Cumanan, A.~G. Burr, H.~Q. Ngo, M.~Debbah, and P.~Xiao,
  ``{Max-min rate of cell-free massive MIMO uplink with optimal uniform
  quantization},'' \emph{IEEE Trans. Wireless Commun.}, vol.~67, no.~10, pp.
  6796--6815, Oct. 2019.

\bibitem{dai2014sparse}
B.~Dai and W.~Yu, ``Sparse beamforming and user-centric clustering for downlink
  cloud radio access network,'' \emph{IEEE Access}, vol.~2, pp. 1326--1339,
  2014.

\bibitem{powercontrol}
G.~J. Foschini, ``{A simple distributed autonomous power control algorithm and
  its convergence},'' \emph{IEEE Trans. Veh. Technol.}, vol.~42, no.~4, pp.
  641--646, Nov. 1993.

\bibitem{Cover}
T.~Cover and J.~Thomas, \emph{{Elements of Information Theory}}.\hskip 1em plus
  0.5em minus 0.4em\relax New York: Wiley, 1991.

\bibitem{DPC}
M.~Costa, ``{Writing on dirty paper},'' \emph{IEEE Trans. Inf. Theory}, vol.
  IT-29, no.~3, pp. 439--441, May 1983.

\bibitem{duality10}
W.~Yu and T.~Lan, ``{Transmitter optimization for the multi-antenna downlink
  with per-antenna power constraints},'' \emph{IEEE Trans. Signal Process.},
  vol.~55, no.~6, pp. 2646--2660, Jun. 2007.

\bibitem{duality12}
L.~Zhang, R.~Zhang, Y.-C. Liang, Y.~Xin, and H.~V. Poor, ``{On Gaussian MIMO
  BC-MAC duality with multiple transmit covariance constraints},'' \emph{IEEE
  Trans. Inf. Theory}, vol.~58, no.~4, pp. 2064--2078, Apr. 2012.

\bibitem{An1}
A.~Liu, Y.~Liu, H.~Xiang, and W.~Luo, ``Polite water-filling for weighted
  sum-rate maximization in {MIMO B-MAC} networks under multiple linear
  constraints,'' \emph{IEEE Trans. Signal Process.}, vol.~60, no.~2, pp.
  834--847, Feb. 2012.

\bibitem{An2}
------, ``{Iterative polite water-filling for weighted sum-rate maximization in
  iTree networks},'' in \emph{Proc. IEEE Globecom}, Dec. 2010.

\bibitem{duality14}
W.~He, B.~Nazer, and S.~Shamai, ``{Uplink-downlink duality for
  integer-forcing},'' \emph{IEEE Trans. Inf. Theory}, vol.~64, no.~3, pp.
  1992--2011, Mar. 2018.

\bibitem{Tsung-Hui18}
T.-H. Chang, Y.-F. Liu, and S.-C. Lin, ``{QoS-based linear transceiver
  optimization for full-duplex multi-user communications},'' \emph{IEEE Trans.
  Signal Process.}, vol.~66, no.~9, pp. 2300--2313, May 2018.

\bibitem{duality16}
S.~A. Jafar, K.~S. Gomadam, and C.~Huang, ``{Duality and rate optimization for
  multiple access and broadcast channels with amplify-and-forward relays},''
  \emph{IEEE Trans. Inf. Theory}, vol.~53, no.~10, pp. 3350--3370, Oct. 2007.

\bibitem{duality13}
K.~S. Gomadam and S.~A. Jafar, ``{Duality of MIMO multiple access channel and
  broadcast channel with amplify-and-forward relays},'' \emph{IEEE Trans. Inf.
  Theory}, vol.~58, no.~1, pp. 211--217, Jan. 2010.

\bibitem{duality15}
Y.~Rong and M.~R.~A. Khandaker, ``{On uplink-downlink duality of multi-hop MIMO
  relay channel},'' \emph{IEEE Trans. Wireless Commun.}, vol.~10, no.~6, pp.
  1923--1931, Jun. 2011.

\bibitem{An3}
A.~Liu, V.~Lau, and Y.~Liu, ``Duality and optimization for generalized
  multi-hop {MIMO} amplify-and-forward relay networks with linear
  constraints,'' \emph{IEEE Trans. Signal Process.}, vol.~61, no.~9, pp.
  2356--2365, May 2013.

\bibitem{Yafeng13}
Y.-F. Liu, M.~Hong, and Y.-H. Dai, ``{Max-Min Fairness linear transceiver
  design problem for a SIMO interference channel is polynomial time
  solvable},'' \emph{IEEE Signal Process. Lett.}, vol.~20, no.~1, pp. 27--30,
  Jan. 2013.

\bibitem{Liang16}
L.~Liu, P.~Patil, and W.~Yu, ``{An uplink-downlink duality for cloud radio
  access network},'' in \emph{Proc. IEEE Int. Symp. Inf. Theory (ISIT)}, Jul.
  2016.

\bibitem{Boyd04}
S.~Boyd and L.~Vandenberghe, \emph{Convex Optimization}.\hskip 1em plus 0.5em
  minus 0.4em\relax Cambridge University Press, 2004.

\bibitem{Gamal}
A.~E. Gamal and Y.-H. Kim, \emph{Network Information Theory}.\hskip 1em plus
  0.5em minus 0.4em\relax Cambridge University Press, 2011.

\bibitem{Yu_CRAN_book}
W.~Yu, P.~Patil, B.~Dai, and Y.~Zhou, ``{Cooperative beamforming and resource
  optimization in C-RAN},'' in \emph{Cloud Radio Access Networks Principles,
  Technologies, and Applications}, T.~Quek, M.~Peng, O.~Simeone, and W.~Yu,
  Eds.\hskip 1em plus 0.5em minus 0.4em\relax Cambridgea, U.K.: Cambridge
  University Press, 2017, ch.~4, pp. 54--81.

\bibitem{Kim_NNC}
S.~H. Lim, Y.-H. Kim, A.~E. Gamal, and S.-Y. Chung, ``Noisy network coding,''
  \emph{IEEE Trans. Inf. Theory}, vol.~57, no.~5, pp. 3132--3152, May 2011.

\bibitem{Kim_DDF}
S.~H. Lim, K.~T. Kim, and Y.-H. Kim, ``Distributed decode-forward for relay
  networks,'' \emph{IEEE Trans. Inf. Theory}, vol.~63, no.~7, pp. 4103--4118,
  Jul. 2017.

\bibitem{Yu19}
P.~Patil and W.~Yu, ``Generalized compression strategy for the downlink cloud
  radio access network,'' \emph{IEEE Trans. Inf. Theory}, vol.~65, no.~10, pp.
  6766--6780, Oct. 2019.

\bibitem{Sanderovich08}
A.~Sanderovich, S.~Shamai, Y.~Steinberg, and G.~Kramer, ``Communication via
  decentralized processing,'' \emph{IEEE Trans. Inf. Theory}, vol.~54, no.~7,
  pp. 3008--3023, Jul. 2008.

\bibitem{Pratik_hybrid}
P.~Patil, B.~Dai, and W.~Yu, ``Hybrid data-sharing and compression strategy for
  downlink cloud radio access network,'' \emph{IEEE Trans. Commun.}, vol.~66,
  no.~11, pp. 5370--5384, Nov. 2018.

\bibitem{Yates95}
R.~D. Yates, ``{A framework for uplink power control in cellular radio
  systems},'' \emph{IEEE J. Sel. Areas Commun.}, vol.~13, no.~7, pp.
  1341--1347, Sep. 1995.

\bibitem{Horn12}
R.~Horn and C.~Johnson, \emph{Matrix Analysis}.\hskip 1em plus 0.5em minus
  0.4em\relax Second Edition, Cambridge University Press, 2012.

\end{thebibliography}

\begin{IEEEbiographynophoto}
{Liang Liu} (S'14-M'15) received the B.Eng. degree from the Tianjin University, China, in 2010, and the Ph.D. degree from the National University of Singapore in 2014. He is currently an Assistant Professor in the Department of Electronic and Information Engineering at the Hong Kong Polytechnic University. Before that, he was a Research Fellow in the Department of Electrical and Computer Engineering at National University of Singapore from 2017 to 2018, and a Postdoctoral Fellow in the Department of Electrical and Computer Engineering at University of Toronto from 2015 to 2017.
	
His research interests include the next generation cellular technologies and machine-type communications for Internet of Things. He was the recipient of the IEEE Signal Processing Society Young Author Best Paper Award, 2017, and a Best Paper Award from IEEE WCSP in 2011. He is recognized by Clarivate Analytics as a Highly Cited Researcher, 2018.
\end{IEEEbiographynophoto}

\begin{IEEEbiographynophoto}
{Ya-Feng Liu} (M'12--SM'18) received the B.Sc. degree in applied mathematics from Xidian University, Xi'an, China, in 2007, and the Ph.D. degree in computational mathematics from the Chinese Academy of Sciences (CAS), Beijing, China, in 2012. During his Ph.D. study, he was supported by the Academy of Mathematics and Systems Science (AMSS), CAS, to visit Professor Zhi-Quan (Tom) Luo at the University of Minnesota (Twins Cities) from 2011 to 2012. After his graduation, he joined the Institute of Computational Mathematics and Scientific/Engineering Computing, AMSS, CAS, Beijing, China, in 2012, where he became an Associate Professor in 2018. His main research interests are nonlinear optimization and its applications to signal processing, wireless communications, and machine learning.
	
Dr. Liu currently serves an Editor for the IEEE Transactions on Wireless Communications and an Associate Editor for the IEEE Signal Processing Letters and the Journal of Global Optimization. He is an elected member of the Signal Processing for Communications and Networking Technical Committee (SPCOM-TC) of the IEEE Signal Processing Society. He received the Best Paper Award from the IEEE International Conference on Communications (ICC) in 2011, the Best Student Paper Award from the International Symposium on Modeling and Optimization in Mobile, Ad Hoc and Wireless Networks (WiOpt) in 2015, the Chen Jingrun Star Award from the AMSS in 2018, the Science and Technology Award for Young Scholars from the Operations Research Society of China in 2018, and the 15th IEEE ComSoc Asia-Pacific Outstanding Young Researcher Award in 2020.
	
\end{IEEEbiographynophoto}

%
\begin{IEEEbiographynophoto}
{Pratik Patil}
	received the B.Tech.~degree in Electronics and Communication Engineering from the Indian Institute of Technology Guwahati, India,
	the M.A.Sc.~and M.Sc.~degrees in Electrical and Computer Engineering and Statistics from the University of Toronto, Canada,
	and the M.S.~degree in Machine Learning from Carnegie Mellon University, U.S.A., where he is currently pursuing Ph.D.~in Statistics and Machine Learning. His research interests lie broadly in statistics, information theory, optimization, machine learning, signal processing, and wireless communications.
\end{IEEEbiographynophoto}

\begin{IEEEbiographynophoto}
{Wei Yu} (S'97--M'02--SM'08--F'14) received the B.A.Sc. degree in computer engineering and mathematics from the University of Waterloo, Waterloo, ON, Canada, in 1997, and the M.S. and Ph.D. degrees in electrical engineering from Stanford University, Stanford, CA, USA, in 1998 and 2002, respectively. Since 2002, he has been with the Electrical and Computer Engineering Department, University of Toronto, Toronto, ON, Canada, where he is currently a Professor and holds the Canada Research Chair (Tier 1) in Information Theory and Wireless Communications.

Prof. Wei Yu is the President of the IEEE Information Theory Society in 2021, and has served on its Board of Governors since 2015. He is a Fellow of the Canadian Academy of Engineering and a member of the College of New Scholars, Artists, and Scientists of the Royal Society of Canada. He received the Steacie Memorial Fellowship in 2015, the IEEE Marconi Prize Paper Award in Wireless Communications in 2019, the IEEE Communications Society Award for Advances in Communication in 2019, the IEEE Signal Processing Society Best Paper Award in 2017 and 2008, the Journal of Communications and Networks Best Paper Award in 2017, and the IEEE Communications Society Best Tutorial Paper Award in 2015. He served as the Chair of the Signal Processing for Communications and Networking Technical Committee of the IEEE Signal Processing Society from 2017 to 2018. He was an IEEE Communications Society Distinguished Lecturer from 2015 to 2016. He is currently an Area Editor of the IEEE Transactions on Wireless Communications, and has in the past served as an Associate Editor for IEEE Transactions on Information Theory, IEEE Transactions on Communications, and IEEE Transactions on Wireless Communications.
\end{IEEEbiographynophoto}

\end{document}